\documentclass{ws-ijmpa}
\usepackage[super,compress]{cite}

\usepackage[utf8]{inputenc}
\usepackage[english]{babel}
\usepackage{graphicx}
\usepackage{amsmath}
\usepackage{amssymb}
\usepackage{subfiles}  
\usepackage{wrapfig}
\usepackage{multicol}

\usepackage{xr-hyper}
\usepackage[pdftex,bookmarks=true]{hyperref}
\hypersetup{colorlinks,urlcolor=black,citecolor=black,linkcolor=black,filecolor=black}
\usepackage{cleveref}
\Crefformat{equation}{Eq.~(#2#1#3)}
\Crefmultiformat{equation}{Eqs.~(#2#1#3)}{ and~(#2#1#3)}{, (#2#1#3)}{ and~(#2#1#3)}
\Crefformat{figure}{Fig.~#2#1#3}
\Crefformat{section}{Sec.~#2#1#3}

\def\sgc{\Sigma_c}
\def\omc{\Omega_c}
\def\occ{\Omega_{cc}}
\def\xcc{\Xi_{cc}}
\def\xccs{\Xi_{cc}^\ast}
\def\om{\Omega}
\def\ocs{\Omega_{c}^\ast}
\def\occs{\Omega_{cc}^\ast}
\def\occc{\Omega_{ccc}}

\def\mcb{{\mathcal{B}}}
\def\mcbp{{\mathcal{B}^\ast}}
\def\pp{p^\prime}
\def\ssp{s^\prime}

\def\spinoh{spin-1/2 }
\def\spinth{spin-3/2 }
\def\1212{\mathrm{\spinoh} \rightarrow \mathrm{\spinoh}}
\def\3232{\mathrm{\spinth} \rightarrow \mathrm{\spinth}}
\def\bsgb{\mcb \gamma \rightarrow \mcbp}
\def\bpgb{\mcb \gamma \rightarrow \mcb^\prime}
\def\ogo{\Omega_c^\ast \rightarrow \Omega_c \gamma}
\def\oog{\Omega_c \gamma \rightarrow \Omega_c^\ast}

\def\octoocs{\Omega_c \gamma \rightarrow \Omega_c^\ast}
\def\occtooccs{\Omega_{cc}^+ \gamma \to \Omega_{cc}^{\ast+}}
\def\xcctoxccs{\Xi_{cc} \gamma \to {\Xi_{cc}^\ast}}
\def\xcctoxccsP{\Xi_{cc}^{+} \gamma \to \Xi_{cc}^{\ast+}}
\def\xcctoxccsPP{\Xi_{cc}^{++} \gamma \to \Xi_{cc}^{\ast++}}

\def\ensh{\texttt{a09k00}}
\def\enh{\texttt{a09k27}}
\def\enm{\texttt{a09k54}}
\def\enl{\texttt{a09k70}}
\def\ensl{\texttt{a09k81}}

\def\pc{\hphantom{c}}
\def\pz{\hphantom{0}}
\def\ppl{\hphantom{(}}
\def\ppr{\hphantom{)}}

\def\twpz{\pz\pz}
\def\tpz{\pz\pz\pz}
\def\fpz{\pz\pz\pz\pz}
\def\thpz{\ppl\pz\pz\pz\ppr}
\def\etpz{\ppl\pz\pz\ppr}

\def\stpz{\pz\ppl\pz\pz\ppr}
\def\spz{\pz\ppl\pz\pz\pz\ppr}
\def\pneg{\hphantom{-}}
\def\pzp{\hphantom{0}\hphantom{.}}
\def\pnpz{$\hphantom{{}^+_-}$}

\newcommand{\ffac}[2]{F_{#1,{\cal #2}}}
\newcommand{\Feyn}[1]{#1\kern-0.55em/}

\begin{document}
\markboth{K. U. CAN}{LQCD STUDY OF THE EM FORM FACTORS OF CHARMED BARYONS}

%
\catchline{}{}{}{}{}
%

\title{ LATTICE QCD STUDY OF THE ELASTIC AND TRANSITION FORM FACTORS OF CHARMED BARYONS}

\author{ K. U. CAN }

\address{Center for the Subatomic Structure of Matter (CSSM), \\ Department of Physics, The University of Adelaide,\\
Adelaide, SA 5000, Australia \\
kadirutku.can@adelaide.edu.au}
\address{Strangeness Nuclear Physics Laboratory, \\ RIKEN Nishina Center, RIKEN,\\
Wako, Saitama 351-0198, Japan\\
kadirutku.can@riken.jp}

\maketitle

\begin{abstract}
Composite nature of a particle can be probed by electromagnetic interactions and information about their structure is embedded in form factors. Most of the experimental and theoretical efforts on baryon electromagnetic form factors have been focused on nucleon while the data on charmed sector is limited to spectroscopy, and weak and strong decays. Forthcoming experiments with a heavy-hadron physics program at major experimental facilities are expected to provide a wealth of information on charmed baryons, which calls for a better understanding of the heavy-sector dynamics from theoretical grounds. 

We review the progress in calculating the elastic and transition form factors of charmed baryons in lattice QCD. A collection of static observables, \emph{e.g.} charge radii, multipole moments, are presented along with the elastic form factors up to $Q^2 \sim 1.5 {\rm GeV}^2$. As one would expect the charmed baryons are compact in comparison to nucleon and this is due to the presence of valence charm quark(s). The elastic and transition magnetic moments are both suppressed. The lattice results provide predictions for the transition magnetic moments, transition and helicity amplitudes and consequentially the decay widths of some singly and doubly charmed baryons.

In general, lattice results are consonant with the qualitative expectations of quark model and heavy-quark symmetry, although there are apparent quantitative differences up to two orders of magnitude in some cases. There are, however, indications that the lattice results can be utilized to improve the model predictions. Nevertheless, discrepancies between the lattice and non-lattice calculations need to be understood better to have a solid insight into the dynamics of the heavy sector.

Furthermore, reliably determined charmed baryon observables would be invaluable input to investigate the nature of exotic states, which further emphasizes the importance of rigorous, first-principles calculations to advance our understanding of the dynamics of the heavy quarks and strong interactions.

\keywords{charmed baryons, elastic form factor, transition form factor, electric form factor, magnetic form factor, magnetic moment, lattice QCD}
\end{abstract}

\ccode{PACS numbers: 14.20.Lq, 12.38.Gc, 13.40.Gp}

\tableofcontents

\section{Introduction} \label{sec:intro}	

	Experimental efforts and observations spanning roughly the last two decades have shown that the hadron physics is still as relevant as it was in the early stages of particle physics. Several candidates of states that do not fall into the conventional quark model categories have taken the center stage where the majority are believed to have a hidden-charm component~\cite{Liu:2014xi,CHEN20161,LEBED2017143}. Additionally, recent experimental results from the LHCb Collaboration on the excited $\Omega_c$~\cite{PhysRevLett.118.182001} and $\Xi_c$ states~\cite{LHCb:2020iby} and the doubly charmed $\Xi_{cc}$ ground state~\cite{PhysRevLett.119.112001} have put a further emphasis on the relevance of the conventional heavy-hadron physics. To this end, charmed hadrons provide a unique laboratory to study the strong interaction and confinement dynamics due to the diverse compositions of the light and charm valence quarks.

	The focus of this review is the dynamics of the conventional charmed baryons from an electromagnetic perspective. Studying the conventional states is the first step to understanding the heavy sector dynamics and a stepping stone to extend the investigations to exotic states. The ground states of the singly-charmed baryons are already established by experimental studies and the lattice QCD results agree remarkably well with those observations~\cite{Alexandrou:2017xwd,Durr:2012dw,Brown:2014ena,Namekawa:2013vu,Chen:2017kxr,Briceno:2012wt,PhysRevD.92.034504,Padmanath:2013zfa,Padmanath:2015jea,PhysRevD.102.054513}. The $\Xi_{cc}$ is the only observed doubly-charmed baryon for the time being. A $\Xi_{cc}^{+}$ state is claimed to be observed by the SELEX Collaboration~\cite{Mattson:2002vu,Ocherashvili:2004hi} but those results were not confirmed by other experiments until the LHCb Collaboration has reported an observation of its isospin partner $\Xi_{cc}^{++}$~\cite{PhysRevLett.119.112001}. Lattice QCD predictions for the mass of $\Xi_{cc}$ lie above the SELEX reported value but agree very well with the LHCb result (see Ref.~\citen{PhysRevD.102.054513} for a recent comparison). Further charmed states, \emph{e.g.} $\occ$, $\occc$, are expected from the viewpoint of the flavor SU(4) multiplet. 
	The SU(4) symmetry is badly broken, however, due to the large mass of the charm quark; and although its predictions cannot be adopted with confidence, it still provides a useful scheme to guide the efforts in search for additional charmed states.   

	Being composite objects, the distribution and dynamics of the components of baryons (and hadron in general) are of special interest in understanding their properties. Studying the elastic form factors of baryons would reveal their electric charge distributions, magnetization densities or the static properties like their magnetic moments. Transition form factors on the other hand allows a probe of the decay mechanisms. Understandably, motivations and bulk of the experimental results are related to the nucleon, however, the same questions and motivations apply to all baryons and heavier sectors present different opportunities.

	An interesting aspect of heavy quarks is the change of dynamics as $m_Q \to \infty$, known as the heavy-quark spin symmetry (HQSS). As a consequence, spin of the heavy quark decouples from the system since it is proportional to $1/m_Q$ and the system is characterized by the light degrees of freedom. 
	One therefore would naively expect the effects of hyperfine splitting to diminish as $m_Q \to \infty$ such that the masses of the HQSS partner particles with the same quark content but different total spins would get closer. Such a trend is most easily seen in certain meson and baryon mass splittings as illustrated below (evaluated using the PDG averages), 
	\begin{alignat*}{5}
		& m_\rho - m_\pi   && = 630 \, {\rm MeV}, 	&& \qquad &&  m_\Delta - m_N 				    && = 290 \, {\rm MeV}, \\
		& m_{K^\ast} - m_K && = 390 \, {\rm MeV}, 	&& \qquad &&  m_{\Sigma^\ast} - m_{\Sigma} 	    && = 195 \, {\rm MeV}, \\
		& m_{D^\ast} - m_D && = 140 \, {\rm MeV},	&& \qquad &&  m_{\Sigma_c^\ast} - m_{\Sigma_c} 	&&  = 65 \, {\rm MeV}, \\
		& m_{B^\ast} - m_B && =  45 \, {\rm MeV}, 	&& \qquad &&  m_{\Sigma_b^\ast} - m_{\Sigma_b}  &&  = 20 \, {\rm MeV},
	\end{alignat*}
	where the mass differences are decreasing as one of the valence quarks gets heavier. Furthermore, a similar trend is shown to hold for excited states in a quark model calculation~\cite{Yoshida:2015tia}. 
	Effects of the HQSS can be identified in structural observables as well. For instance, the electromagnetic observables of pseudoscalar and vector mesons tend to coincide as the light quark gets heavier~\cite{Can:2012tx}, which one would expect since the contribution of the spin-spin interaction $\left( \vec{\sigma}_Q \cdot \vec{\sigma}_q / m_Q m_q \right)$ would decrease with increasing $m_q$.

	Forthcoming experiments with a heavy-hadron physics program at major experimental facilities, \emph{e.g.} J-PARC, SuperKEKB, BES-III \emph{etc.}, are expected to provide a wealth of information which calls for a better understanding of the heavy-sector dynamics from theoretical grounds. However, since perturbative QCD breaks down at the typical hadronic energy scale, other methods that can incorporate the non-perturbative effects are needed. 

	Lattice QCD~\cite{PhysRevD.10.2445} is currently the only known method that is free from any assumptions or model-dependencies at Lagrangian level to quantitatively and systematically study the non-perturbative aspects of strong interactions from first principles. It has matured to a state-of-the-art level thanks to the evolution of the computer technology and significant improvements on numerical algorithms. In a nutshell, the space-time continuum is discretized and the QCD Lagrangian is reformulated in Euclidean space to compute the correlation functions defined in a Feynman path integral formalism via numerical Monte Carlo methods. There is a close connection between the lattice approach and the statistical mechanics where one obtains the observables by averaging over QCD vacuum configurations. Currently, it is possible to make fully dynamical simulations, \emph{i.e.} including the fermionic vacuum fluctuations, with physical quark masses and controlled extrapolations to continuum limit. 

	From a hadron physics point of view, most of the modern lattice QCD computation efforts converge on the spectroscopy of light and heavy hadrons and hadron-hadron interactions~\cite{Prelovsek:2014zga,Aoki:2012tk,Doi:2011gq,Gongyo:2020pyy,HALQCD:2015qmg,HALQCD:2016ofq,HALQCD:2017xsa,HALQCD:2018qyu,HALQCD:2019wsz,Ikeda:2017mee,Inoue:2010es,Inoue:2010hs,Inoue:2011ai,PhysRev.60.61,Lyu:2021qsh} and structure calculations of light pseudoscalar and vector mesons and nucleon~\cite{Constantinou:2014tga}. There is also a tremendous effort on calculating the meson decay constants and CKM matrix elements~\cite{Aoki:2019cca}. Elastic and transition form factors of low-lying octet~\cite{PhysRevD.74.093005,Shanahan:2014uka,Shanahan:2014cga} and decuplet~\cite{PhysRevD.80.054505,Alexandrou:2008bn,Alexandrou:2010jv} baryons are also studied. Lattice QCD calculations in the charmed baryon sector is mostly limited to spectroscopy calculations~\cite{Durr:2012dw,Briceno:2012wt,Namekawa:2013vu,Padmanath:2013bla,Padmanath:2013zfa,Brown:2014ena,Alexandrou:2014sha,Padmanath:2014bxa,Padmanath:2015bra,Padmanath:2015jea,PhysRevD.92.034504,Chen:2017kxr,Alexandrou:2017xwd} and semileptonic decays~\cite{Meinel:2016dqj,Zhang:2021oja}, with recent interest in tetraquarks~\cite{HALQCD:2016ofq,Ikeda:2017mee} and hadron-hadron interactions~\cite{Lyu:2021qsh}. Electromagnetic interactions of charmed mesons, on the other hand, have already been studied extensively in lattice QCD~\cite{Abada:2002xe, Becirevic:2002vp, Becirevic:2009xp, Becirevic:2012dc, Becirevic:2012pf, Becirevic:2013bsa, Becirevic:2014rda}. 

	What has been missing, however, was a systematic study on the charmed baryon electromagnetic form factors and observables. Several works~\cite{Can:2013zpa,Can:2013tna,Bahtiyar2015281,Can:2015exa,Bahtiyar:2016dom,PhysRevD.98.114505}, which form the basis of this review, by the members of the TRJQCD Collaboration investigate this sector and provide valuable information on the internal dynamics of observed and yet unobserved charmed baryons. Widths of the experimentally seen but unmeasured electromagnetic decays of positive parity $\Xi_c^\prime$ and $\Omega_c^\ast$ baryons are predicted. 
	There are many early and recent non-lattice works available in the literature investigating the electromagnetic interactions of charmed baryons. These include variations of quark model approaches~\cite{Lichtenberg:1976fi,Barik:1984tq,Oh:1991ws,Dey:1994qi,Ivanov:1996fj,SilvestreBrac:1996bg,Ivanov:1999bk,JuliaDiaz:2004vh,Kumar:2005ei,Faessler:2006ft,Albertus:2006ya,Patel:2007gx,Branz:2010pq,Sharma:2010vv,Wang:2017hej,Wang:2017kfr,Gandhi:2018lez}; bag model~\cite{Hackman:1977am,Bernotas:2012nz,Bernotas:2013eia,Simonis:2018rld}; QCD sum rules~\cite{Zhu:1997as,Zhu:1998ih,Sharma:2010vv,Aliev:2014bma,Aliev:2016xvq,Ozdem:2018uue,Aliev:2021hqq}; (heavy baryon) chiral perturbation theory~\cite{Cheng:1992xi,Banuls:1999br,Jiang:2015xqa,Li:2017pxa,Li:2017cfz,Meng:2017dni,Liu:2018euh,Meng:2018zbl,Wang:2018cre,Li:2020uok}; effective field theory~\cite{Wang:2018gpl,Meng:2018gan,Xiabng:2018qsd}, and chiral quark-soliton model~\cite{Yang:2018uoj,Kim:2018nqf,Kim:2019wbg,Yang:2019tst,Kim:2020uqo,Kim:2021xpp} calculations regarding the elastic and transition form factors.
	A striking outcome of all the lattice works~\cite{Can:2013zpa,Can:2013tna,Bahtiyar2015281,Can:2015exa,Bahtiyar:2016dom,PhysRevD.98.114505} is the uncovered discrepancies between the lattice and non-lattice methods for charmed observables.      

	We have organized this review as follows: a short introduction to lattice QCD formulation is given in \Cref{sec:lqcd}. The theoretical formalism for an essential part of the hadronic calculations, the two-point correlation functions, is presented in \Cref{sec:mass}. The theoretical formalism for accessing the electromagnetic matrix elements and the elastic and transition form factors from the three-point functions are given in \Cref{sec:emff,sec:rad_ff}. Results are discussed extensively and compared to available non-lattice methods with comments on possible reasons of discrepancies. \Cref{sec:sum} summarizes the findings and presents the conclusions of the reviewed works.

\section{Lattice QCD} \label{sec:lqcd}

	\subsection{Euclidean formulation} \label{sec:ea}
	In Feynman's path integral formalism, expectation value of a physical observable is given by a functional integral,
\begin{equation}
	\label{eq:obs_pthi}
	\langle \hat{\mathcal{O}} \rangle = \frac{1}{Z} \int \mathcal{D}\left[ \bar\psi, \psi, A \right] e^{i\mathcal{S}_{QCD}[\bar\psi, \psi, A]} \mathcal{O}\left[ \bar\psi, \psi, A \right],
\end{equation}
where the integration measure $\mathcal{D}[\cdots]$ the product measure of quark and gauge fields.
The QCD action in the Minkowski space reads,
\begin{equation}
	\mathcal{S}^M_{QCD}[\psi, \bar{\psi}, A]= \int d^4x \left\{ \sum_q \bar{\psi}_q \left(i  \gamma^\mu D_\mu - m_q \right) \psi_q - \frac{1}{4} F^a_{\mu \nu}F_a^{\mu\nu} \right\},
\end{equation}
where the covariant derivative is $D_\mu = \partial_\mu - ig A^a_\mu t^a$ and the field strength tensor is $-igF_{\mu\nu} = [D_\mu, D_\nu]$. The partition function is defined as,
\begin{equation}
	Z = \int \mathcal{D}\left[ \bar\psi, \psi, A \right] e^{i\mathcal{S}_{QCD}[\bar\psi, \psi, A]}.
\end{equation}

In the lattice approach, one evaluates the integral in \Cref{eq:obs_pthi} numerically by means of Monte Carlo methods. However, the sampling weight, $e^{i\mathcal{S}_{QCD}[\bar\psi, \psi, A]}$, is highly oscillatory which renders a reliable numerical treatment rather challenging. A Wick rotation, \emph{i.e.} $t \to it$, transforms the action $i S^M \to - S^E$ allowing a connection to the statistical mechanics. Corresponding Euclidean QCD action in continuum formalism then takes the form,
\begin{equation}
	S^E_{QCD}[\psi, \bar{\psi}, A] = \int d^4x \left\{ \sum_q \bar{\psi}_q \left(\gamma^\mu D_\mu + m_q \right) \psi_q + \frac{1}{4} F^a_{\mu \nu}F_a^{\mu\nu} \right\},
\end{equation}
where the Dirac gamma matrices satisfy $\left\{ \gamma_\mu, \gamma_\nu \right\} = 2 \delta_{\mu\nu}$. 

The gauge and fermionic parts of the QCD action needs to be discretized and improved to keep the discretizations error at a controllable level while ensuring that the continuum limit is left intact. There are several discretized forms of the action each with its own merit but to be concise we only cover the ones that are relevant to this paper. 
 	
	\subsection{Gauge action} \label{sec:ga}
	An improved action for the gauge sector is formed by adding the dimension-6\footnote{In principle one should start with adding the dimension-5 operators however no gauge-invariant dimension-5 operator can be formed by using the link variables only.} operators to the standard Wilson plaquette action,
\begin{equation}
	S_G[U] = \frac{\beta}{6} \left( c_0 \sum_{n, \mu < \nu} U^{(4)}_{\mu\nu}(n) + \sum_{i=1}^3 c_i \sum_{n, \mu < \nu} U^{(6)}_{\mu\nu}(n) \right),
\end{equation} 
where the first term is the dimension-4 Wilson loop (plaquette) and the summed terms are the dimension-6 rectangular, twisted and L-shaped Wilson loops, respectively. $\beta = 6/g^2$ is the inverse coupling. 

L\"{u}scher and Weisz show that in the tree-level approximation, coefficients of the twisted and L-shaped contributions can be set to $c_2 = c_3 = 0$. Furthermore, they study the 1-loop corrections and introduce a 1-loop improved action~\cite{Luscher:1984xn} with $\mathcal{O}(g^4 a^4)$ corrections which has the coefficients $c_0(g^2) = 5/3 + 0.2370 g^2$, $c_1(g^2) = -1/12 - 0.02521g^2$, $c_2(g^2) = -0.00441 g^2$ and $c_3(g^2) = 0$. 

On the other hand, Iwasaki~\cite{Iwasaki:2011jk} includes the rectangular term only and uses an approximate block-spin renormalization group analysis of Wilson loops to determine $c_1 = -0.331$. Coefficient of the plaquette term satisfies the normalization condition, $c_0 = 1 - 8c_1 = 3.648$. 
 
	\subsection{Light and heavy quark actions} \label{sec:fa}
	Wilson's formulation of the fermion action has $\mathcal{O}(a)$ discretization errors, where $a$ is the lattice spacing. An improved discretized action is necessary to demote these errors to $\mathcal{O}(a^2)$. Following Symanzik's improvement program~\cite{CURCI1983205,Symanzik1982,SYMANZIK1983187,SYMANZIK1983205,WEISZ19831,WEISZ1984397}, the $\mathcal{O}(a)$-improved Wilson quark action is written as,
\begin{align}\label{eq:faction}
	\begin{split}
		S(x) =& a^4 \sum_x \bar{\psi}(x) \left[ m_0 + \gamma_4 D_4 + \nu \sum_i \gamma_i D_i + \frac{a}{2} r_t D_4^2 - \frac{a}{2} r_s \sum_i D_i^2 \right. \\
		&\left. \hspace{20mm} - \frac{a}{4} c_E \sum_i \sigma_{4i} igF_{4i}(x) - \frac{a}{4} c_B \sum_{i < j} \sigma_{ij} ig F_{ij}(x) \right] \psi(x),
	\end{split}
\end{align}
where $\psi(x)$ a quark field operator, $m_0$ is the bare quark mass, $\sigma_{\mu\nu} \equiv [\gamma_\mu, \gamma_\nu]/2i$ and $-igF_{\mu\nu} \equiv [D_\mu, D_\nu]$ is the clover field strength tensor~\cite{Luscher:1996sc}. The covariant derivatives and the Laplace operator are defined as,
\begin{align}
	D_{\mu} &= \frac{1}{2} \left( \nabla_\mu + \nabla^\ast_\mu \right), \\
	\nabla_\mu \, \psi(x)&= \frac{1}{a} \left[ U_\mu(x) \, \psi(x+a\hat{\mu}) - \psi(x) \right], \\
	\nabla^\ast_\mu \, \psi(x)&= \frac{1}{a} \left[ \psi(x) - U^\dag_\mu(x-a\hat{\mu}) \, \psi(x-a\hat{\mu}) \right], \\
	D_{\mu}^2 &= \nabla^\ast_\mu \nabla_\mu = \frac{1}{a} \sum_\mu \left( \nabla_\mu - \nabla^\ast_\mu \right) = \nabla_\mu \nabla^\ast_\mu,
\end{align}
where $U_\mu$ are the link variables, $a$ is the lattice spacing and $\hat\mu$ is the unit vector in the $\mu$ direction. We have written the action in this form to ease the discussion for a relativistic heavy quark prescription. Setting $r_s = r_t = 1$, $\nu = 1$ and $c_E = c_B = c_{SW}$, gives the familiar isotropic Sheikholeslami-Wohlert (Clover) action~\cite{Sheikholeslami:1985ij}.

The Clover action has $\mathcal{O}((m_q a)^n)$ discretization errors which would be significant, i.e. $m_q a \gtrsim 1$, for the heavy (charm and bottom) quarks. If the lattice spacing is fine enough, i.e. $m_q a < 1$, then one can simply use an anisotropic Clover action with its spatial and temporal clover coefficients, $c_B$ and $c_E$, determined non-perturbatively. However, a further tuning is required for coarser lattices where the five free parameters need to be determined to suppress the leading discretization errors. A choice of $r_t = 1$ is allowed and the parameters $r_s$, $c_E$ and $c_B$ can be calculated in 1-loop perturbation theory~\cite{PTP.109.383,Aoki:2004271}. In order to produce the correct relativistic dispersion relation, the parameter $\nu$ needs to be determined non-perturbatively. After a proper adjustment of the parameters, the leading discretization errors are reduced to $\mathcal{O}((a \Lambda_{QCD})^2)$, which is expected to be negligibly small and can be further removed via a continuum extrapolation. This formulation is known as the Tsukuba action~\cite{PTP.109.383} and the outlined procedure produces a relativistic quark action equivalent to another widely used action in the literature---the Fermilab action~\cite{ElKhadra:1996mp}. 

	\subsection{Gauge configurations} \label{sec:gc}
	Within the context of this review, all of the simulations are performed on the gauge configurations generated by the PACS-CS collaboration~\cite{Aoki:2008sm}. These ensembles have a volume of $32^3\times 64$ and the fermionic vacuum contains the isospin symmetric $up$ and $down$, and the $strange$ dynamical quarks. An Iwasaki gauge action with $\beta=1.90$ and a Clover action with $c_{SW}=1.715$ have been used to generate the configurations. The scale is set via the physical masses of the $\pi$, $K$ and $\Omega$ particles and the lattice spacing is determined to be $a=0.0907(13)$ fm corresponding to the physical scale $a^{-1} = 2.176(31)$ GeV. The full set consists of configurations with five different light-sea-quark masses yielding $m_\pi \sim 700$, $570$, $410$, $300$ and $156$~MeVs. Mass of the strange quark is reported to be tuned to its physical value, although a mistuning has been noted in the literature~\cite{PhysRevD.98.114505,Tiburzi:2008bk,PhysRevLett.108.112001}. 

Details of the gauge configurations are given in \Cref{tab:ens}. In order to ease the discussions of the following sections we have assigned \texttt{ID} numbers to the ensembles. Since different ensembles of gauge configurations or different quark actions/parameters for the valence quarks are used in the quoted works, we clarify further details explicitly when necessary.    

\begin{table}[ph]
	\tbl{Details of the gauge configurations. We list the volume~($V$), the number of flavors~($N_f$), the lattice spacing~($a$), the spatial lattice extent~($L$), inverse gauge coupling~($\beta$), and the clover coefficient~($c_{SW}$). Corresponding pion masses~($m_\pi$) as calculated by PACS-CS Collaboration are also given as a reference. The entry in italics has $m_\pi L < 4$. }
	{
	\begin{tabular}{@{}ccccccccc|cc@{}} \toprule
	\texttt{ID} & $V$  & $N_f$ & $a$ [fm] &  $L$ [fm] & $\beta$ & $c^{u/d,s}_{SW}$ & $\kappa^{s}_{sea}$ & $\kappa^{u,d}_{sea}$ & $m_\pi$ [GeV] & $m_\pi L$ \\
	\colrule
	\ensh &                  &       &              &        &        &         &           & $0.13700$ & $0.702(11)$   & $10.34$ \\
	\enh  &                  &       &              &        &        &         &           & $0.13727$ & $0.570(10)$   & $\pz8.40$ \\
	\enm  & $32^3 \times 64$ & $2+1$ & $0.0907(13)$ & $2.90$ & $1.90$ & $1.715$ & $0.13640$ & $0.13754$ & $0.411(8)\pz$ & $\pz6.06$ \\
	\enl  &                  &       &              &        &        &         &           & $0.13770$ & $0.296(7)\pz$ & $\pz4.36$ \\
	\textit{\ensl} &                  &       &              &        &        &         &           & $0.13781$ & $0.156(9)\pz$ & $\mathit{\pz2.29}$ \\ \botrule
	\end{tabular} \label{tab:ens}
	}
\end{table}

\section{Two-point Correlation Functions} \label{sec:mass}

	In the continuum formulation, a two-point correlation function describes the freely propagating baryon along the time direction. In a lattice simulation, the correlation function is computed at the quark level on an ensemble of gauge configurations and matched to its baryon-level expression which gives access to the energy levels of the baryon. We give the definitions of a two point correlation function at the baryon and the quark level in the following discussion since both formulae are essential for form factor calculations.

	\subsection{Baryon-level correlation functions}
Consider the two-point correlation function of a \spinoh baryon projected to a definite momentum,
\begin{equation}
	\label{eq:2pt_corr}
	\langle \mathcal{G}^{\mcb \mcb} (t; {\bf p}; \Gamma) \rangle = \sum_x e^{-i {\bf p} \cdot {\bf x}} \Gamma^{\beta \alpha} \langle \Omega | \mathcal{T} \{ \chi^\alpha_{\mcb}({\bf x},t) \bar{\chi}^\beta_\mcb(0) \}| \Omega \rangle,
\end{equation}
where $\chi_\mcb$ is the interpolating field of the baryon, $\Gamma$ is the spin-parity projection matrix, and $\alpha$ and $\beta$ are Dirac indices. $\Omega$ denotes the vacuum. By inserting a complete set of states, $\sum_{B,s}| B(p,s) \rangle \langle B(p,s)|$, \Cref{eq:2pt_corr} is written as a sum over all possible states that correspond to the quantum numbers defined by the $\chi_\mcb$ interpolating field,   
\begin{align} 
	\label{eq:2pt_sum}
	\langle \mathcal{G}^{\mcb \mcb} (t; {\bf p}; \Gamma) \rangle &= \sum_{B, s} e^{-E_B({\bf p}) t} \Gamma^{\beta \alpha} \langle \Omega |\chi^\alpha_{\mcb}({\bf p}) | B(p,s) \rangle \langle B(p,s) | \bar{\chi}^\beta_\mcb (0) | \Omega \rangle \\
	\label{eq:2pt_sum_simp}
	&=  A_0({\bf p}) e^{-E_0({\bf p}) t} + A_1({\bf p}) e^{-E_1(\bf p) t} + \dots ,
\end{align}
where the time dependence of the fields are factored out, acted upon the states properly and have been collected into overlap factors, $A_n({\bf p})$, in writing the second line. $E_0$ corresponds to the energy of the respective state starting from the $n=0$ ground state. In the large Euclidean time limit, $t \gg a$, where $a$ is the lattice spacing, only the lowest-lying state in the specral sum of \Cref{eq:2pt_sum} contributes,
\begin{equation} \label{eq:2pt}
	\langle \mathcal{G}^{\mcb \mcb} (t; {\bf p}; \Gamma) \rangle \simeq \sum_s e^{-E_\mcb({\bf p}) t} \Gamma^{\beta \alpha} \langle \Omega|\chi^\alpha_{\mcb}(p) | \mcb(p,s) \rangle \langle \mcb(p,s) | \bar{\chi}^\beta_\mcb (0) | \Omega \rangle.
\end{equation}
The overlap between the interpolating field and the physical state is defined as,
\begin{equation} \label{eq:overlap}
	\langle 0|\chi^\alpha_{\mcb}(0) | \mcb({\bf p},s) \rangle = Z_\mcb \sqrt{\frac{M_\mcb}{E_\mcb({\bf p})}} u(p,s),
\end{equation} 
where $Z_\mcb$ is the overlap factor, $M_\mcb$ is the mass of the baryon and $u(p,s)$ is the Dirac spinor. Using the Dirac spinor sum,
\begin{equation}
	\sum_s u(p,s) \bar{u}(p,s) = \frac{\gamma_\mu p^\mu + M_\mcb}{2 M_\mcb},
\end{equation} 
and inserting \Cref{eq:overlap} into \Cref{eq:2pt}; the two-point function takes the form,
\begin{equation} \label{eq:2pt_hl}
	\langle \mathcal{G}^{\mcb \mcb} (t; {\bf p}; \Gamma) \rangle \simeq \left|Z_\mcb\right|^2 \frac{M_\mcb}{E_\mcb({\bf p})} e^{-E_\mcb({\bf p}) t} \text{Tr}[\Gamma \frac{\gamma_\mu p^\mu + M_\mcb}{2 M_\mcb}],
\end{equation}
where $\left| Z_\mcb \right|^2 = Z_\mcb Z^\ast_\mcb$ as long as the source and the sink operators are smeared in the same manner on the lattice.
Finally, using the definitions of the $\Gamma$ matrices as,
\begin{equation} \label{eq:Gamma}
	\Gamma_i=\frac{1}{2}\left(\begin{matrix}\sigma_i & 0 \\ 0 & 0 \end{matrix}\right), \qquad \Gamma_4=\frac{1}{2}\left(\begin{matrix}I & 0 \\ 0 & 0 \end{matrix}\right),
\end{equation}
only the $\Gamma_4$ component survives. Setting ${\bf p} = (p,0,0)$, the two-point correlation function reduces to, 
\begin{align} \label{eq:2pt_fit}
	\langle \mathcal{G}^{\mcb \mcb} (t; {\bf p}; \Gamma_4) \rangle &= \left|Z_\mcb\right|^2 \left( \frac{E_\mcb({\bf p}) + M_\mcb}{2E_\mcb({\bf p})} \right)^{1/2} e^{-E_\mcb({\bf p}) t}.
\end{align}

For a \spinth baryon, its two-point correlation function projected to a definite momentum is given as,
\begin{align}
	\label{eq:deltacf}
	\langle \mathcal{G}_{\sigma\tau}^{\mcbp \mcbp}(t; {\bf p};\Gamma)\rangle & \equiv \sum_{\bf x}e^{-i{\bf p}\cdot {\bf x}}\Gamma^{\beta \alpha} \langle \Omega | \mathcal{T} \{ \chi_\sigma^\alpha(x) \bar{\chi}_\tau^\beta(0) \} | \Omega \rangle,
\end{align}
where $\chi_{\sigma, \tau}$ are the interpolating fields of the spin-3/2 baryon $\mcbp$ with $\sigma, \tau$ the Lorentz indicies. It can be shown that the large Euclidean time limit ($t \gg a$) of the two-point correlation function with momentum ${\bf p} = (p,0,0)$ is,
\begin{align} \label{eq:deltacf_hl}	
	\langle \mathcal{G}_{\sigma\tau}^{\mcbp \mcbp}(t; {\bf p};\Gamma)\rangle &\simeq \left| Z_\mcbp \right|^2 \frac{M_\mcbp}{E_\mcbp({\bf p})} e^{- E_\mcbp({\bf p}) t} \operatorname{Tr}[\Gamma \Lambda_{\sigma\tau}],
\end{align}
where $Z_\mcbp$ is the overlap factor of the interpolating field to the corresponding baryon state and the trace acts in the Dirac space. The Rarita-Schwinger spin sum, $\Lambda_{\sigma \tau}$, is defined as,
\begin{align}
	\begin{split}
		\sum_s u_\sigma(p,s) \bar{u}_\tau(p,s) &= -\frac{\gamma_\mu p^\mu + M_\mcbp}{2M_\mcbp}  \left[g_{\sigma\tau} - \frac{1}{3}\gamma_\sigma \gamma_\tau - \frac{2p_\sigma p_\tau}{3M_\mcbp^2} + \frac{p_\sigma \gamma_\tau-p_\tau \gamma_\sigma}{3M_\mcbp}\right] \\
		& \equiv \Lambda_{\sigma \tau}(p).
	\end{split}
\end{align} 
	
	\subsection{Quark-level correlation functions}
At the quark level, correlation functions are obtained via contracting the pairs of quark fields of the interpolating operators. After this Wick contraction, correlation functions are expressed in terms of quark propagators which are then evaluated on gauge configurations. Since this procedure is abundantly detailed in the literature (\emph{e.g.} see Ref. \citen{Leinweber:1990dv}), we only present the main results. 

Consider an interpolating operator of a \spinoh baryon with the quark fields $q_1$, $q_2$, and $q_3$,
\begin{equation} \label{eq:genint}
	\chi_\mcb(x) = \varepsilon^{abc} \, [q_1^{T a}(x) \, C \gamma_5 \, q_2^b(x)] \, q_c^k(x),
\end{equation}
where $a,b,c$ are the color indices and $C=\gamma_4\gamma_2$. One can recover the standard nucleon operator simply by replacing $(q_1, q_2, q_3) \to (u, d, u)$. Two-point correlation function associated with $\chi_\mcb$ is,
\begin{align}\label{eq:gencf_ql}
	\langle \mathcal{G}^{\mcb \mcb}(t; {\bf p};\Gamma_4)\rangle &= \sum_{\bf x} e^{-i{\bf p} \cdot {\bf x}} \Gamma_4^{\beta\alpha} \langle \Omega | {\cal T} \{ \chi_{\cal B}^\alpha(x) \bar{\chi}_{{\cal B}}^{\beta}(0) \} | \Omega \rangle \nonumber \\
	&= \sum_{\mathbf{x}} e^{-i{\mathbf{p}} \cdot {\mathbf{x}}} \varepsilon^{abc} \varepsilon^{a^\prime b^\prime c^\prime} 
	\left\{ \text{Tr} \left[ \Gamma_4 S_{q_1}^{a a^\prime}(x,0) \underbar{S}_{q_2}^{b b^\prime}(x,0) S_{q_3}^{c c^\prime}(x,0) \right] \right. \nonumber \\
	& \left. \hspace{3.3cm} + \text{Tr}\left[ \Gamma_4 S_{q_1}^{a a^\prime}(x,0)\right] \text{Tr}\left[ \underbar{S}_{q_2}^{b b^\prime}(x,0) S_{q_3}^{c c^\prime}(x,0) \right] \right\},
\end{align}
where $S_{q_i}(x,0)$ is the propagator of the quark field $q_i$. Here, $\tilde{C} \equiv C \gamma_5$ and $\underbar{S} \equiv (\tilde{C} S \tilde{C}^{-1})^T$. 

For a \spinth baryon interpolating field consider the decuplet $\Delta^+$-like baryon with the quark fields $q_1$, $q_2$, and $q_3$ as,
\begin{align} \label{eq:deltaint}	
	\chi_\mu(x)=\frac{1}{\sqrt{3}} \varepsilon^{abc} \{2[q_1^{T a}(x) C \gamma_\mu q_2^b(x)]q_3^c(x) +[q_1^{T a}(x) C \gamma_\mu q_3^b(x)]q_2^c(x)\}.
\end{align} 

Two-point correlation function written in terms of quark propagators associated with $\chi_\mu$ is,
\begin{align} \label{eq:deltacf_ql}
	&\langle \mathcal{G}_{\sigma\tau}^{\mcbp \mcbp}(t; {\bf p};\Gamma) \rangle \equiv \sum_{\bf x} e^{-i{\bf p} \cdot {\bf x}} \Gamma^{\beta\alpha} 
	\langle \Omega | {\cal T} \{ \chi_{\sigma, \alpha}(x) \bar{\chi}_{\tau, \beta}(0) \} | \Omega \rangle \\	
	\begin{split}
	&= \frac{1}{3} \sum_{\bf x}e^{-i{\bf p}\cdot {\bf x}} \varepsilon^{abc} \varepsilon^{a^\prime b^\prime c^\prime} 
	\left\{ 4 \text{Tr}\left[ \Gamma S_{q_2}^{a a^\prime}(x,0) \gamma_\tau C S_{q_3}^{T b b^\prime}(x,0) C \gamma_\sigma S_{q_1}^{c c^\prime}(x,0) \right] \right. \\
	& \hspace{3.565cm} + 4 \text{Tr}\left[ \Gamma S_{q_2}^{a a^\prime}(x,0) \gamma_\tau C S_{q_1}^{T b b^\prime}(x,0) C \gamma_\sigma S_{q_3}^{c c^\prime}(x,0) \right] \\
	& \hspace{3.565cm} + 4 \text{Tr}\left[ \Gamma S_{q_3}^{a a^\prime}(x,0) \gamma_\tau C S_{q_1}^{T b b^\prime}(x,0) C \gamma_\sigma S_{q_2}^{c c^\prime}(x,0) \right] \\
	& \hspace{3.565cm} + 2 \left( \text{Tr}\left[ \Gamma S_{q_2}^{a a^\prime}(x,0) \right] \text{Tr}\left[\gamma_\tau C S_{q_3}^{T b b^\prime}(x,0) C \gamma_\sigma S_{q_1}^{c c^\prime}(x,0) \right] \right) \\
	& \hspace{3.565cm} + 2 \left( \text{Tr}\left[ \Gamma S_{q_2}^{a a^\prime}(x,0) \right] \text{Tr}\left[\gamma_\tau C S_{q_1}^{T b b^\prime}(x,0) C \gamma_\sigma S_{q_3}^{c c^\prime}(x,0) \right] \right) \\
	& \left. \hspace{3.565cm} + 2 \left( \text{Tr}\left[ \Gamma S_{q_3}^{a a^\prime}(x,0)\right] \text{Tr}\left[ \gamma_\tau C S_{q_1}^{T b b^\prime}(x,0) C \gamma_\sigma S_{q_2}^{c c^\prime}(x,0) \right] \right) \right\}.
	\end{split}
\end{align}

	Apart from extracting the energy levels of a given baryon, two-point correlation functions have a certian importance for the calculation of the interaction matrix elements. In order to isolate the matrix element from a three point correlation function in a lattice simulation one needs to cancel out the exponential behaviour of the three point function by using the two point function. This procedure is detailed in the following sections as it is an essential element of the analysis.

\section{Elastic Form Factors} \label{sec:emff}

	Studying the elastic form factors gives access to the physical observables of a single baryon such as its electric charge radius or magnetic moment. In the following sections we summarize the formalism and the lattice methodology to extract the form factors before presenting some selected results.   

	\subsection{Baryon form factors} \label{sec:baryonff}
	Matrix element of a \spinoh baryon is parametrized by two form factors,
\begin{align} \label{eq:s12_me}
	\langle \mcb(\pp, \ssp) | \mathcal{V}_\mu(q) | \mcb(p, s) \rangle = \bar{u}(\pp, \ssp) &\left[\gamma_\mu \ffac{1}{B}(Q^2) +i \frac{\sigma_{\mu\nu} q^\nu}{2m_{\mcb}} \ffac{2}{B}(Q^2) \right]u(p,s),
\end{align}
where $\mathcal{V}_\mu = \sum_q e_q \bar{q} \gamma_\mu q$ is the electromagnetic current, $q = p - \pp$ and $\sigma_{\mu\nu} = \frac{1}{2} \{\gamma_\mu, \gamma_\nu\}$. Here $u(p,s)$ is the Dirac spinor of a baryon with four-momentum $p^\mu$, spin $s$, and mass $m_{\mcb}$. The Dirac, $\ffac{1}{B}(Q^2)$, and Pauli, $\ffac{2}{B}(Q^2)$, form factors are related to the Sachs electric and magnetic form factors by the relations~\cite{Ernst:1960zza},
\begin{align}
	\label{eq:s12_sachs_e}
	G_{E, \mcb}(Q^2) &= \ffac{1}{B}(Q^2) - \tau \ffac{2}{B}(Q^2),\\
	\label{eq:s12_sachs_m}
	G_{M, \mcb}(Q^2) &= \ffac{1}{B}(Q^2) + \ffac{2}{B}(Q^2),
\end{align}
where $\tau = Q^2/4m^2_{\mcb}$.

In the case of a \spinth baryon, there are four form factors that parameterize the matrix element,
\begin{align}\label{eq:s32_me}
	\langle \mathcal{B}^\ast_\sigma(\pp,\ssp) | \mathcal{V}_\mu(q)| \mathcal{B}^\ast_\tau(p,s) \rangle = \sqrt{\frac{m^2_\ast}{E_\ast(p) E_\ast(\pp)}} \bar{u}_\sigma(\pp,\ssp) \, \mathcal{O}^{\sigma}{}_\mu{}^\tau \, u_\tau(p,s),
\end{align}
where $p(s)$ and $\pp (\ssp)$ denote the four-momentum (spin) of the initial and final states, respectively. Here, $m_\ast \equiv m_\mcbp$ and $E_\ast \equiv E_\mcbp$ are the mass and energy of the baryon state and $u_\alpha(p,s)$ is the baryon spinor in the Rarita-Schwinger formalism~\cite{PhysRev.60.61}. The tensor, $\mathcal{O}^{\sigma}{}_\mu{}^\tau$, is given in a Lorentz-covariant form~\cite{Nozawa:1990gt},
\begin{align}
	\mathcal{O}^{\sigma}{}_\mu{}^\tau = -g^{\sigma\tau} \left\{ a_1 \gamma_\mu + \frac{a_2}{2 m_\ast} P_\mu \right\}- \frac{q^\sigma q^\tau}{(2 m_\ast)^2} \left\{ c_1 \gamma_\mu + \frac{c_2}{2 m_\ast} P_\mu \right\},
\end{align}  
where $P = p + \pp$ and $q = p - \pp$. The electric-charge ($E0$), electric-quadrupole ($E2$), magnetic-dipole ($M1$) and magnetic-octupole ($M3$) multipole form factors are consequently defined in terms of the covariant vertex functions $a_1,\,a_2,\,c_1$ and $c_2$,
\begin{align}
	G_{E0}(q^2)  &= (1+\frac{2}{3}\tau) \left\{ a_1 + (1 + \tau) a_2 \right\}- \frac{1}{3}\tau (1+\tau) \left\{c_1 + (1+\tau) c_2 \right\}, \\
	G_{E2}(q^2)  &=  \left\{ a_1 + (1 + \tau) a_2 \right\}- \frac{1}{2}(1+\tau) \left\{c_1 + (1+\tau) c_2 \right\}, \\
	G_{M1}(q^2) &= (1 + \frac{4}{3}\tau)a_1 - \frac{2}{3}\tau(1+\tau)c_1, \\ 
	G_{M3}(q^2) &= a_1 - \frac{1}{2}(1+\tau)c_1,
\end{align}
with $\tau = Q^2/4m^2_\mcbp$.

	\subsection{Lattice methodology for the three-point functions} \label{sec:diaglatmet}
	In a lattice simulation, the matrix element describing a baryon interacting with a single external current is extracted from the three-point correlation function,
\begin{align}\label{eq:3pt}
	\langle \mathcal{G}^{(3)}(t_2,t_1; {\bf p}^\prime, {\bf p}; \mathbf{\Gamma}; \mu) \rangle \equiv 
		\sum_{{\bf x_2},{\bf x_1}} e^{-i{\bf p}\cdot {\bf x_2}} e^{i{\bf q}\cdot {\bf x_1}} \Gamma^{\beta\alpha} 
		\langle \Omega | {\cal T} \{ \chi^\alpha(x_2) {\cal V}_\mu(x_1) \bar{\chi}^\beta(0) \} | \Omega\rangle,
\end{align}
where $\chi(x)$ is the interpolating field of the baryon, $\alpha$ and $\beta$ are the Dirac indices, $\Gamma$ is the spin-parity projection matrix in Dirac space. Note that for a \spinth baryon the interpolating fields have a Lorentz index as well. $x_i = (t_i, {\bf x_i})$ are four-vectors, where $t_1$ is the time slice at which the external current interacts with the baryon and $t_2$ is the time slice when the baryon is annihilated. $\mathbf{p}$ and $\mathbf{\pp}$ are the three-momenta of the respective initial and final states, and $q = p - \pp$. Here, $\langle \dots \rangle$ denotes the ensemble average. The electromagnetic current can be either taken as a local bilinear operator $\mathcal{V}_\mu(x) = \sum_q \bar{q}(x) \gamma_\mu q(x)$, where $q$ is the quark flavor or in a non-local form,
\begin{equation}\label{eq:lat_current}
	\mathcal{V}_\mu(x) = \frac{1}{2}[\bar{q}(x+\hat\mu)U^\dagger_\mu(1+\gamma_\mu)q(x) -\bar{q}(x)U_\mu(1-\gamma_\mu)q(x+\hat\mu)].
\end{equation}
The local lattice current needs to be renormalized but the non-local formulation is conserved by the Wilson fermions so a renormalization is not necessary. The non-local discretization is adopted for the rest of the paper. 
Compared to using a local lattice current this choice has no disadvantage---it is just another discretization of the electromagnetic current. On the contrary, it eliminates the renormalization step, thus a source of a possible systematic error.

On the baryon level and for large Euclidean time separations, $t_2 - t_1\gg a$ and $t_1 \gg a$, the three-point function in \Cref{eq:3pt} takes the limit,
\begin{align}\label{eq:3pt_had}
	\begin{split}
		\langle \mathcal{G}^{(3)}(t_2,t_1; {\bf p}^\prime, {\bf p}; \mathbf{\Gamma}; \mu) \rangle &\simeq 
		e^{-E({\bf \pp}) \tau} e^{-E({\bf p}) t_1} \Gamma^{\beta\alpha} \\
		&\times \langle \Omega | \chi^\alpha | \mcb \rangle \, \langle \mcb | \mathcal{V}_\mu | \mcb \rangle \, \langle \mcb | \bar{\chi}^\beta | \Omega \rangle, 
	\end{split}
\end{align}
where we have defined $\tau = t_2 - t_1$ and $\mcb$ denotes the ground state of the baryon that corresponds to the chosen interpolating field $\chi(x)$. On the quark level, expressions for the connected piece of the three-point functions for \spinoh and \spinth baryons are obtained by inserting the current $\mathcal{V}_\mu$ to each quark~\cite{Leinweber:1990dv} in \Cref{eq:gencf_ql,eq:deltacf_ql} respectively.    

The exponential time dependence and the overlap factors in \Cref{eq:3pt_had} must be canceled in order to extract the matrix element. This is achieved by forming the appropriate ratios of three-point to two-point functions since the extra factors are common to both. Two of the most widely used ratios are,
\begin{align}\label{eq:em_lat_ratio1}
	R_1(t_2,t_1; {\bf \pp}, {\bf p}; \mathbf{\Gamma}; \mu) & =
	\frac{\langle \mathcal{G}^{(3)}(t_2,t_1; {\bf \pp}, {\bf p}; \mathbf{\Gamma}; \mu) \rangle}{\langle \mathcal{G}^{(2)}(t_2; {\bf \pp}; \Gamma_4) \rangle} \nonumber \\ 
	&\times 
	\left[\frac{\langle \mathcal{G}^{\mcb \mcb}(\tau; {\bf p}; \Gamma_4) \rangle 
				\langle \mathcal{G}^{\mcb \mcb}(t_1; {\bf \pp}; \Gamma_4) \rangle 
				\langle \mathcal{G}^{\mcb \mcb}(t_2; {\bf \pp}; \Gamma_4)\rangle}
			   {\langle \mathcal{G}^{\mcb \mcb}(\tau; {\bf \pp}; \Gamma_4)\rangle 
				\langle \mathcal{G}^{\mcb \mcb}(t_1; {\bf p}; \Gamma_4)\rangle 
				\langle \mathcal{G}^{\mcb \mcb}(t_2; {\bf p}; \Gamma_4)\rangle}
	\right]^{1/2},
\end{align}
and,
\begin{align}\label{eq:em_lat_ratio2}
	R_2^{\sigma \tau}(t_2,t_1; {\bf \pp}, {\bf p}; {\bf \Gamma}; \mu) &= 
	\left[\frac{\langle \mathcal{G}^{(3)}_{\sigma \tau}(t_2, t_1; {\bf \pp}, {\bf p}; {\bf \Gamma}; \mu) \rangle 
				\langle \mathcal{G}^{(3)}_{\tau \sigma}(t_2, t_1; {\bf p}, -{\bf \pp}; {\bf \Gamma}; \mu) \rangle}
				{\langle \mathcal{G}^{\mcbp \mcbp}_{\sigma \sigma}(t_2; {\bf \pp}; \Gamma_4) \rangle 
				\langle \mathcal{G}^{\mcbp \mcbp}_{\tau \tau}(t_2; -{\bf p}; \Gamma_4) \rangle} \right]^{1/2},
\end{align}
where $\sigma$ and $\tau$ are Lorentz indices and the two-point correlation functions for \spinoh and \spinth baryons are given in \Cref{eq:2pt_corr,eq:deltacf} respectively. We have explicitly assigned \Cref{eq:em_lat_ratio1} to \spinoh and \Cref{eq:em_lat_ratio2} to \spinth baryons in consideration of their use in the reviewed works, however both ratios are equally applicable to both sectors. In the large Euclidean-time limit, these ratios reduce to,
\begin{align}
 	R_{1,2}^{(\sigma \tau)}(t_2,t_1; {\bf \pp}, {\bf p}; \mathbf{\Gamma}; \mu) &\xrightarrow[t_2-t_1\gg a]{t_1\gg a} \Pi_{1,2}^{(\sigma \tau)}({\bf \pp}, {\bf p}; \mathbf{\Gamma}; \mu).
\end{align} 
The form factors are consequently extracted by the relations,
\begin{align} 
	\label{eq:12_ff_e}
	G_{E,\mcb}(Q^2) &= \sqrt{\frac{2E_\mcb}{E_\mcb + m_\mcb}} \, \Pi_1({\bf p}^\prime,{\bf p}; \Gamma_4; \mu=4),\\
	\label{eq:12_ff_m}
	\varepsilon_{ijk}\, {\bf p}_k \, G_{M,\mcb}(Q^2) &= \sqrt{2E_\mcb (E_\mcb+m_\mcb)} \, \Pi_1({\bf p}^\prime,{\bf p}; \Gamma_j; \mu=i),
\end{align}
for the \spinoh baryons where $G_E$ and $G_M$ are the Sachs form factors, and,
\begin{align}
    \label{eq:E0lat}
	G_{E0}(Q^2) &= \frac{1}{3} \left( \Pi_{2}^{11}({\bf p}^\prime_i,{\bf p}_i;\Gamma_4; 4) +\Pi_{2}^{22}({\bf p}^\prime_i,{\bf p}_i;\Gamma_4; 4) + \Pi_{2}^{33}({\bf p}^\prime_i,{\bf p}_i;\Gamma_4; 4) \right), \\
	\begin{split} \label{eq:E2lat}
	{\bf p}_i^2 \, G_{E2}(Q^2) &= 2 m_\ast(E_\ast+m_\ast) \left( \Pi_{2}^{11}({\bf p}^\prime_i,{\bf p}_i;\Gamma_4;4) +\Pi_{2}^{22}({\bf p}^\prime_i,{\bf p}_i;\Gamma_4;4) \right. \\ &\left.- 2\Pi_{2}^{33}({\bf p}^\prime_i,{\bf p}_i;\Gamma_4;4) \right),
	\end{split}\\
	\begin{split} \label{eq:M1lat}
	{\bf p}_1^2 \, G_{M1}(Q^2) &= -\frac{3}{5} E_\ast+m_\ast \left( \Pi_{2}^{11}({\bf p}^\prime_1,{\bf p}_1;\Gamma_2;3) +\Pi_{2}^{22}({\bf p}^\prime_1,{\bf p}_1;\Gamma_2;3) \right. \\ &\left.+ \Pi_{2}^{33}({\bf p}^\prime_1,{\bf p}_1;\Gamma_2;3) \right),
	\end{split}\\
	\begin{split} \label{eq:M3lat}
	{\bf p}_1^3 \, G_{M3}(Q^2) &= -4 m_\ast(E_\ast+M_\ast)^2 \left( \Pi_{2}^{11}({\bf p}^\prime_1,{\bf p}_1;\Gamma_2;3) +\Pi_{2}^{22}({\bf p}^\prime_1,{\bf p}_1;\Gamma_2;3) \right. \\ &\left.- \frac{3}{2} \Pi_{2}^{33}({\bf p}^\prime_1,{\bf p}_1;\Gamma_2;3) \right),
	\end{split}
\end{align}
for the \spinth baryons, where $i=1,2,3$ and ${\bf p}_i$ are the momentum vectors in three spatial directions. 

	\subsection{Results} \label{sec:diagres}
In analogy to non-relativistic physics, a form factor can be considered as a three-dimensional Fourier transform of the density distribution of a quantity of a baryon under certain conditions. 
Considering a small momentum transfer, $Q^2$, between the baryon and the external current or a limit where the mass of the baryon is much larger than the transferred momentum, $Q^2 \ll M^2$, the assumption is that the initial- and final-state baryons are fixed at the same location and that they have the same internal structure. Then, the physical interpretation that the Fourier transforms of the form factors lead to density distributions holds. However, with increasing $Q^2$, recoil effects become significant so that the initial and final baryon states no longer have the same momentum, thus their wavefunctions differ (i.e. there is a relative Lorentz contraction), and it is no longer possible to have a probability or density interpretation~\cite{Miller:2007uy}. One frame where the physical interpretation of the form factors are realized is the Breit frame in which the magnitude of the initial and final momenta of the baryon are equal, i.e. $|{\bf p}| = |{\bf p^\prime}|$. The exchanged photon is space-like, it carries momentum ${\bf q}$ but no energy so the four-momentum transfer is $q^\mu = (0, {\bf q})$ so that $Q^2={\bf q}^2$. 
For each $Q^2$ there is a Breit frame in which form factors are represented as $\mathcal{F}(Q^2)$. The corresponding electric and magnetic form factors are related to the charge and magnetization densities of the baryon~\cite{PhysRev.126.2256} via a Fourier transform. For instance a multipole expansion for the electric form factor reveals the correspondence, 
\begin{equation}
	\mathcal{F}(Q^2) = \int d^3 {\bf x} \, e^{i {\bf x} \cdot {\bf q}} \, \rho({\bf x}) \simeq \mathcal{F}(0) \left( 1 - \frac{1}{6} Q^2 \langle r^2 \rangle + \cdots \right), 
\end{equation}
where $\rho({\bf x})$ is interpreted as the electric charge density. The slopes of the form factors at $Q^2=0$ determine the mean-square radii,
\begin{equation}\label{eq:radii_deriv}
 	\langle r^2 \rangle = -\frac{6}{\mathcal{F}(0)} \left. \frac{d}{dQ^2} \mathcal{F}(Q^2) \right|_{Q^2=0}.
\end{equation} 
Experimental evidence indicate that the functional forms of the form factors can be approximated by a multipole form,
\begin{equation}\label{eq:multipole_ansatz}
	\mathcal{F}(Q^2) = \frac{\mathcal{F}(0)}{\left( 1 + Q^2/\Lambda^2\right)^n},
\end{equation}
where $n=1$ gives a monopole, and $n=2$ gives a dipole ansatz. $\Lambda$ is the multipole mass---a free fit parameter to be determined. Inserting \Cref{eq:multipole_ansatz} into \Cref{eq:radii_deriv}, the mean-squared radii can be estimated via,
\begin{equation}
	\label{eq:radii}
	\langle r^2 \rangle = \frac{6n}{\Lambda^2}.
\end{equation}

The magnetic moment of a baryon, on the other hand, is defined as,
\begin{equation} \label{eq:magmom}
	\mu = \mathcal{F}_M(0) \left( \frac{e}{2m_\mcb} \right) = \mathcal{F}_M(0) \left( \frac{m_N}{m_\mcb} \right) \mu_N,
\end{equation}
in units of nuclear magnetons, where $m_N$ is the physical nucleon mass and $m_{\mcb}$ is the baryon mass as obtained on the lattice. In order to estimate the magnetic moment, the $Q^2=0$ value of the magnetic form factor, $\mathcal{F}_M(0)$, is needed. 
Due to its explicit momentum dependence however (c.f. \Cref{eq:12_ff_m,eq:M1lat}), the magnetic form factor cannot be extracted at $Q^2=0$ directly in a lattice simulation using a conventional three-point approach but needs to be estimated via an extrapolation for which a dipole ansatz \Cref{eq:multipole_ansatz} can be used. Note that it is possible to extract the form factors at zero spatial momentum transfer by considering momentum derivatives of the correlation functions~\cite{Aglietti:1994nx,deDivitiis:2012vs} as well.
An alternative approach to fitting with respect to $Q^2$ is to assume a similar $Q^2$ dependence for the electric and magnetic form factors in the low-$Q^2$ region and consider a scaling~\cite{Litt:1969my} of the form,
\begin{equation} \label{eq:scaling}
	\mathcal{F}_M(0) = \mathcal{F}_M(Q^2) \frac{\mathcal{F}_E(0)}{\mathcal{F}_E(Q^2)},
\end{equation} 
for each quark sector separately since each sector has a different scaling property. This procedure has been utilized in Refs.~\citen{PhysRevD.46.3067, Leinweber:1992pv, PhysRevD.74.093005, PhysRevD.80.054505} to study the magnetic form factors of octet and decuplet baryons. 

Higher orders in the multipole expansion of the form factors can be evaluated given that they are allowed by the angular momentum selection rule. Considering the electromagnetic transition in a non-relativistic setting, $\langle J^\prime | J_\gamma| J \rangle$, where $J^{(\prime)} = L^{(\prime)} + S^{(\prime)}$ is the total angular momentum of the initial (final) state and $J_\gamma$ is the total angular momentum of the photon, the selection rule states, 
\begin{equation}
	| J - J^\prime | \leq J_\gamma \leq J + J^\prime.
\end{equation} 
For a ground state $\1212$ electromagnetic transition there are at most two form factors, which are the Dirac and Pauli form factors. Higher order multipole form factors vanish when $L=0$. In the case of $\3232$ transition, selection rule allows for four multipole form factors contributing to the interaction. The electric-quadrupole form factors provide information about the shape of the electric charge distribution of the baryon and a non-zero value is also an indication of a tensor, or a $D$-wave, interaction. In the Breit frame, the quadrupole form factor and the electric charge distribution are related as~\cite{PhysRevD.46.3067},
\begin{equation}
	\mathcal{F}_{Q}(0) = m^2_\mcb \int d^3 r \bar{\psi}(r) (3z^2 - r^2) \psi (r),
\end{equation}
where $3z^2 - r$ is the quadrupole moment operator. A non-zero value of the quadrupole moment indicates a deviation from a spherically symmetric charge distribution and the shape is determined by the sign of the moment where a positive (negative) value is assigned to that a prolate (oblate) shape for a positively charged baryon. 

\subsubsection{Spin-1/2 baryons} \label{sec:spin12_em}
Electric and magnetic Sachs form factors (\Cref{eq:s12_sachs_e,eq:s12_sachs_m}) of the positive parity $\sgc$, $\xcc$, $\omc$, and $\occ$ baryons are studied in Refs.~\citen{Can:2013zpa} and \citen{Can:2013tna} using the ensembles \ensh, \enh, \enm, and \enl. Masses of all the light valence quarks are set to be equal to that of the sea quarks, $\kappa^{u/d,s}_{val.} = \kappa^{u/d,s}_{sea}$ and the final results are extrapolated to the physical light-quark mass. For the charm quark, the spatial and temporal improvement coefficients of the Clover action are taken to be equal to each other, $c_{B} = c_{E} = c_{SW}$ and $c_{SW}$ is determined from the tadpole-improved value $1/u^3_0$, where $u_0$ is the average link which is estimated to be the fourth root of the average plaquette. Mass of the $c$-quark is tuned with respect to the experimental $1S$ spin-averaged static masses of charmonium and open-charm mesons. 

\begin{figure}[htb]
	\centerline{\includegraphics[width=\textwidth]{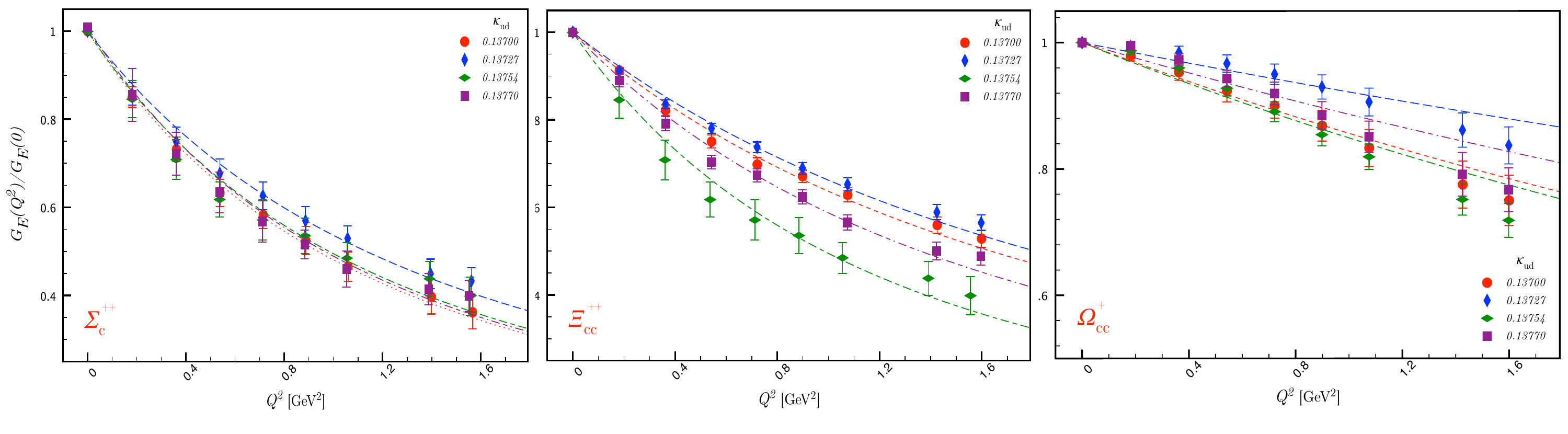}}
	\centerline{\includegraphics[width=\textwidth]{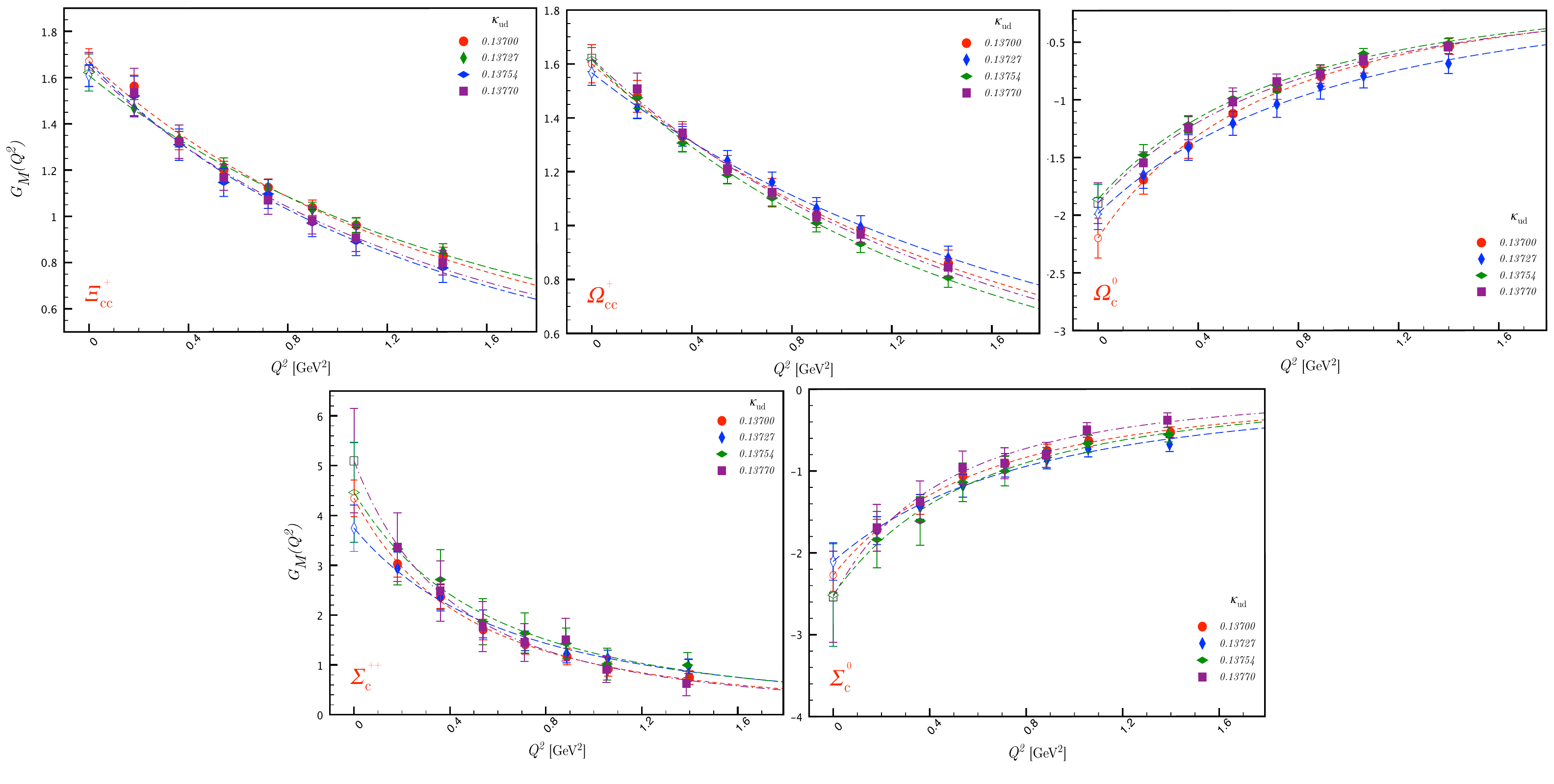}}
	\caption{Electric (top row) and magnetic Sachs form factors of the \spinoh baryons as obtained on four different lattice ensembles with varying pion mass down to $m_\pi \sim 300 \; {\rm MeV}$ ($\kappa_\text{ud} = 0.13770$). Open symbols at $G_M(Q^2=0)$ are estimated via an extrapolation using a dipole ansatz. Figure taken from Ref.~\citen{Can:2013tna}. \label{fig:em_s12}}
\end{figure}

Electric and magnetic form factors of the baryons are shown in \Cref{fig:em_s12}. They are evaluated at different momenta yielding a coverage up to $Q^2 \simeq 1.6 \; {\rm GeV}^2$. Each point is extracted via the relations in \Cref{eq:12_ff_e,eq:12_ff_m} using the lattice ratio of \Cref{eq:em_lat_ratio1} by a plateau method. Effects of any excited state contamination is checked independently by changing the source-sink separation, performing multi-exponential fits and a summed operator insertions analysis where the results are found to be consistent with each other and those of the plateau method~\cite{Can:2013tna}. Functional forms of the form factors are approximated by a dipole form (\Cref{eq:multipole_ansatz}) and the charge radii is estimated via \Cref{eq:radii}, while the magnetic moments are evaluated through \Cref{eq:magmom} where an extrapolated $G_M(Q^2=0)$ value is used. It is evident that the dipole form describes the lattice data quite successfully.  

\begin{table}[ht]
	\tbl{Electric and magnetic charge radii and the magnetic moments of the \spinoh charmed baryons as obtained on ensembles \ensh-\enl \;and their extrapolated values to the physical-mass point. Charge radii and magnetic moments are given in units of fm$^2$. Magnetic moments are given in units of natural, $(\mu_{\mcb} \equiv e/2m_{\mcb})$, and nuclear magnetons, $[\mu_N]$, both, in the $G_{M,\mcb}(0)$ and $\mu_B$ columns respectively. }
	{
	\begin{tabular}{@{}cccccc@{}} \toprule
	\texttt{ID} & $\langle r_{E,\sgc^{++}}^2 \rangle$ & $\langle r_{M,\sgc^{++}}^2 \rangle$ & $G_{M,\sgc^{++}}(0)$ & $\mu_{\sgc^{++}}$ \\
	\colrule
	\ensh     & 0.206(23) & 0.492(66)\pz & 4.343(371)\pzp & 1.447(125) \\
	\enh      & 0.170(19) & 0.360(56)\pz & 3.747(466)\pzp & 1.289(161) \\
	\enm      & 0.196(27) & 0.419(77)\pz & 4.462(1.003)   & 1.591(358) \\
	\enl      & 0.195(34) & 0.574(133)   & 5.098(1.050)   & 1.867(388) \\
	Lin. Fit  & 0.192(22) & 0.410(81)\pz & 4.295(700)\pzp & 1.569(253) \\
	Quad. Fit & 0.234(37) & 0.696(153)   & 6.017(1.385)   & 2.220(505) \\
	\colrule
	%
	\texttt{ID} & & $\langle r_{M,\sgc^0}^2 \rangle$ & $G_{M,\sgc^0}(0)$ & $\mu_{\sgc^0}$ \\
	\colrule
	\ensh     & & 0.379(47)\pz & -2.272(199) & -0.757(67)\pz \\
	\enh      & & 0.287(44)\pz & -2.105(230) & -0.724(80)\pz \\
	\enm      & & 0.391(87)\pz & -2.516(627) & -0.897(223)   \\
	\enl      & & 0.507(111)   & -2.537(557) & -0.929(206)   \\
	Lin. Fit  & & 0.377(75)\pz & -2.330(368) & -0.852(133)   \\
	Quad. Fit & & 0.650(126)   & -2.891(736) & -1.073(269)   \\
	\colrule
	%
	\texttt{ID} & $\langle r_{E,\xcc^{+}}^2 \rangle$ & $\langle r_{E,\xcc^{++}}^2 \rangle$ & $\langle r_{M,\xcc^+}^2 \rangle$ & $G_{M,\xcc^+}(0)$ & $\mu_{\xcc^+}$ \\
	\colrule
	\ensh     & 0.035(6) & 0.118(8)\pz & 0.141(9)\pz & 1.672(53)\pz & 0.412(13) \\ %
	\enh      & 0.017(5) & 0.107(6)\pz & 0.127(10)   & 1.609(47)\pz & 0.404(12) \\ %
	\enm      & 0.018(6) & 0.127(8)\pz & 0.136(12)   & 1.622(80)\pz & 0.410(20) \\ %
	\enl      & 0.032(8) & 0.142(9)\pz & 0.141(13)   & 1.635(74)\pz & 0.416(19) \\ %
	Lin. Fit  & 0.017(6) & 0.136(8)\pz & 0.135(10)   & 1.602(58)\pz & 0.411(15) \\ %
	Quad. Fit & 0.042(9) & 0.165(12)   & 0.154(19)   & 1.670(110)   & 0.425(29) \\ %
	\colrule
	%
	\texttt{ID} & & $\langle r_{M,\omc}^2 \rangle$ & $G_{M,\omc}(0)$ & $\mu_{\omc}$ \\
	\colrule
	\ensh     & & 0.346(43) & -2.199(173) & -0.701(56) \\
	\enh      & & 0.247(39) & -1.987(138) & -0.658(46) \\
	\enm      & & 0.313(30) & -1.863(129) & -0.621(44) \\
	\enl      & & 0.303(29) & -1.896(176) & -0.640(55) \\
	Lin. Fit  & & 0.297(33) & -1.773(141) & -0.608(45) \\
	Quad. Fit & & 0.354(54) & -1.903(276) & -0.639(88) \\ 
	\colrule
	%
	\texttt{ID} & $\langle r_{E,\occ}^2 \rangle$ & $\langle r_{M,\occ}^2 \rangle$ & $G_{M,\occ}(0)$ & $\mu_{\occ}$  \\
	\colrule
	\ensh     &  0.038(8)\pz & 0.122(12) & 1.600(71) & 0.389(18) \\
	\enh      &  0.019(6)\pz & 0.109(12) & 1.567(47) & 0.386(11) \\
	\enm      &  0.040(6)\pz & 0.138(11) & 1.616(45) & 0.400(11) \\
	\enl      &  0.029(6)\pz & 0.130(13) & 1.621(50) & 0.402(15) \\
	Lin. Fit  &  0.032(6)\pz & 0.135(11) & 1.625(47) & 0.405(13) \\
	Quad. Fit &  0.043(11)   & 0.148(21) & 1.662(87) & 0.413(24) \\ \botrule
	\end{tabular} \label{tab:res_s12}
	}
\end{table}
\begin{table}[ht]
	\tbl{Individual quark sector contributions to the electric charge radii, magnetic charge radii and the magnetic moments of the charmed baryons. Note that the numbers are given independently from the electric charge of the individual quarks that compose the baryons. Charge radii and magnetic moments are given in units of ${\rm fm}^2$ and nuclear magnetons $\mu_N$ respectively. }
	{
	\begin{tabular}{@{}cccccccc@{}} \toprule
	Baryon & \texttt{ID} & $\langle r_{E}^2 \rangle_q$  & $\langle r_{E}^2 \rangle_Q$ & $\langle r_{M}^2 \rangle_q$ & $\langle r_{M}^2 \rangle_Q$  & $\mu_q$ & $\mu_Q$ \\
	\colrule
	$\Sigma_c^{0,++}$	& \ensh & 0.289(49) & 0.091(22)   & 0.444(55) & 0.067(43)\pz   & \pneg2.178(188)    & -0.080(16)  \\
						& \enh  & 0.273(41) & 0.054(12)   & 0.346(52) & 0.063(25)\pz   & \pneg2.046(248)    & -0.096(16)  \\
						& \enm 	& 0.353(65) & 0.042(18)   & 0.394(69) & 0.185(152)     & \pneg2.427(572)    & -0.115(26)  \\				
	 					& \enl  & 0.338(60) & 0.057(25)   & 0.506(99) & 0.141(165)     & \pneg2.581(555)    & -0.061(31)  \\	
	\\		
	$\Xi_{cc}^{+,++}$	& \ensh & 0.264(29) & 0.071(5)\pz & 0.471(38) & 0.079(7)\pz\pz & -0.474(34)\pz      & \pneg0.414(11)   \\
						& \enh  & 0.282(25) & 0.056(5)\pz & 0.371(40) & 0.078(8)\pz\pz & -0.416(33)\pz      & \pneg0.421(9)\pz \\
						& \enm  & 0.379(38) & 0.063(5)\pz & 0.473(58) & 0.076(8)\pz\pz & -0.434(52)\pz      & \pneg0.420(9)\pz \\				
	 					& \enl  & 0.358(38) & 0.080(7)\pz & 0.477(85) & 0.085(8)\pz\pz & -0.471(86)\pz      & \pneg0.432(11)   \\						
	\\
	$\Omega_{c}^0$		& \ensh & 0.253(35) & 0.074(20)   & 0.424(53) & 0.090(37)\pz   & \pneg2.080(155)    & -0.071(17)  \\
						& \enh  & 0.199(34) & 0.049(15)   & 0.300(54) & 0.064(21)\pz   & \pneg1.833(144)    & -0.098(13)  \\
						& \enm  & 0.320(28) & 0.076(13)   & 0.405(44) & 0.096(30)\pz   & \pneg1.785(144)    & -0.088(10)  \\				
	 					& \enl  & 0.313(36) & 0.061(10)   & 0.405(38) & 0.053(18)\pz   & \pneg1.838(183)    & -0.099(18)  \\
	\\
	$\Omega_{cc}^+$		& \ensh & 0.249(29) & 0.071(7)\pz & 0.405(49) & 0.077(11)\pz   & -0.402(32)\pz      & \pneg0.412(17)   \\
						& \enh  & 0.198(18) & 0.051(6)\pz & 0.253(25) & 0.073(9)\pz\pz & -0.356(19)\pz      & \pneg0.411(11)   \\
						& \enm  & 0.276(22) & 0.082(6)\pz & 0.367(40) & 0.096(9)\pz\pz & -0.370(27)\pz      & \pneg0.432(9)\pz \\				
	 					& \enl  & 0.316(48) & 0.074(9)\pz & 0.385(47) & 0.088(10)\pz   & -0.393(33)\pz      & \pneg0.436(15)   \\	\botrule
	\end{tabular} \label{tab:res_s12_q}
	}
\end{table}

%
Extracted charge radii and magnetic moments are collected in \Cref{tab:res_s12}. Values at the physical light-quark mass are estimated via extrapolations linear and quadratic in pion mass-squared,
\begin{align}
	\label{eq:lin}
	f^{L}(m^2_\pi) &= a_L m^2_\pi + b_L, \\
	\label{eq:quad}
	f^{Q}(m^2_\pi) &= a_Q (m^2_\pi)^2 + b_Q m^2_\pi + c_Q,
\end{align}
where $a_{L,Q}$, $b_{L,Q}$, and $c_Q$ are free fit parameters. We observe that there are one to two sigma deviations between the values obtained by these extrapolations, in particular for $\sgc$. In the case of the charmed-strange baryons $\omc$ and $\occ$, the pion-mass dependence is solely due to sea-quark effects. Their charge radii and magnetic moments show slight fluctuations or a linear behavior as they approach to the physical quark mass limit. These fluctuations however might as well be due to limited statistics or uncontrolled systematics. 

Charmed baryons that have the same electric charge can be compared to each other to understand the effects of their internal dynamics. Considering the quadratic fit values, the electric charge radii of $\occ^+$ and $\xcc^+$ are about the same size, $\langle r^2_{E,\occ^+} \rangle = 0.043(11) \, {\rm fm}^2$ and $\langle r^2_{E,\xcc^+} \rangle = 0.042(9) \, {\rm fm}^2$, which is much smaller compared to proton ($\langle r^2_{E,p} \rangle = 0.707 \, {\rm fm}^2$~\cite{PDG:2020}). The only internal difference between $\omc^+$ and $\xcc^+$ is that a light quark is changed to a strange quark or vice versa, yet the $s$-quark in $\occ^+$ seems to have no extra effect on charge radius with respect to the light quark in $\xcc^+$. Of all the four charged baryons that has been studied, $\sgc^{++}$ has the largest charge radius, due to it having two units of electric charge and a single charm quark. 

A deeper insight into quark dynamics can be obtained by investigating the individual quark sector contributions. In \Cref{tab:res_s12_q} light- ($u/d$, $s$) and $c$-quark distributions within the baryons are shown. It is evident that the light quark contributions are systematically larger than those of the charm quark. From a naive quark model perspective one may interpret the smaller values of the $c$ quark as it acting as a heavy core to shift the center of mass towards itself thus reducing the size of the baryon. The difference in the $c$-quark distributions between the singly and the doubly charmed baryons is small. Similarly, the $u/d$- and $s$-quark distributions are roughly the same. In this case, much of the difference arises due to the electric charges and the representation of the valence quarks in the baryon. For instance $\occ^+$ and $\xcc^+$ have almost the same sizes whereas the charge radius of $\sgc^{++}$ is slightly larger than that of $\xcc^{++}$ since the doubly represented $u$ quark has larger contribution than the $c$ quark.
Note that the light-quark contributions to the charge radii in $\xcc$ show a tendency of being slightly larger than the strange-quark contributions in $\occ$. Although it is possible to interpret these values as an indication of light-quarks having a larger spread around the $cc$ piece, or vice versa, this difference might decrase or vanish after taking the chiral, infinite volume and continuum limits. 

Pion-mass dependence of the charge radii is also of interest. As the $u/d$ quark in $\sgc$ and $\xcc$ baryons becomes lighter, the radius of the light quark increases. This can be understood by a shift in the center of mass towards the heavy $c$ quark leading the light quark to have a larger distribution. The behavior is, however, rather peculiar when the mass of the $u/d$ quark increases. Initially, the charge radii decrease but they increase as we approach to $s$-quark mass. Unlike the nucleon, it is interesting that the charge radii do not systematically decrease as the pion mass increases. This behavior might be due to a modification of the confinement force in baryons such that the two charm quarks assume a compact nature in the $\xcc$ and the effect of the extra light quark modifies the string tension between the two-charm component~\cite{Yamamoto:2007nn}. Here it is interesting to note that these electric charge radii results have been utilized to investigate the charge symmetry breaking in the $\xcc$ baryon masses~\cite{Cushman:2018zza} to further understand the peculiarities of $\xcc$ baryons. Quark sectors of the $\occ^+$ baryon show a somewhat unstable quark-mass dependence making it harder to interpret. A detailed analysis might be rather informative for sea-quark dynamics since the variation is solely due to the effects of the $u/d$ quarks in the sea, assuming it is not related to lattice systematics. 

Magnetic charge radii of the charmed baryons have a similar pattern as their electric charge radii. $\sgc^{++}$ has the largest magnetic radii amongst other charmed baryons. While $\sgc^{++}$ and $\sgc^0$ have magnetic radii,$\langle r_{M,\sgc^{++}}^2\rangle = 0.696(153) \, {\rm fm}^2$ and $\langle r_{M,\sgc^0}^2\rangle = 0.650(126) \, {\rm fm}^2$, comparable to that of the proton, $\langle r_{M,p}^2\rangle = 0.724 \, {\rm fm}^2$~\cite{PDG:2020}, others have much smaller values. $\xcc^+$ and $\occ^+$ have the smallest magnetic charge radii $\langle r_M^2\rangle \sim 0.15 \, {\rm fm}^2$ ,comparable to each other. Unlike the electric sector, magnetic moments of $\omc$ and $\occ$ are rather stable with respect to pion mass but show a mild dependence, which suggests that there might be subtle sea-quark effects.

It is instructive to investigate the individual quark sector contributions to the magnetic moments of the baryons. Looking at the last two columns of \Cref{tab:res_s12_q} one again sees that the charm quark contributions are suppressed for the singly charmed $\sgc$ and $\omc$ baryons where their magnetic moments are dominantly determined by the light degrees of freedom. There is a pronounced effect for the doubly charmed $\xcc$ and $\occ$ baryons where the quark sector contributions are similar in magnitude. It follows that the heavy quark plays an equivalent role to the light quark only when it is doubly represented in the baryons. 

Signs of the magnetic moments reveal the interplay of the spin degrees of freedom of the quarks further. The opposite signs of the light- and heavy-quark magnetic moments indicate that their spins are anti-aligned in the baryon. The spins of the singly charmed $\sgc$ and $\omc$ baryons are mainly determined by the doubly represented light quarks. Generally speaking, when a quark is doubly represented, the quarks are paired in a spin-1 state with their spins aligned. In the case of the doubly charmed baryons, this leads to a larger heavy-quark contribution to the total spin and magnetic moment. $\sgc^{++}$ has the largest magnetic moment of all and the strange-charmed baryons $\omc$ and $\occ$ have smaller moments. Overall, charmed baryon magnetic moments are smaller than proton's, $\mu_p=2.793 \, \mu_N$~\cite{PDG:2020}.

\begin{table}[ht]
	\tbl{Comparison of the lattice magnetic moments to various other non-lattice calculations. All values are given in nuclear magnetons [$\mu_N$]. Numbers in parentheses indicate the error on the last digits. }
	{
	\begin{tabular}{@{}lccccc@{}} \toprule
	& $\mu_{\Sigma_{c\pc}^{0\pz}}$ & $\mu_{\Sigma_{c\pc}^{++}}$ & $\mu_{\xcc^{+\pz}}$ & $\mu_{\Omega_{c\pc}^{0\pz}}$ & $\mu_{\occ^{+\pz}}$ \\
	\colrule
	Latt. (Lin. fit)              & -0.852(133)  & 1.569(253)   & 0.411(15)\pnpz        & -0.608(45)    & 0.405(13)\pnpz \\
	Latt. (Quad. fit)             & -1.073(269)  & 2.220(505)   & 0.425(29)\pnpz        & -0.639(88)    & 0.413(24)\pnpz \\
	Ref.~\citen{JuliaDiaz:2004vh} & -1.78\spz    & 3.07\spz     & 0.94\stpz\pnpz        & -0.90\stpz    & 0.74\stpz\pnpz \\
	Ref.~\citen{Faessler:2006ft}  & -1.04\spz    & 1.76\spz     & 0.72\stpz\pnpz        & -0.85\stpz    & 0.67\stpz\pnpz \\
	Ref.~\citen{Albertus:2006ya}  & ---\spz      & ---\spz      & $0.785(^{+50}_{-30})$ & ---\stpz      & $0.635(^{+12}_{-15})$ \\
	Ref.~\citen{Bernotas:2012nz}  & -1.043\thpz  & 1.679\thpz   & 0.722\etpz\pnpz       & -0.774\etpz   & 0.668\etpz\pnpz \\ 
	Ref.~\citen{Sharma:2010vv}    & -1.60\spz    & 2.20\spz     & 0.84\stpz\pnpz        & -0.90\stpz    & 0.697\etpz\pnpz \\
	Ref.~\citen{Barik:1984tq}     & -1.391\thpz  & 2.44\spz     & 0.774\etpz\pnpz       & -0.85\stpz    & 0.639\etpz\pnpz \\
	Ref.~\citen{Kumar:2005ei}     & -1.17\spz    & 2.18\spz     & 0.77\stpz\pnpz        & -0.92\stpz    & 0.70\stpz\pnpz  \\
	Ref.~\citen{Patel:2007gx}     & -1.015\thpz  & 2.279\thpz   & ---\spz               & -0.960\etpz   & 0.785\etpz\pnpz \\
	Ref.~\citen{Zhu:1997as}       & -1.6(2)\fpz  & 2.1(3)\fpz   & ---\spz               & ---\stpz      & ---\spz         \\
	Ref.~\citen{Li:2017cfz}       & ---\spz      & ---\spz      & 0.85\stpz\pnpz        & ---\stpz      & 0.78\stpz\pnpz  \\
	Ref.~\citen{Yang:2018uoj}     & -1.24(5)\tpz & 2.15(10)\twpz& ---\spz               & -0.85(5)\twpz & ---\spz         \\
	Ref.~\citen{Simonis:2018rld}  & -1.31\spz    & 2.28\spz     & 0.719\etpz\pnpz       & -0.950\etpz   & 0.645\etpz\pnpz \\
	Ref.~\citen{Kim:2018nqf}      & -0.79\spz    & 1.58\spz     & ---\spz               & -0.79\stpz    & ---\spz         \\
	Ref.~\citen{Ozdem:2018uue}    & ---\spz      & ---\spz      & 0.43(9)\twpz\pnpz     & ---\stpz      & 0.39(9)\twpz\pnpz \\
	Ref.~\citen{Gandhi:2018lez}   & -1.095\thpz  & 1.835\thpz   & ---\spz               & -1.127\etpz   & ---\spz         \\
	Ref.~\citen{Li:2020uok}       & ---\spz      & ---\spz      & 0.62\stpz\pnpz        & ---\stpz      & 0.41\stpz\pnpz  \\
	\botrule
	\end{tabular} \label{tab:res_s12_modelcomp}
	}
\end{table}

Finally, a comparison of the lattice magnetic moments to those obtained from various other non-lattice methods is given in \Cref{tab:res_s12_modelcomp}. Note that we have extended upon the comparison given in Ref.~\citen{Can:2013tna} to include the results appeared after its publication. While the signs of the magnetic moments are correctly determined, there is a significant discrepancy among results. For all the baryons, majority of the moments obtained via other methods overestimate the lattice results.  

This discrepancy is rather curious and it is persistent where such discrepancies arise for \spinth baryons and several transition form factors, which will be discussed in the following sections. Although the origin of this evident discrepancy is still an open issue, the most plausible reason seems like the treatment of the heavy quark sector in model approaches. It has been shown that it is possible to improve the model predictions by treating the heavy and light quark sectors separately and determining the hyperfine splittings more accurately~\cite{Simonis:2018rld}, although a discrepancy remains for the doubly charmed sector. Additionally, there are indications that when the lattice data, rather than the quark model predictions, are used as input to fix some low-energy constants of effective theories their results get smaller than quark model's and closer to lattice predictions~\cite{Wang:2018gpl,Meng:2018gan,Xiabng:2018qsd}. It is found that the form factors of doubly charmed baryons are described well by the chiral perturbation theory calculations~\cite{Blin:2018pmj}. In Ref.~\citen{Kim:2019wbg} authors systematically calculate the electromagnetic form factors of singly-charmed \spinoh baryons with varying pion masses in the framework of the chiral quark-soliton model and compare to the present lattice data. Their scheme describes the lattice data well, hinting the applicability of the pion mean-field approximation in this sector.

%
\subsubsection{Spin-3/2 baryons} \label{sec:spin32_em}
Electromagnetic form factors of \spinth positive parity $\om$, $\ocs$, $\occs$, and $\occc$ and in comparison the \spinoh positive parity $\ocs$ and $\occ$ baryons are studied in Ref.~\citen{Can:2015exa}. Here, inclusion of the fully strange $\Omega$ baryon provides an opportunity to assess the strange and charm sectors independently. 
The quantum numbers of $\ocs \, (ssc)$, $\occs \, (scc)$, and $\occc \, (ccc)$ baryons are not measured but assigned according to the flavor SU(4) classification. They are isospin singlets, $I=0$, and have the charm-strangeness quantum number $C+S=3$. With respect to the SU(4) decomposition of the fully symmetric representation, ${\bf 20}_S \to {\bf 10} + {\bf 6} + {\bf \bar{3}} + {\bf 1}$, their ground states are expected to belong to the sextet (${\bf 6}$), antitriplet (${\bf \bar{3}}$) and singlet (${\bf 1}$) flavor SU(3) layers where ${\bf 10}$ is the familiar light baryon decuplet. The $\ocs$ baryon is observed while no experimental evidence of $\occs$ and $\occc$ exist as of yet. 

Lattice simulations are done on the near-physical-point ensemble \ensl. Light valence and sea quark masses are set equal $\kappa^{u/d,s}_{val.} = \kappa^{u/d,s}_{sea}$. Since this ensemble has almost physical light quark masses yielding $m_\pi \simeq 156 \, {\rm MeV}$, no extrapolations to the chiral limit are performed for any of the extracted quantities. Because this is an extension of the \spinoh study, charm quark parameters are kept the same. 

\begin{figure}[htb]
	\centerline{\includegraphics[width=.5\textwidth]{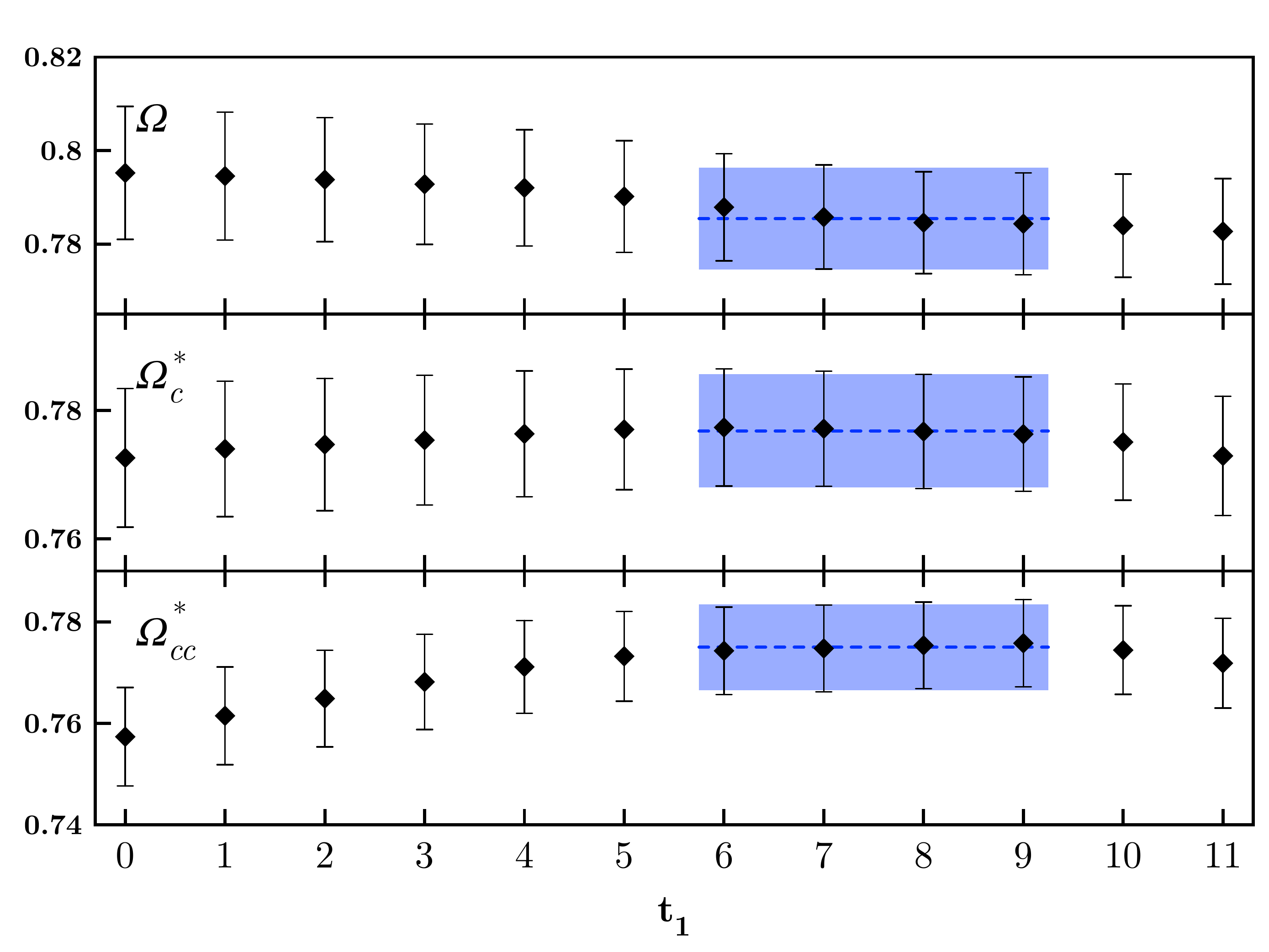} \includegraphics[width=.5\textwidth]{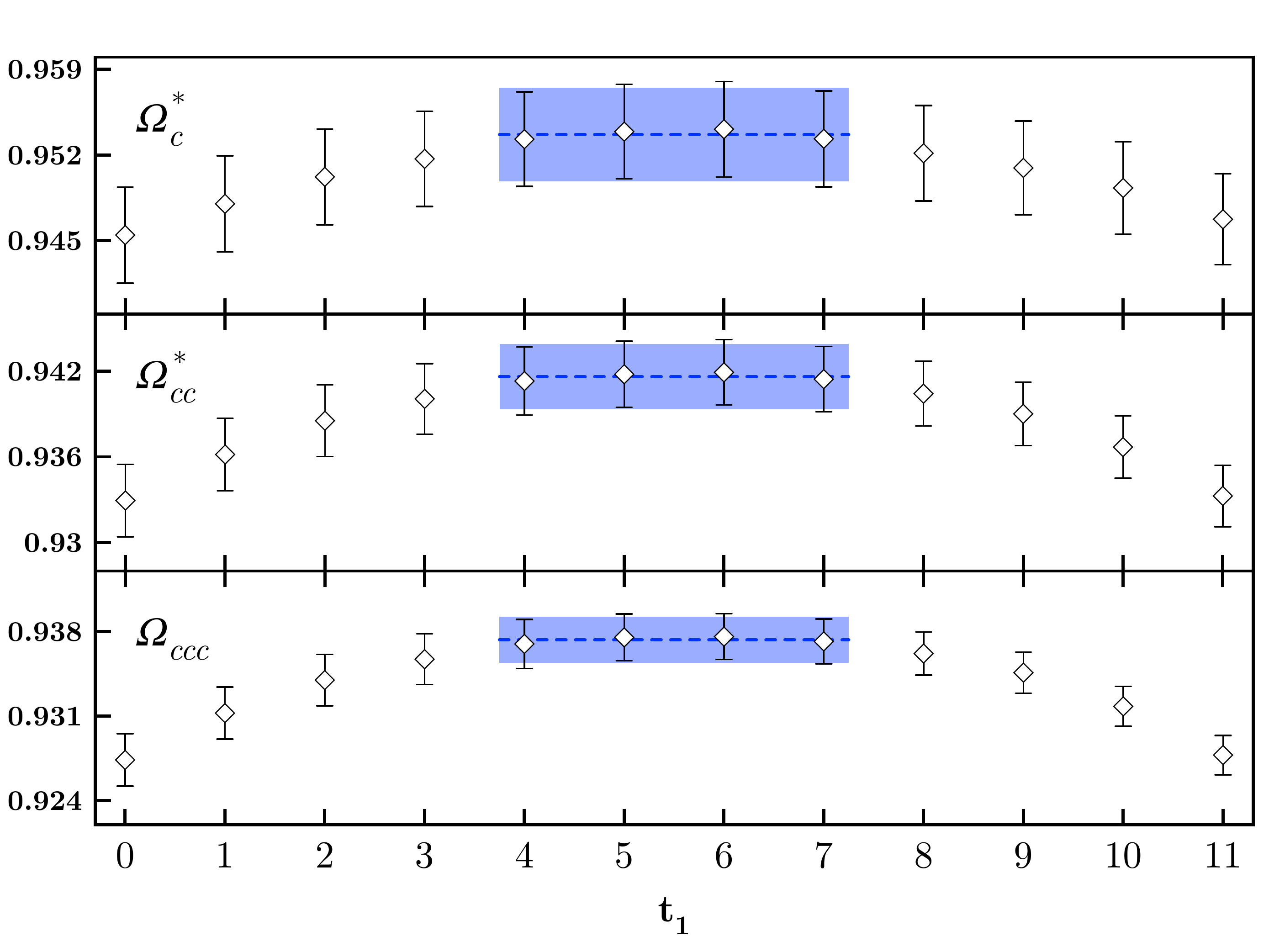}}
	\centerline{\includegraphics[width=.5\textwidth]{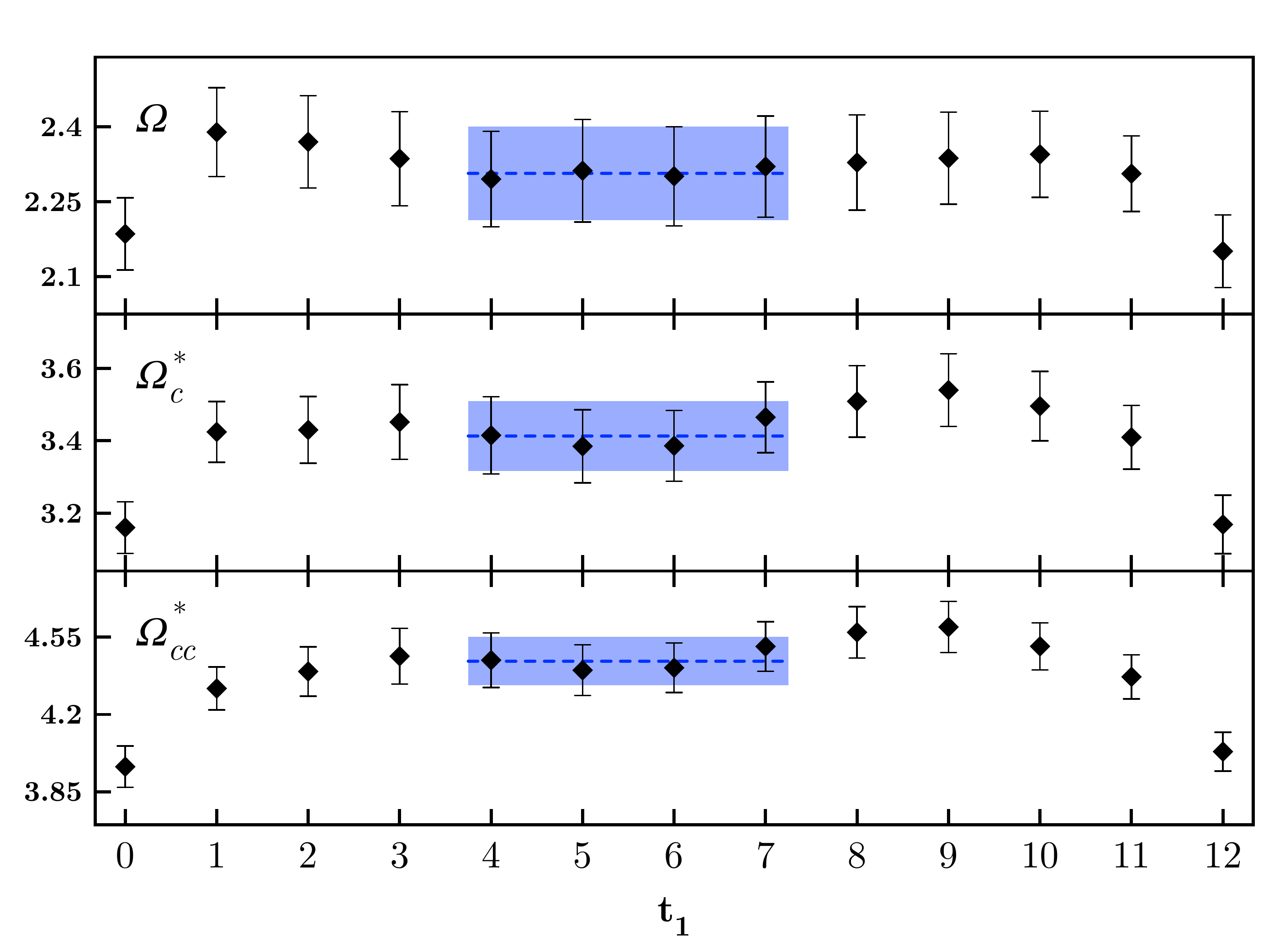} \includegraphics[width=.5\textwidth]{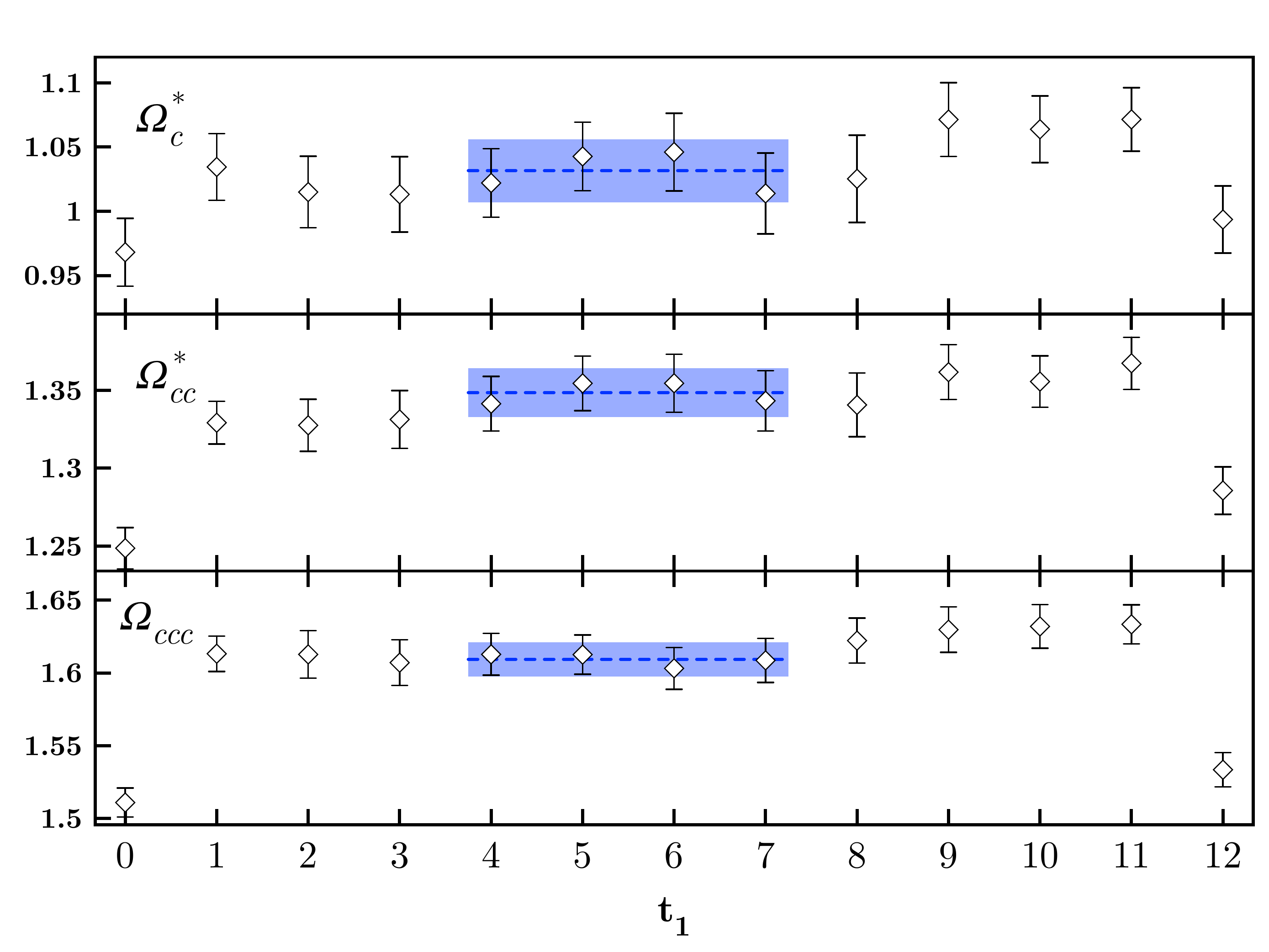}}
	\caption{Strange (left) and charm quark (right) contributions to the $E0$ and $M1$ form factors at the lowest allowed three-momentum transfer ($\mathbf{q}^2$=0.183 GeV$^2$) with respect to the current insertion time $t_1$. Contributions are shown for a single quark and normalized to unit charge. Bands indicate the fit regions. Figure taken from Ref.~\citen{Can:2015exa}. \label{fig:em_s32}}
\end{figure}

Electric charge, $E0$, form factors are evaluated at zero and lowest allowed lattice three-momentum transfer ${\bf q}^2_\text{MIN} = 0.183 \, {\rm GeV}^2$, while the magnetic dipole, $M1$, and electric quadrupole, $E2$, form factors are only evaluated at ${\bf q}^2_\text{MIN}$. Signals for the strange and charm quark contributions to $E0$ and $M1$ form factors obtained via \Cref{eq:E0lat,eq:M1lat} using the ratio given in \Cref{eq:em_lat_ratio2} are shown in \Cref{fig:em_s32}. Their values extracted via a plateau analysis are given in \Cref{tab:s32_ff}. Excited states contamination is checked by employing a summed operator insertions (SOI) method and the results are found to be consistent with each other suggesting negligible excited states effects~\cite{Can:2015exa}.

\begin{figure}[htb]
	\centerline{\includegraphics[width=.5\textwidth]{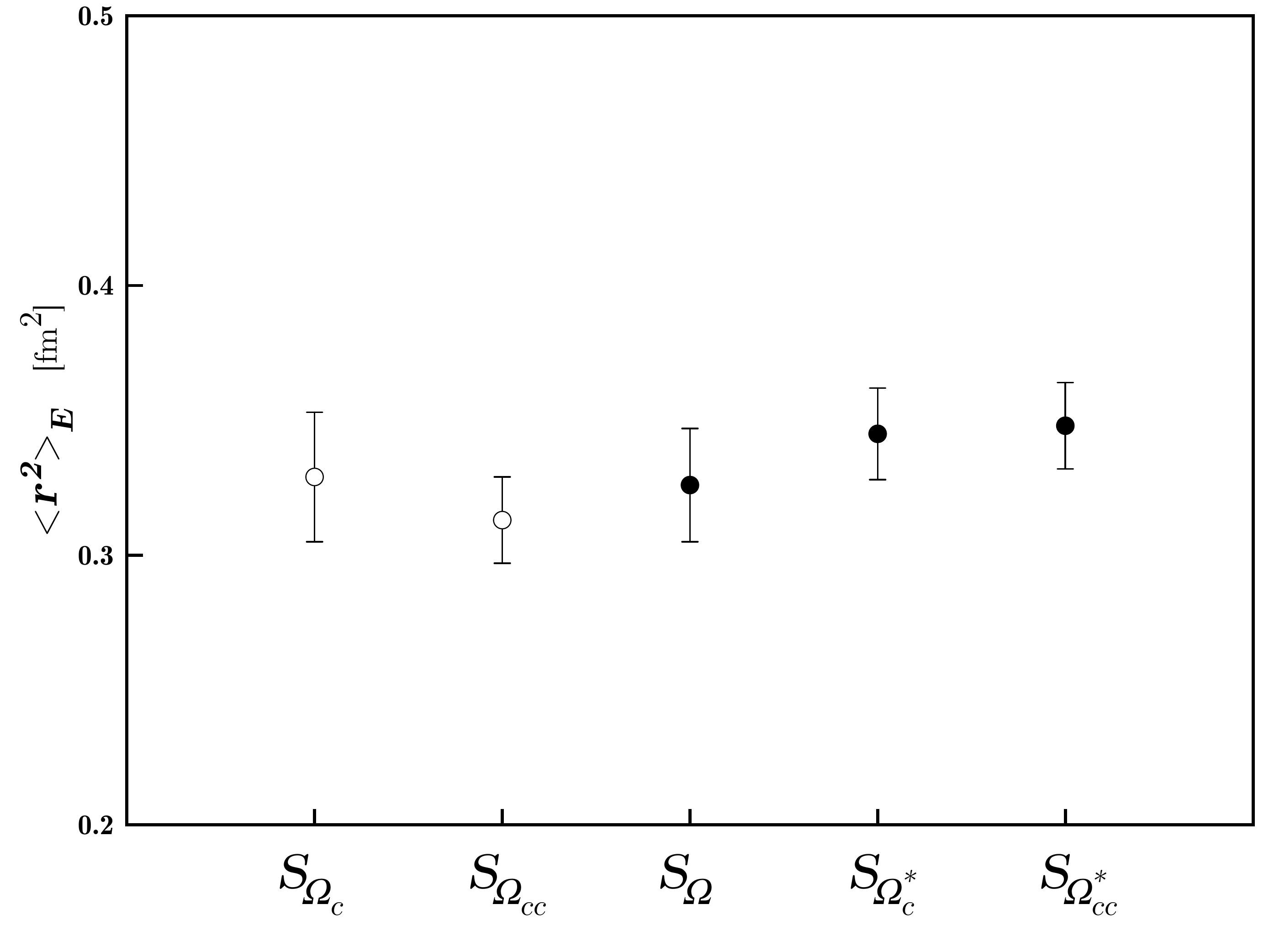} \includegraphics[width=.5\textwidth]{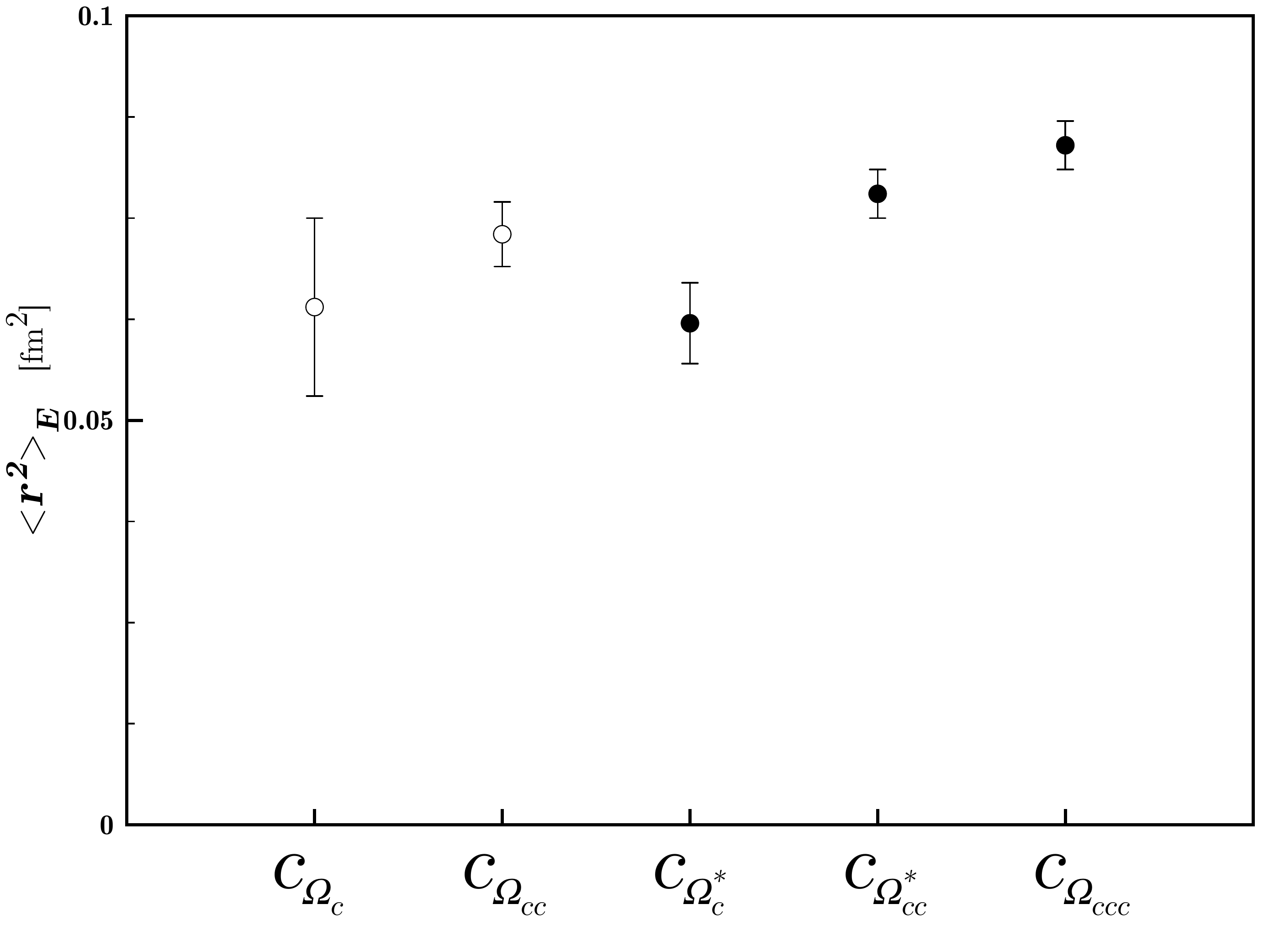}}
	\centerline{\includegraphics[width=.5\textwidth]{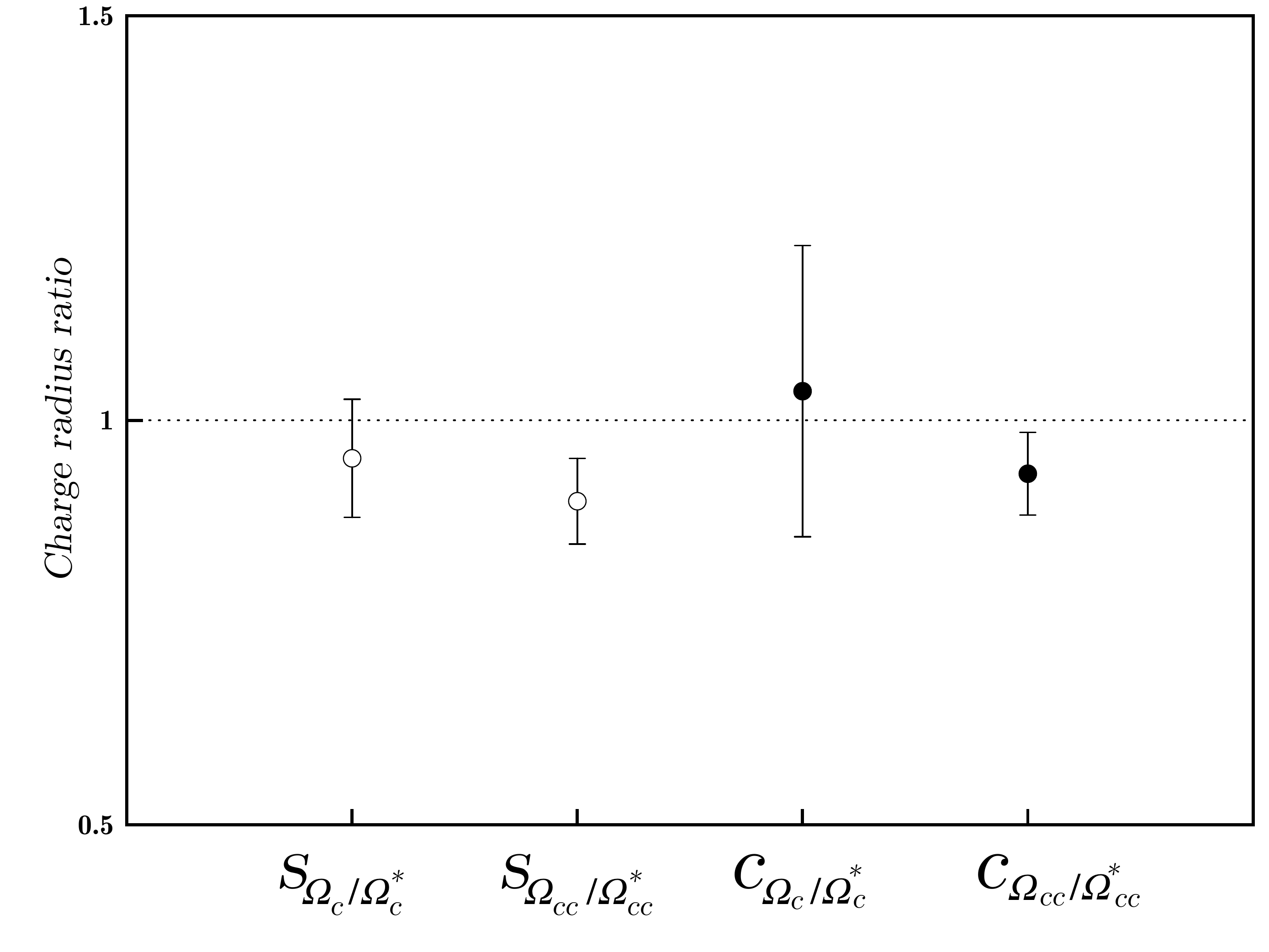} \includegraphics[width=.5\textwidth]{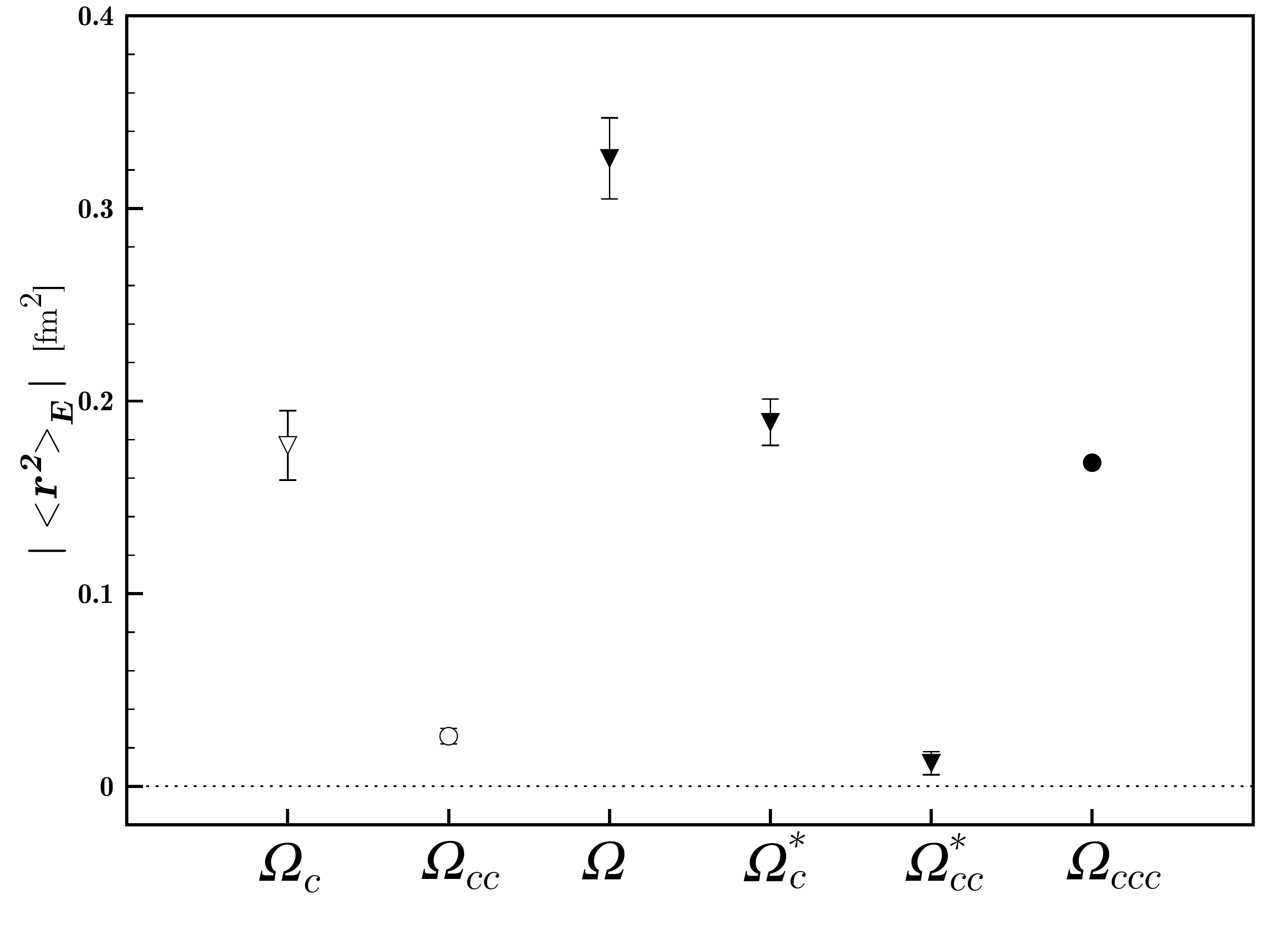}}
	\caption{Electric charge radii of \spinoh (open) and \spinth (filled) charmed-strange baryons. Upper plots show the strange and charm quark contributions while the bottom row holds the ratios of quark contributions, $Q_{B/B^\ast} \equiv \langle r^2_E \rangle^B_q/\langle r^2_E \rangle^{B^\ast}_q$, and the total radii. Absolute values are shown for a better comparison. Data points denoted by a triangle indicate a negative value. Figures taken from Ref.~\citen{Can:2015exa}. \label{fig:radii_s32}}
\end{figure}

\begin{figure}[htb]
	\centerline{\includegraphics[width=.5\textwidth]{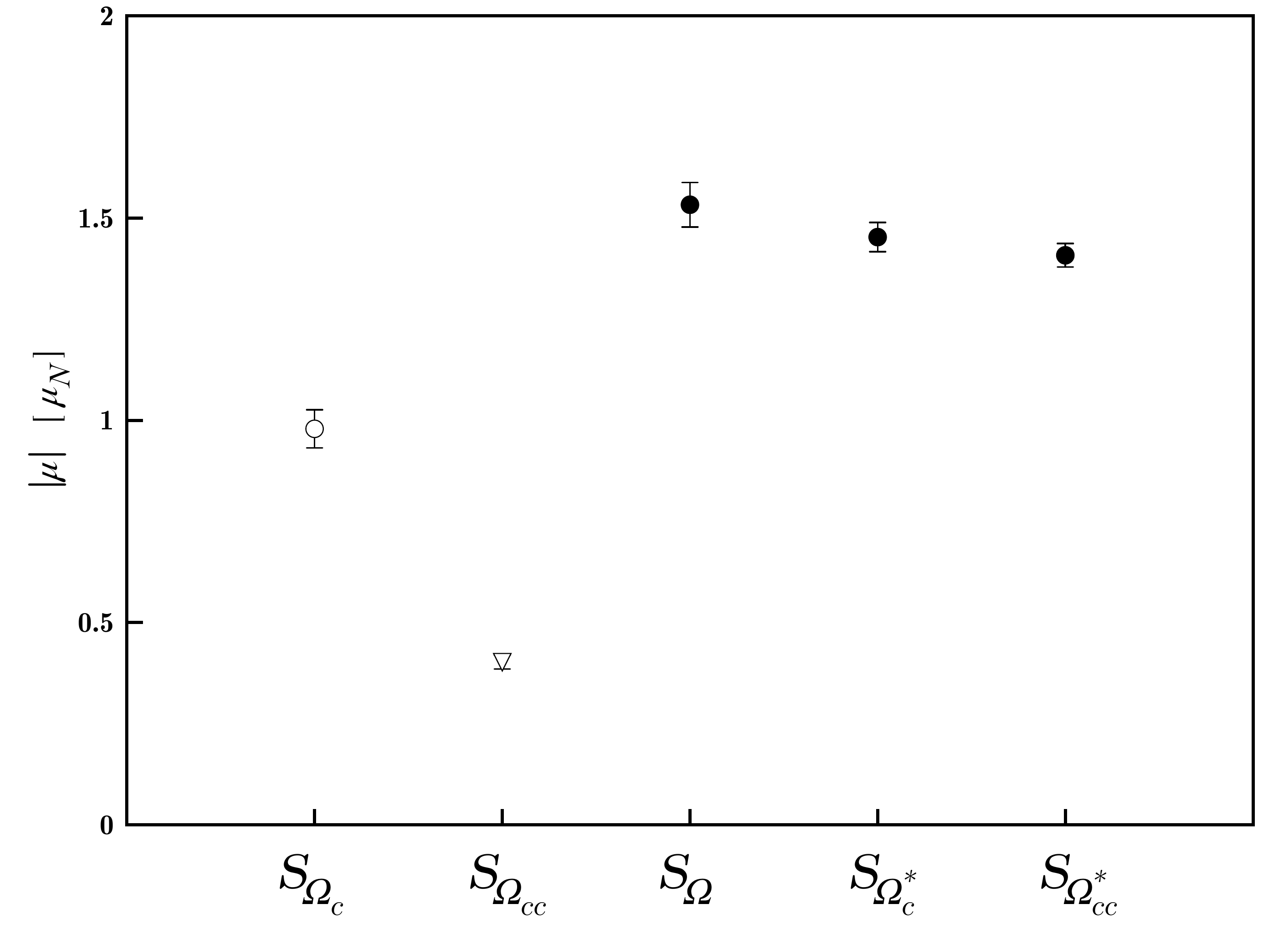} \includegraphics[width=.5\textwidth]{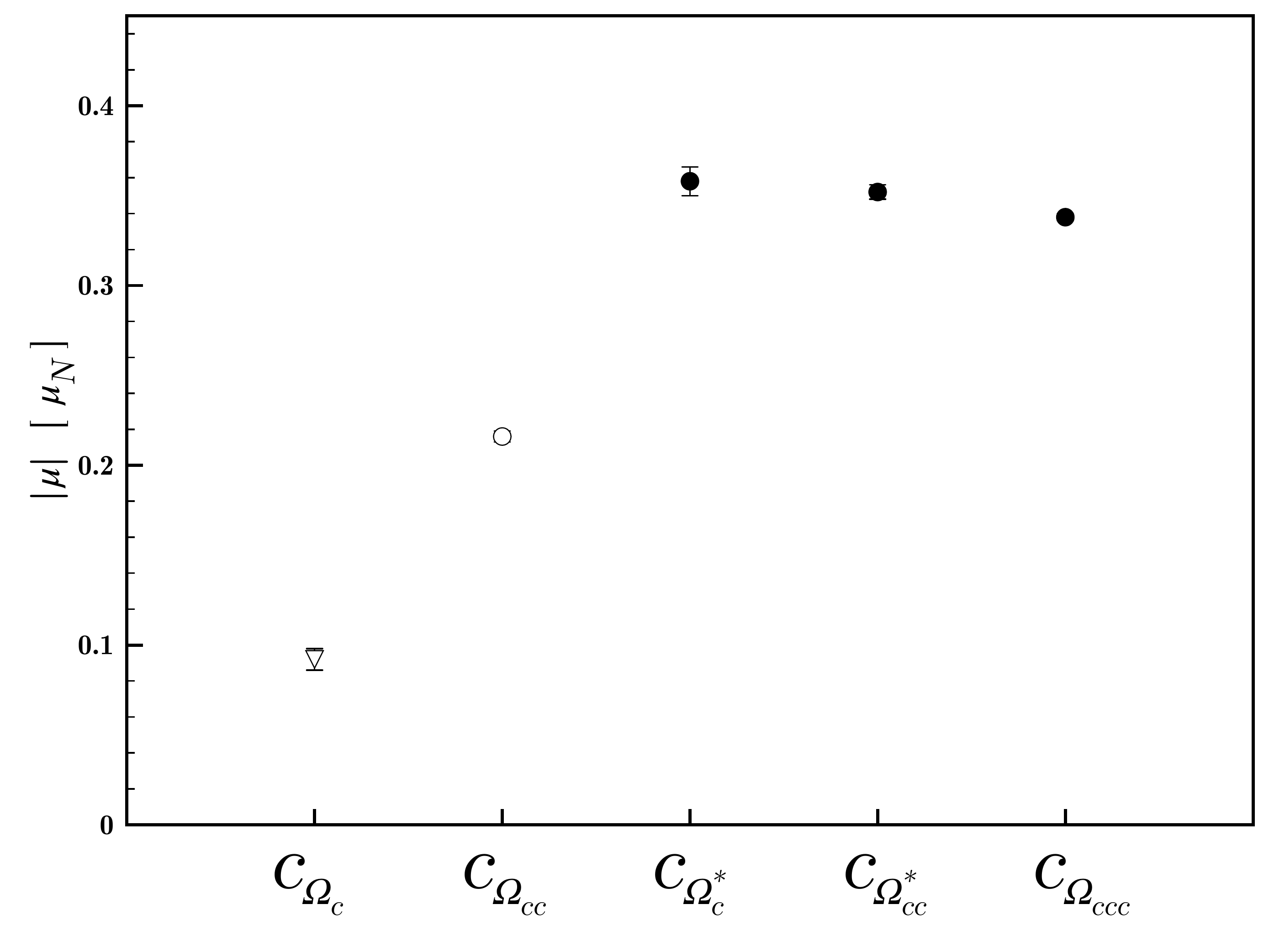}}
	\centerline{\includegraphics[width=.5\textwidth]{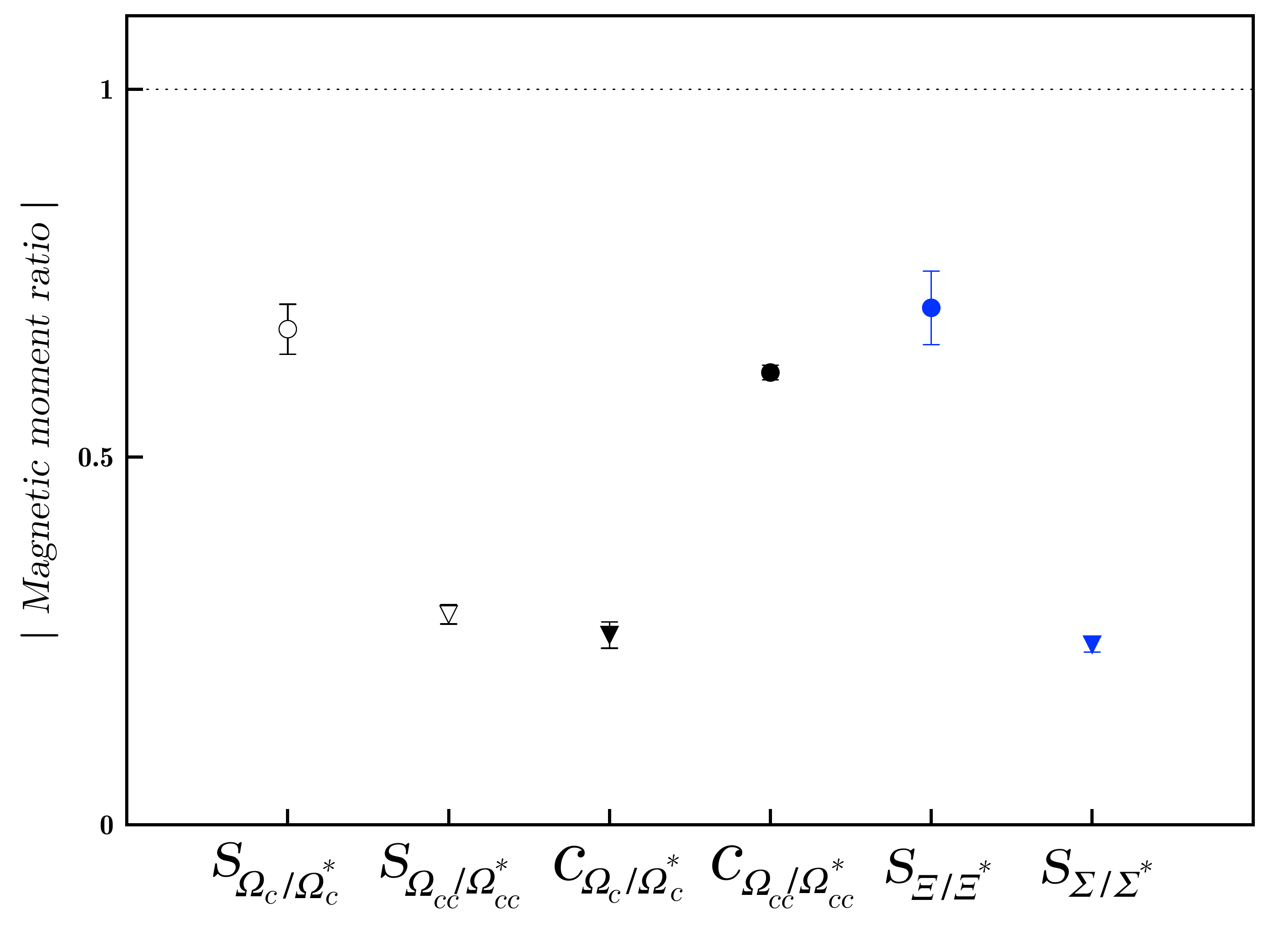} \includegraphics[width=.5\textwidth]{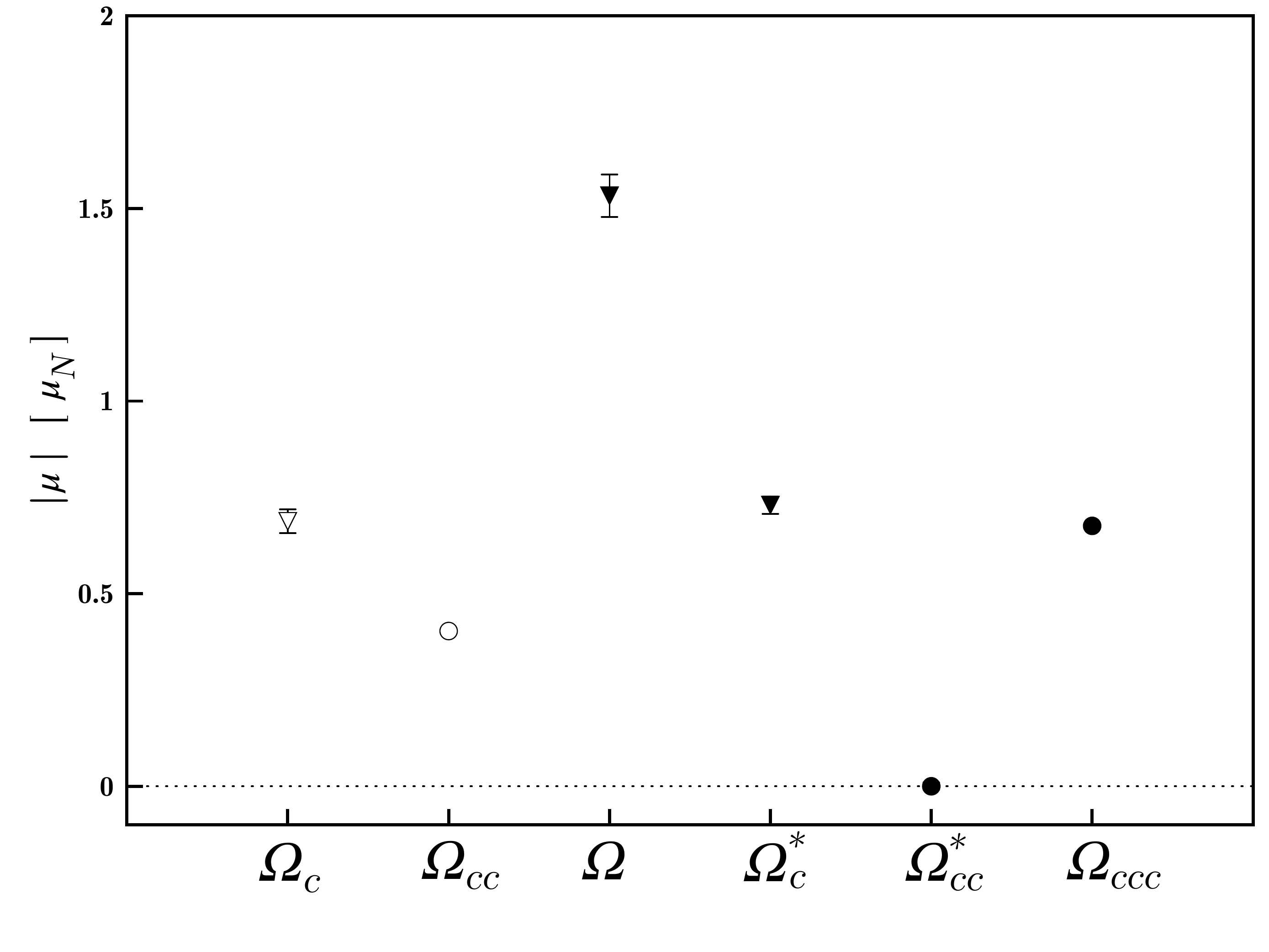}}
	\caption{ Same as \Cref{fig:radii_s32} but for the magnetic moments. Rightmost blue data points are octet/decuplet ratios
calculated using the $m_\pi = 300 \, {\rm MeV}$ quenched simulation results of Refs.~\citen{PhysRevD.74.093005,PhysRevD.80.054505}. Figures taken from Ref.~\citen{Can:2015exa}. \label{fig:moms_s32}}
\end{figure}

In this case, a fit to a dipole form is not possible however assuming a dipole ansatz describes the $E0$ and $M1$ form factors, one can extract the charge radii via,
\begin{equation} \label{eq:32radii}
	\frac{\langle r^2_{E} \rangle}{G_{E0}(0)} = \frac{12}{Q^2_\text{MIN}} \left( \sqrt{\frac{G_{E0}(0)}{G_{E0}(Q^2_\text{MIN})}} - 1\right).
\end{equation}
The magnetic moments can be estimated by a scaling approach given in \Cref{eq:scaling}. Extracted electric charge radii and magnetic moments are given in \Cref{tab:res_s32_e0m1} and illustrated in \Cref{fig:radii_s32,fig:moms_s32} for the ease of discussion.

\begin{table}[ht]
	\tbl{$E0(Q^2)$ and $M1(Q^2)$ at ${\bf q}^2_\text{MIN}$ for $\Omega$, $\Omega_c^\ast$, $\Omega_{cc}^\ast$, and $\Omega_{ccc}$ baryons. Results are given for a single quark contribution and normalized to unit charge. $M1(Q^2)$ are given in units of natural magnetons $(\mu_\mcb \equiv e/2m_\mcb)$.}
	{
	\begin{tabular}{@{}ccccc@{}} \toprule
			                & $E0^s(Q^2)$ & $E0^c(Q^2)$	& $M1^s(Q^2)$  & $M1^c(Q^2)$ \\
	\colrule
	$\Omega_{\pc\pc\pc}$  	& 0.789(12)   & ---		    & 2.307(94)\pz & ---         \\
	$\Omega_{c\pc\pc}^\ast$	& 0.778(9)\pz & 0.954(4)    & 3.413(96)\pz & 1.032(25)   \\
	$\Omega_{cc\pc}^\ast$   & 0.775(8)\pz & 0.942(2)    & 4.442(110)   & 1.349(16)   \\
	$\occc$                 & ---	      & 0.937(2)    & ---          & 1.609(12)   \\ \botrule
	\end{tabular} \label{tab:s32_ff}
	}
\end{table}

%
\begin{table}[ht]
	\tbl{Electric charge radii and magnetic moments of the charmed-strange baryons. Results are given in fm$^2$ and nuclear magnetons ($\mu_N$) respectively. Quark sector contributions are for a single quark and normalized to unit charge. Electric charge radii of spin-1/2 baryons are estimated through form factor fits~\cite{Can:2015exa}. }
	{
	\begin{tabular}{@{}ccccccc@{}} \toprule
        & $\langle r_E^2 \rangle_s$ 	& $\langle r_E^2 \rangle_c$	& $\langle r_E^2 \rangle$ & $\mu_s$ & $\mu_c$ & $\mu$ \\
		\colrule
		$\Omega_{c\pc\pc}$	    & 0.329(25) & 0.064(11)   & -0.177(18)       & \pneg0.979(47) & -0.092(6)	  & -0.688(31)		 \\
		$\Omega_{cc\pc}$	    & 0.313(16) & 0.073(4)\pz & \pneg0.026(4)\pz & -0.402(17) 	  & \pneg0.216(3) & \pneg0.403(7)\pz \\		
		$\Omega_{\pc\pc\pc}$ 	& 0.326(21) & ---	      & -0.326(21)       & \pneg1.533(55) & ---	   	      & -1.533(55) 	 	 \\
		$\Omega_{c\pc\pc}^\ast$ & 0.345(17) & 0.062(5)\pz & -0.189(12)       & \pneg1.453(36) & \pneg0.358(8) & -0.730(23) 		 \\
		$\Omega_{cc\pc}^\ast$  	& 0.348(16) & 0.078(3)\pz & -0.012(6)\pz     & \pneg1.408(29) & \pneg0.352(4) & \pneg0.000(10) 	 \\
		$\occc$ 	            & ---	    & 0.084(3)\pz & \pneg0.168(5)\pz & ---		      & \pneg0.338(2) & \pneg0.676(5)\pz \\ \botrule
	\end{tabular} \label{tab:res_s32_e0m1}
	}
\end{table}

It is evident from \Cref{fig:radii_s32} that strange quark contribution to the electric charge radius in all baryons is similar implying that its effect is almost independent of the quark-flavor composition of the baryon. For the charm quark, although its contribution seems to increase slightly with the increasing number of the valence $c$-quarks in the baryon, smallness of the scale makes this change negligible compared to the $s$-quark sector. 

Effects of the alignment of the spins of quarks to the charge radii are revealed in comparing the \spinoh and \spinth sectors. Apparent agreement between the contributions in \spinoh and \spinth sectors shows that the effect of the spin on the $s$- and $c$-quarks is almost non-existent. One may form the ratios of the individual quark-flavor contributions in \spinoh to that in the \spinth sector as $\left< r^2_E\right>^q_B / \left< r^2_E\right>^q_{B^\ast}$, in order to study the effect of the spin alignment more systematically. In the lower left plot of \Cref{fig:radii_s32} this ratio is constructed for the $s$- and $c$-quarks in singly- and doubly-charmed baryons. In the singly-charmed $\Omega_c$ baryon the $s$- and $c$-quark charge distributions are seen to be insensitive to the spin-flip of the $c$-quark while a deviation from one is an indication of an increase in their contributions in case of the doubly-charmed $\Omega_{cc}$ baryon.

Total electric charge radii of the baryons are shown in the lower right plot of \Cref{fig:radii_s32}. In magnitude, $\Omega$ baryon has the largest electric charge radius amongst all. Note that its value $\langle r_E^2 \rangle_{\Omega^-} = -0.326(21)$ fm$^2$, is in good agreement with other lattice determinations \cite{Alexandrou:2010jv,PhysRevD.80.054505}. $\omc$, $\ocs$ and $\occc$ have rather similar charge radii while the charge radii of the $\occ$ and $\occs$ almost vanish. Under a naive assumption based on the similarity of the quark contributions to the charge radii one may consider the quark sector contributions in the \spinth sector to be same so that, $\langle {r_E^2} \rangle^s_\om = \langle r_E^2 \rangle^s_{\ocs} = \langle r_E^2 \rangle^s_{\occs} = R^2_s$ and $\langle r_E^2 \rangle^c_{\ocs} = \langle r_E^2 \rangle^c_{\occs} = \langle r_E^2 \rangle^c_{\occc} = R^2_c$. With these simplified definitions one can relate the electric charge radii of the \spinth baryons to each other as, $\left( \langle {r_E^2} \rangle_{\ocs} + \langle {r_E^2} \rangle_{\occc} \right)/2 = \langle {r_E^2} \rangle_{\occs}$. Inputting the results given in \Cref{tab:res_s32_e0m1}, charge radius of $\occs$ evaluates to $\left(\langle {r_E^2} \rangle_{\ocs} + \langle {r_E^2} \rangle_{\occc} \right)/2 = -0.011(8)$, which compares nicely to the calculated charge radius of $\occs$, $\langle {r_E^2} \rangle_{\occs} = -0.012(6)$. This good agreement implies that the contribution to the charge radii from each flavor is similar for these baryons and their radii differ due to their different quark compositions. 

One insight we have gained from the \spinoh sector was that a single quark's contribution to the magnetic moment increased significantly if it is doubly represented in the baryon. From the plots shown in \Cref{fig:moms_s32} one concludes that this is true for the \spinth sector as well where doubly or triply represented quark sectors dominate the total magnetic moment. Comparing the quark sector contributions in \spinoh $\omc$ and $\occ$ baryons to that of the \spinth $\ocs$ and $\occs$ baryons, a sign change is evident due to the spin flip. Contributions in the \spinth sector are larger compared to \spinoh, which can be understood by the change in the configuration of the baryons. In order to compose a \spinoh baryon, one of the quark sectors should be anti-aligned with the total spin of the baryon causing an overall decrease in the contributions unlike the \spinth baryons, where the spin of all the valence quarks and the total spin of the baryon are aligned which enhances the contributions. Within the \spinth baryons on the other hand, $s$-quark contributions have a slight tendency to decrease with the decreasing number of valence $s$-quarks. Contributions of the $c$-quark, however, tend to decrease as the number of valence $c$-quarks increase. 

Effects of the quark spin configurations on the quark magnetic moments can be studied further by forming the ratios of the quark-sector contributions in \spinoh and \spinth baryons, $\mu^q_B / \mu^q_{B^\ast}$, where $q$ is the quark flavor and $B$ is the baryon. Such ratios together with the octet-decuplet ratios extracted from Refs.~\citen{PhysRevD.74.093005}~and~\citen{PhysRevD.80.054505} are illustrated in the lower left plot of \Cref{fig:moms_s32}. It is evident that the contributions from both of the quark sectors are enhanced in \spinth baryons. Notice that the strange quark sector in $\omc$ and $\ocs$ and the charm quark sector in $\occ$ and $\occs$ are doubly represented and the difference is that a strange quark is exchanged by a charm quark or vice versa. A comparison of the ratios, $S_{\Omega_c/\Omega_c^\ast}$ and $C_{\Omega_{cc}/\Omega_{cc}^\ast}$, would reveal the effect of such a quark flavor exchange. Same exercise can be done for the singly strange and singly charmed baryon ratios. These ratios are respectively consistent with each other suggesting that the flavor of the quark has almost no role in the difference between the \spinoh and the \spinth baryons.  

By including the magnetic moment of the $s$-quark sectors of the octet and decuplet baryons, namely the $\Sigma$, $\Xi$, $\Sigma^\ast$ and $\Xi^\ast$ baryons, one can probe the behavior of the strange sector further with respect to its environment. Shown in the lower left plot of \Cref{fig:moms_s32}, ratios for the doubly strange baryons $S_{\Omega_c / \Omega_c^\ast}$ and $S_{\Xi / \Xi^\ast}$ agree with each other within one-sigma suggesting that the strange quark is insensitive to its accompanying quark---be it a light quark or a heavy charm quark. However, there is a slight discrepancy in singly-strange baryons when one compares the $S_{\Omega_{cc} / \Omega_{cc}^\ast}$ ratio to $S_{\Sigma / \Sigma^\ast}$. The smaller value of the $S_{\Sigma / \Sigma^\ast}$ ratio suggests that going from \spinoh to \spinth affects the $s$-quark more when its accompanying quarks are light flavored. Including the $C_{\Omega_c/\Omega^\ast_c}$ ratio into the comparison of the contributions of the singly-represented quark sectors, the effect of the environment seems less pronounced for the charm quark. However, these interpretations should be considered with caution since the compared lattice studies have very different setups where one is a full-QCD simulation with $m_\pi \sim 156 \, {\rm MeV}$ and the others are quenched simulations with $m_\pi \sim 300 \, {\rm MeV}$.  

Finally we discuss the total magnetic moments. First, the total magnetic moment of the $\Omega^-$ baryon is estimated to be $\mu_{\Omega^-} = -1.533 \pm 0.055$ $\mu_N$, smaller compared to the current experimental value of, $\mu^{exp}_{\Omega^-} = -2.02 \pm 0.05 $ $\mu_N$~\cite{PDG:2020}. This is most probably due to the mistuning of the strange quark mass on these lattice ensembles~\cite{PhysRevD.98.114505,Tiburzi:2008bk,PhysRevLett.108.112001}. Another plausible possibility is that different simulation methods might lead to different results. For instance, a determination in Ref.~\citen{PhysRevD.79.051502} by a background field method finds $\mu_{\Omega^-} = -1.93 \pm 0.08$ $\mu_N$ on $m_\pi=366$ MeV lattices while other determinations (using the same method outlined in this paper) find $\mu_{\Omega^-} = -1.697 \pm 0.065 \, \mu_N$~\cite{PhysRevD.80.054505} and $\mu_{\Omega^-} = -1.875 \pm 0.399 \, \mu_N$~\cite{Alexandrou:2010jv}, in agreement with the value in \Cref{tab:res_s32_e0m1}.

Magnetic moments of the \spinoh $\omc$ and \spinth $\ocs$ baryons are almost the same, which indicates that the spin flip of the charm quark has minimal effect, in agreement with the heavy-quark spin symmetry expectations. From a quark-model perspective one would expect the magnetic moments of the $\omc$ ($\occ$) and $\ocs$ ($\occs$) to be similar to each other and such an expectation holds for the $\omc$ and $\ocs$ while there is a striking discrepancy for the $\occ$ and $\occs$ baryons where the latter has a vanishing magnetic moment. $\omc$ and $\ocs$ have the same quark content but their spin configurations differ from each other in a way that the spin of the single $c$-quark is anti-aligned with that of the ($ss$) component in $\omc$ in contrast to $\ocs$ where the spin of all the quarks are aligned. When the quark sectors are combined relative to their electric charges, they add constructively for the $\omc$ baryon but destructively for the $\ocs$. Although their behavior is different, quark sectors combine in a balanced way leading to similar magnetic moments for $\omc$ and $\ocs$. This balance is broken in the case of doubly-charmed $\occ$ and $\occs$ and the interplay of the electric charge and valence quarks leads to a significant difference in their magnetic moments.

\begin{figure}[htb]
	\centerline{\includegraphics[width=.75\textwidth]{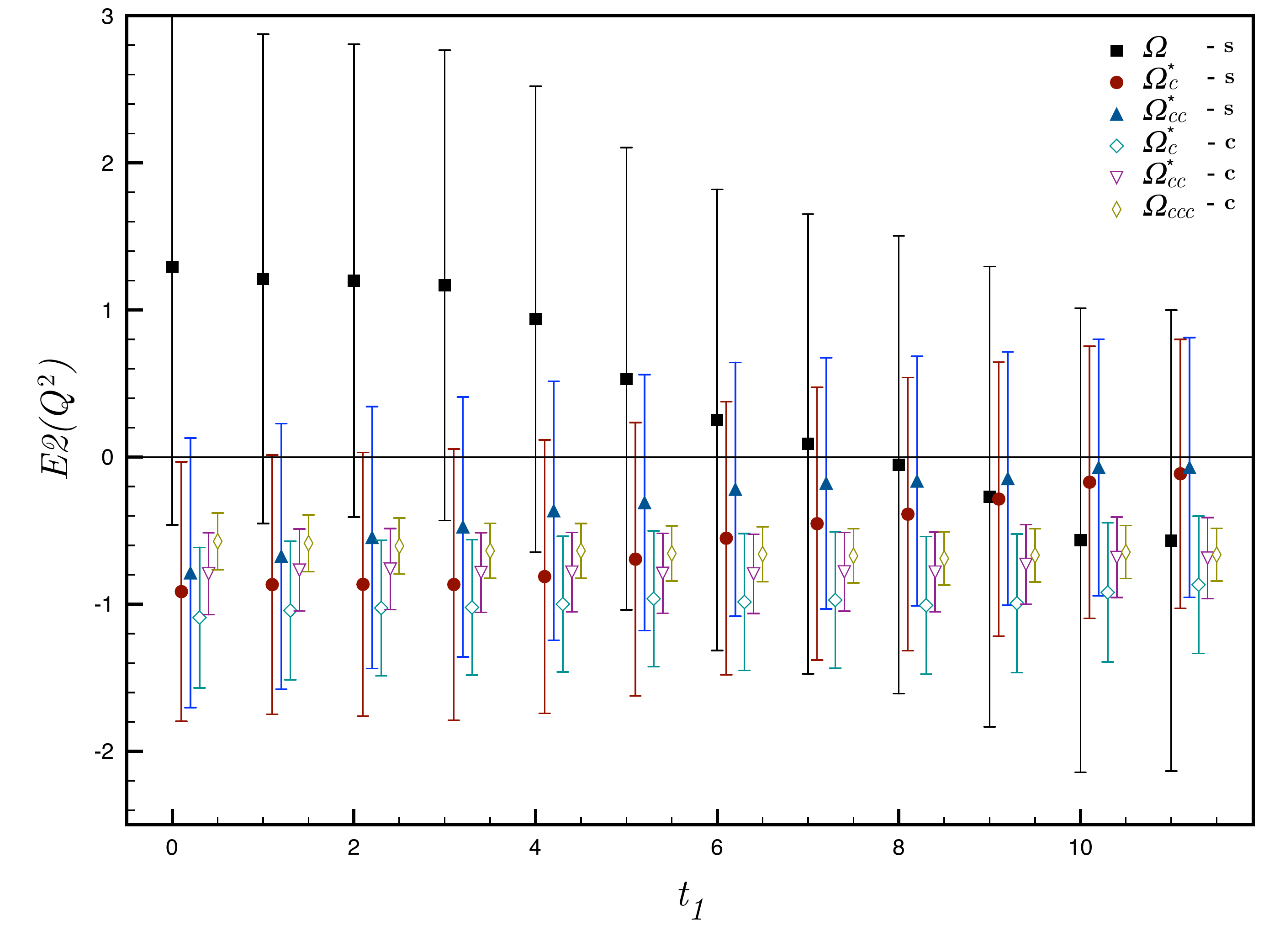}}
	\caption{Strange (filled) and charm quark (open) contributions to the $E2$ form factor at the lowest allowed three-momentum transfer (${\bf q}^2$=0.183 GeV$^2$) with respect to the current insertion time $t_1$. Data points are slightly shifted for clarity. Contributions are shown for a single quark and normalized to unit charge. Figure taken from Ref.~\citen{Can:2015exa}. \label{fig:em_s32_e2}}
\end{figure} 

The electric-quadrupole moment gives a measure of the deviation of the electric charge from a spherically symmetric distribution. It also hints to the tensor force or $D$-wave interactions. The $E2$ form factor results are estimated by the plateau approach from the signal shown in \Cref{fig:em_s32_e2}. Rather than a precise determination of the quadrupole moments, distinguishing their signs is of more interest so that the deformation of the electric charge distribution can be estimated. The form factor values quoted in \Cref{tab:s32_ff} are converted to the physical units of $[e/m^2]$ and given in \Cref{tab:res_s32_e2}. The evident poor signal in \Cref{fig:em_s32_e2} is reflected in the estimated values, especially for the strange quark contributions, where it is not possible to isolate the sign of the quadrupole moments of the $\om$ and $\ocs$ baryons. Better statistical precision of the heavy quarks, however, weights in when the heavier $\occs$ and $\occc$ baryons are considered. ${\occs}^+$ and $\occc^{++}$ have negative $E2$ moments thus their charge distributions deform to an \emph{oblate} shape. 

Available literature on $E2$ moments of charmed baryons is rather limited. A chiral quark-soliton model calculation investigates the $E2$ moments of the \spinth singly-charmed baryon sextet where a comparison to the ${\occs}$ baryon is made. Unfortunately, the large error of the lattice result prevents us from having a clear comparison, however, the central value of the lattice $E2$ moment is negative while the chiral quark-soliton model prediction~\cite{Kim:2020uqo} is positive and visibly excludes the negative values for $Q^2 < 1$.   
%
\begin{table}[ht]
	\tbl{$E2(Q^2)$ results for the $\Omega$, $\Omega_c^\ast$, $\Omega_{cc}^\ast$ and $\Omega_{ccc}$ at ${\bf q}^2_\text{MIN}$. Values are given in units of $[e/m^2]$. Quark sector contributions are for a single quark and normalized to unit charge. Last column is calculated by combining the quark sector contributions weighted by their electric charges. }
	{
	\begin{tabular}{@{}cccc@{}} \toprule
				& $E_2(Q^2)_s$ & $E_2(Q^2)_c$ 	& $E_2(Q^2)$ \\
		\colrule
		$\Omega_{\pc\pc\pc}$ 	& -0.337(1.142)  & ---			 & \pneg0.337(1.142) \\
		$\Omega_{c\pc\pc}^\ast$ & -0.371(539)\pz & -0.577(269)	 & -0.137(352)\pz    \\
		$\Omega_{cc\pc}^\ast$ 	& -0.091(277)\pz & -0.255(87)\pz & -0.310(128)\pz    \\
		$\occc$ 	            & ---		     & -0.136(38)\pz & -0.273(76)\pz\pz  \\ \botrule
	\end{tabular} \label{tab:res_s32_e2}
	}
\end{table}

\section{Radiative Transition Form Factors} \label{sec:rad_ff}

	Although the strong decay is the preferred decay mode when the phase space permits, decreased mass differences between the states in the charmed sector due to heavy-quark symmetry allow for dominantly electromagnetic decays. Therefore examining the radiative transitions of charmed baryons is a crucial element of under-standing the heavy quark dynamics.

	\subsection{Transition form factors} \label{sec:radff}
	Electromagnetic matrix element of a $\bpgb$ transition is studied with the same formalism discussed in \Cref{sec:baryonff} with an adjustment of the kinematic factors that appear in front of the form factors. Kinematic factors appearing in \Cref{eq:s12_sachs_e,eq:12_ff_e,eq:12_ff_m} are replaced by
\begin{align}
	\frac{Q^2}{4m^2_{\mcb}} &\to \frac{Q^2}{(m_{\mcb} + m_{\mcb^\prime})^2} \\
	\sqrt{\frac{2E_\mcb}{E_\mcb + m_\mcb}} &\to 2\sqrt{\frac{E_{\mcb} E_{\mcb^\prime}}{(E_{\mcb} + m_{\mcb})(E_{\mcb^\prime} + m_{\mcb^\prime})}} \\
	\sqrt{2E_\mcb (E_\mcb+m_\mcb)} &\to 2\sqrt{\frac{E_{\mcb} E_{\mcb^\prime} (E_{\mcb^\prime} + m_{\mcb^\prime})}{E_{\mcb} + m_{\mcb}}},
\end{align}
respectively, where $\mcb$ ($\mcb^\prime$) is the initial (final) baryon. Instead of repeating the same procedure here, we refer the reader to Ref.~\citen{Bahtiyar:2016dom} and references therein.

Electromagnetic transition form factors for a $\bsgb$ process is encoded into baryon matrix elements written in the following form: 
\begin{equation}\label{eq:3212_me}
	\langle \mathcal{B}^\ast_\sigma (p^\prime,s^\prime)|\mathcal{V}_\mu|\mcb(p,s)\rangle= i \sqrt{\frac{2}{3}}\left(\frac{m_\mcbp \
	m_\mcb}{E_\mcbp({\bf p^\prime})E_\mcb({\bf p})}\right)\bar{u}_\sigma(p^\prime,s^\prime) {\cal O}^{\sigma\mu} u(p,s),
\end{equation}
where $\mcb$ and $\mcbp$ denote \spinoh and \spinth baryons, respectively. $p$ and $p^\prime$ denote the initial and final four momenta and, $s$ and $s^\prime$ denote the spins. $u(p,s)$ is the Dirac spinor and $u_\sigma(p,s)$ is the Rarita-Schwinger spin vector.
Operator $\cal{O}^{\sigma\mu}$ can be parameterized in terms of Sachs form factors \cite{Jones:1972ky},
\begin{equation}\label{eq:sachs}
	{\cal O}^{\sigma\mu}=G_{M1}(Q^2) \mathcal{K}_{M1}^{\sigma\mu}+G_{E2}(Q^2) \mathcal{K}_{E2}^{\sigma\mu}+G_{C2}(Q^2) \mathcal{K}_{C_2}^{\sigma\mu},
\end{equation}
where $G_{M1}$, $G_{E2}$ and $G_{C2}$ denote the magnetic dipole, the electric quadrupole and the electric charge quadrupole transition form factors, respectively. The kinematical factors are defined as
\begin{align}
	\mathcal{K}_{M1}^{\sigma\mu}&=-3\left[ (m_\mcbp+m)^2-q^2 \right]^{-1}i\epsilon^{\sigma\mu\alpha\nu}P^\alpha q^\nu~(m_\mcbp+m_\mcb)/2m_\mcb,\\
	\mathcal{K}_{E2}^{\sigma\mu}&=-\mathcal{K}_{M1}^{\sigma\mu}-6\Omega^{-1}(q^2)~ i\epsilon^{\sigma\beta\alpha\nu}P^\alpha q^\nu~ \epsilon^{\mu\beta\rho\theta}p^{\prime\rho} q^{\theta}~\gamma_5 (m_\mcbp+m_\mcb)/m_\mcb,\\
	\mathcal{K}_{C_2}^{\sigma\mu}&=-3\Omega^{-1}(q^2)~q^\sigma (q^2 P^\mu-q\cdot P~ q^\mu)~i\gamma_5(m_\mcbp+m_\mcb)/m_\mcb.
\end{align}
Here $q=p^\prime-p$ is the transferred four\--momentum, $P=(p^\prime+p)/2$ and 
\begin{equation}
	\Omega(q^2)=\left[ (m_\mcbp+m_\mcb)^2-q^2 \right] \left[ (m_\mcbp-m_\mcb)^2-q^2 \right].
\end{equation}

	\subsection{Lattice methodology for the three-point functions} \label{sec:radlatmet}
	The transition matrix element is extracted from the three-point function,
\begin{align} \label{eq:thrpcf}
	&\langle \mathcal{G}_{\sigma}^{(3)}(t_2,t_1; {\bf p}^\prime, {\bf p};\mathbf{\Gamma}; \mu)\rangle = \sum_{{\bf x_2},{\bf x_1}} e^{-i{\bf p}\cdot {\bf x_2}} e^{i{\bf q}\cdot {\bf x_1}} \mathbf{\Gamma}^{\beta\alpha} \langle \Omega | \mathcal{T} \{ \chi_\sigma^\alpha(x_2) \mathcal{V}_\mu(x_1) \bar{\chi}^{\beta}(0) \} | \Omega \rangle,
\end{align}
where the spin projection matrices are defined in \Cref{eq:Gamma}. Here, $\alpha$, $\beta$ are the Dirac indices, $\sigma$ and $\mu$ are the Lorentz indices of the \spinth interpolating field and the external current respectively. Interpolating fields are given in \Cref{eq:genint,eq:deltaint}. The lattice electromagnetic current, $\mathcal{V}_\mu$, is defined in \Cref{eq:lat_current}.

Matrix elements are accessed using the ratio of the two- and three-point functions,
\begin{equation} \label{eq:rad_ratio}
	R_\sigma(t_2,t_1;{\bf p}^\prime,{\bf p};\mathbf{\Gamma};\mu) = \cfrac{\langle \mathcal{G}_{\sigma}^{(3)}(t_2,t_1; {\bf p}^\prime, {\bf p};\mathbf{\Gamma}; \mu)\rangle}{\langle \delta_{ij} \mathcal{G}_{ij}^{ \mcbp \mcbp}(t_2; {\bf p}^\prime;\Gamma_4)\rangle} \left[\cfrac{ \delta_{ij} \mathcal{G}_{ij}^{ \mcbp \mcbp} (2t_1; {\bf p}^\prime;\Gamma_4)\rangle }{ \mathcal{G}^{\mcb \mcb} (2t_1; {\bf p};\Gamma_4)\rangle } \right]^{1/2},
\end{equation}
where the two-point correlations functions are defined in \Cref{eq:2pt_corr,eq:deltacf}. In the large Euclidean time limit, $t_2-t_1\gg a$ and $t_1\gg a$, time dependence of the correlators are eliminated so that the ratio in \Cref{eq:rad_ratio} reduces to the desired form
\begin{equation}\label{eq:des_ratio}
	R_\sigma(t_2,t_1;{\bf p^\prime},{\bf p};\Gamma;\mu) \xrightarrow[t_2-t_1\gg a]{t_1\gg a} \Pi_\sigma({\bf p^\prime},{\bf p};\Gamma;\mu),
\end{equation}
which isolates the interaction matrix element. The ratio in \Cref{eq:rad_ratio} is found to yield a good plateau region and signal quality~\cite{Bahtiyar2015281} in comparison to several other alternatives~\cite{Leinweber:1992pv, Alexandrou:2003ea, Alexandrou:2004xn, Alexandrou:2007dt}.

Sachs form factors $G_{M1}$, $G_{E2}$ and $G_{C2}$ can be singled out choosing appropriate combinations of the Lorentz direction $\mu$ and projection matrices $\Gamma$. When $\mcb$ is produced at rest and momentum is inserted in one spatial direction, ${\bf q} = (q,0,0)$, we have~\cite{Leinweber:1992pv},
\begin{align}
	\begin{split}\label{eq:cff}
	G_{C2}(q^2) &= C({\bf q}^2) \frac{2m_{\ast}}{{\bf q}^2} \Pi_k({\bf q},{\bf 0};i \Gamma_k;4)
	\end{split}
	\\
	\begin{split}\label{eq:mff}
	G_{M1}(q^2) &= C(\mathbf{q}^2) \frac{1}{|{\bf q}|} \left[\Pi_l(q_k,{\bf 0}; \Gamma_k;l) - \frac{m_{\ast}}{E_{\ast}} \Pi_k(q_k,{\bf 0}; \Gamma_l;l) \right], 
	\end{split}\\
	\begin{split}\label{eq:qff}
	G_{E2}(q^2) &= C(\mathbf{q}^2) \frac{1}{|{\bf q}|} \left[\Pi_l(q_k,{\bf 0}; \Gamma_k;l)+\frac{m_{\ast}}{E_{\ast}}\Pi_k(q_k,{\bf 0}; \Gamma_l;l)\right],  
	\end{split}
\end{align}
where the common kinematical factor is
\begin{equation}
	C(\mathbf{q^2})= 2\sqrt{6} \frac{E_{\mcb} m_{\mcb}}{m_{\mcbp}+m_{\mcb}}\left( 1+\frac{m_{\mcb}}{E_{\mcb}}\right)^{1/2} \left(1+\frac{\mathbf{q}^2}{3m_{\mcbp}^2} \right)^{1/2}.
\end{equation}
If $\mcbp$ is produced at rest, $m_\ast=E_\ast$ in \Cref{eq:mff,eq:qff} and the common factor becomes,
\begin{equation}
	C(\mathbf{q^2})= 2\sqrt{6} \frac{E m}{m_\ast+m}\left( 1+\frac{m}{E}\right)^{1/2} \left(1+\frac{\mathbf{q}^2}{3m_\ast^2} \right)^{1/2},
\end{equation}
where $m_\ast \equiv m_{\mcbp}$ and $E_\ast \equiv E_{\mcbp}$.

Here, $k$ and $l$ are two distinct indices running from 1 to 3. For real photons, only $G_{M1}$ and $G_{E2}$ contribute. $G_{C2}$ does not play any role since it is proportional to the longitudinal helicity amplitude.
A sum of all possible kinematics of the correlation-function ratios can be defined as
\begin{equation}\label{eq:pi1pi2}
	\Pi_1=\frac{C(q^2)}{|\bf{q}|}\frac{1}{6}\sum_{k,l}\Pi_l(q_k,{\bf 0}; \Gamma_k;l), \quad
	\Pi_2=\frac{C(q^2)}{|\bf{q}|}\frac{1}{6}\sum_{k,l}\Pi_k(q_k,{\bf 0}; \Gamma_l;l),
\end{equation}
so that \Cref{eq:mff,eq:qff} become,
\begin{align}
	\label{eq:mffavg}
	G_{M1}(q^2)&=\Pi_1-\frac{m_\ast}{E_\ast}\Pi_2, \\
	\label{eq:qffavg}
	G_{E2}(q^2)&=\Pi_1+\frac{m_\ast}{E_\ast}\Pi_2.
\end{align}

	\subsection{Results} \label{sec:radres}
\subsubsection{\texorpdfstring{$\Xi_c \gamma \rightarrow \Xi_c^\prime$}{Xic gamma -> Xic*}} 
Amongst the heavy baryons, $\Xi_c$ and $\Xi_c^\prime$ are particularly interesting as they both are composed of the valence quarks $u$, $s$ and $c$ which are different flavors. Their distinct places on the mass scale--where up sits close to the chiral symmetric point, strange quark is at the same order as $\Lambda_{QCD}$ and charm very heavy compared to the other two but not heavy enough to warrant for a strict heavy-quark symmetry approximation--opens up intriguing possibilities for confinement dynamics. Additionally,
 $\Xi_c$ has an anti-symmetric flavor wavefunction while $\Xi_c^\prime$ has a symmetric one under the exchange of the light quarks. Since the mass difference $m_{\Xi_c^\prime} - m_{\Xi_c} \sim 107 \; {\rm MeV}$\cite{Jessop:1998wt} does not allow for a strong decay, $\Xi_c^\prime \rightarrow \Xi_c \gamma$ is the dominant decay mode. 

$\Xi_c \gamma \rightarrow \Xi_c^\prime$ transition is studied in Ref.~\citen{Bahtiyar:2016dom} on ensemble \ensl. The valence light-quark masses are taken equal to that of sea-quarks. Charm quark treatment is the same as the one discussed in \Cref{sec:diagres}. These allow for a reliable calculation of the transition form factors and a first-principles estimation of the decay width which is not yet measured experimentally. 

\begin{figure}[htb]
	\centerline{\includegraphics[width=.5\textwidth]{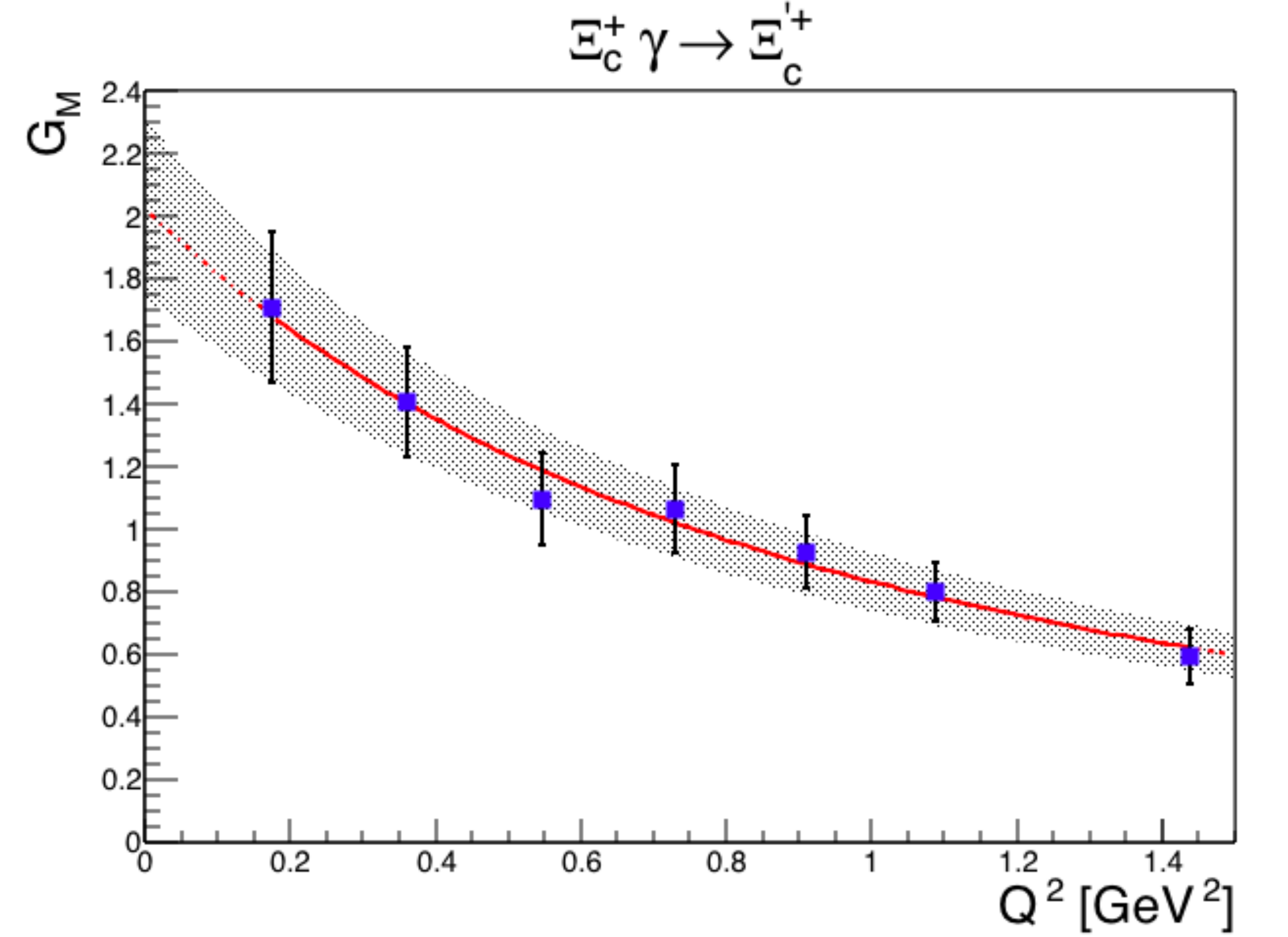} \includegraphics[width=.5\textwidth]{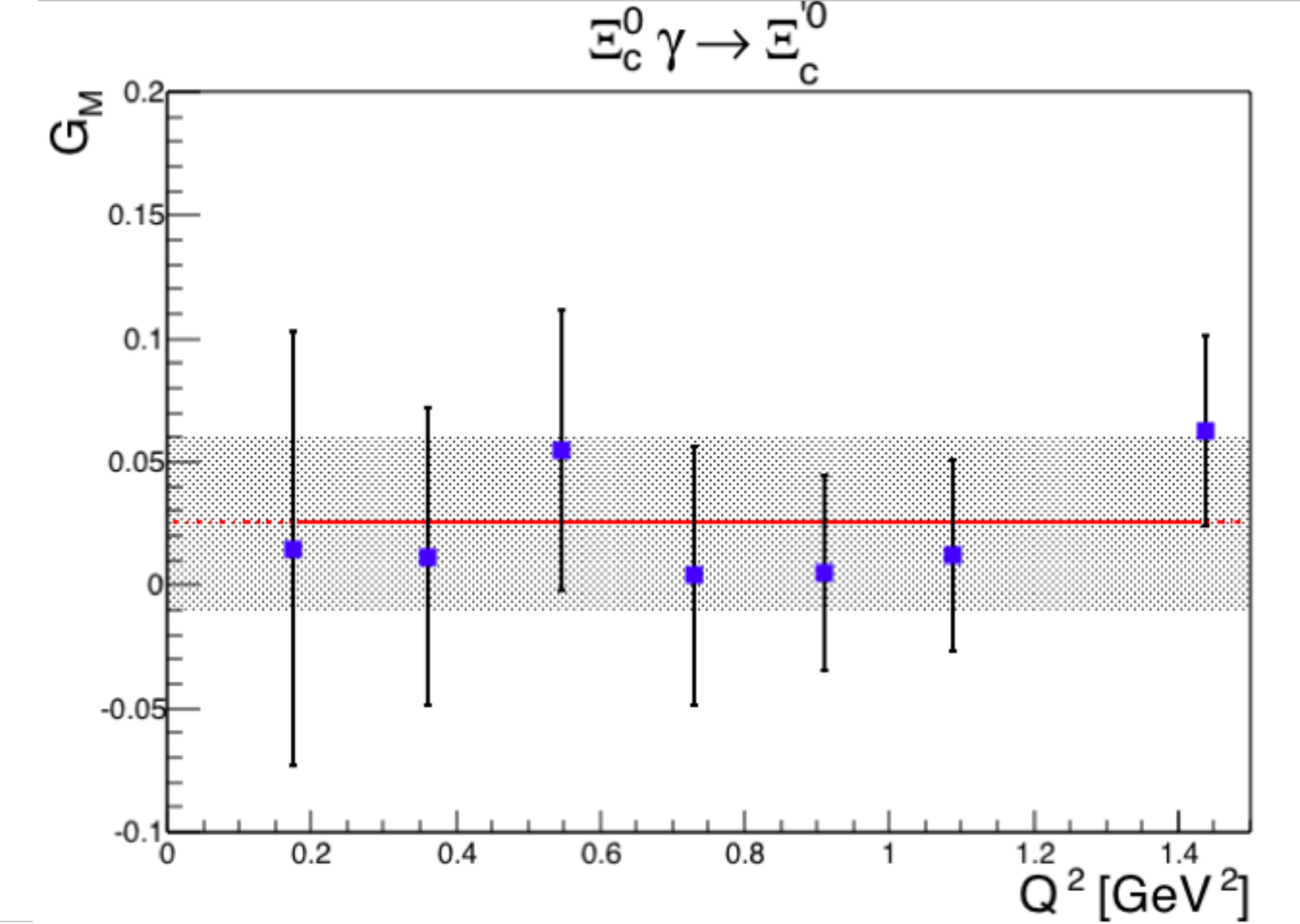}}
	\caption{Magnetic Sachs form factors of the charged and neutral $\Xi_c \gamma \rightarrow \Xi_c^\prime$ transitions. Figure taken from Ref.~\citen{Bahtiyar:2016dom}. \label{fig:radxic_ff}}
\end{figure}

\begin{figure}[htb]
	\centerline{\includegraphics[width=1\textwidth]{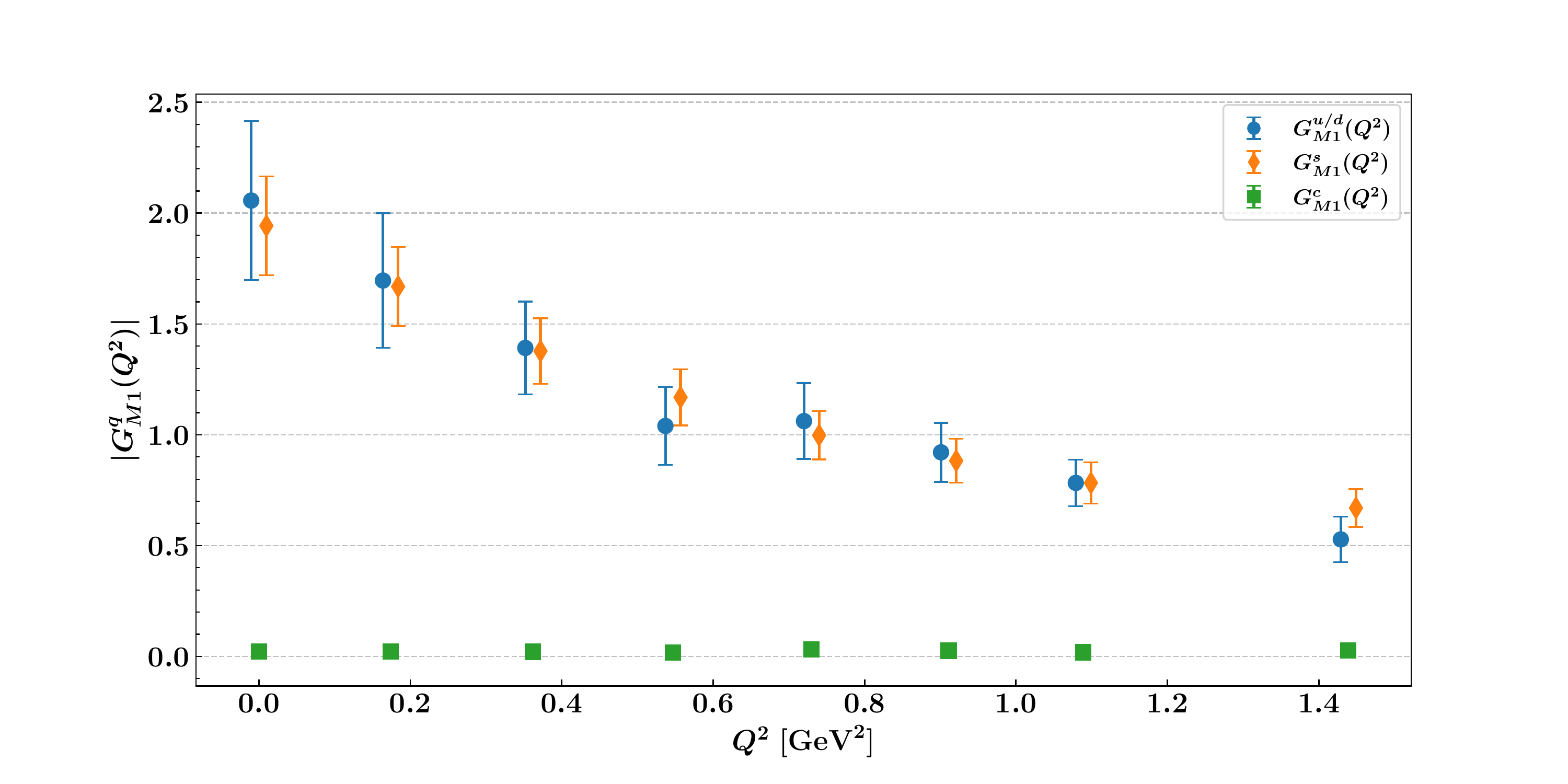}}
	\caption{Individual quark contributions to the transition magnetic form factor. Contributions are given for unit electric charge. Values are taken from Table 3 of Ref.~\citen{Bahtiyar:2016dom}. Note that the absolute values are plotted for an easier comparison. Strange quark contribution is negative while light and charm quark contributions are positive. \label{fig:radxic_ff_q}}
\end{figure}

Magnetic Sachs form factors for the charged and neutral transitions are extracted up to $Q^2 \simeq 1.5 \; {\rm GeV}^2$ and shown in \Cref{fig:radxic_ff}. We also plot the individual quark contributions to the total form factor in \Cref{fig:radxic_ff_q} with the values taken from Table 3 of Ref.~\citen{Bahtiyar:2016dom}.
As evident in \Cref{fig:radxic_ff_q}, light $u/d$- and $s$-quark contributions to the transition magnetic form factors of $\Xi_c \gamma \rightarrow \Xi_c^\prime$ are equal in magnitude and opposite in sign. On the other hand, the $c$ quark has almost no effect. When the quark contributions are combined, the $u$ and $s$ contributions to $\Xi_c^+ \gamma \rightarrow \Xi_c^{'+}$ form factors are multiplied with electric charges of opposite sign and add constructively. In contrast, the neutral transition $\Xi_c^0 \gamma \rightarrow \Xi_c^{\prime 0}$ is highly suppressed as a result of equal electric charges of the $d$ and $s$ quarks. According to conserved U-spin flavor symmetry, which assumes a degeneracy between two equally charged $d$ and $s$ quarks, a transition from $\Xi_c^0$ to $\Xi_c^{\prime 0}$ is forbidden. The results are in agreement with what U-spin flavor symmetry predicts. As shown in Fig.~\ref{fig:radxic_ff} magnetic form factor of the neutral $\Xi_c^0 \gamma \rightarrow \Xi_c^{\prime 0}$ transition is consistent with zero.

Since the $G_{M1}(Q^2=0)$ value cannot be calculated directly, it is determined via an extrapolation by assuming a dipole form for the form factor. Subsequently the transition magnetic moments are estimated from \Cref{eq:magmom}. They are given in units of natural and nuclear magnetons in \Cref{tab:radxic_mom}. 
%
\begin{table}[ht]
	\tbl{The magnetic form factor extrapolated to $Q^2=0$, together with the magnetic moments in units of nuclear magneton. $G_{M1}(0)$ can be considered as the magnetic moment in natural units $(e/2m_\mcb)$. }
	{
	\begin{tabular}{@{}ccc@{}} \toprule
	Transition                                    & $G_{M1}(0)$    & $\mu$ [$\mu_N$]  \\  
	\colrule
	$\Xi_c^+ \gamma \rightarrow \Xi_c^{\prime +}$ & 2.027(286)   & 0.729(103)   \\ 
	$\Xi_c^0 \gamma \rightarrow \Xi_c^{\prime 0}$ & 0.025(36)\pz & 0.009(13)\pz \\ \botrule
	\end{tabular} \label{tab:radxic_mom}
	}
\end{table}
%

The decay width of $\Xi_c^\prime$ baryon is related to the Pauli form factor $F_2(0)$ of $\Xi_c \gamma \rightarrow \Xi_c^\prime$ as 
\begin{equation}
		 \Gamma_{\mcb \gamma \rightarrow \mcb^\prime } = \frac{4 \alpha |\vec{q}|^3}{(m_{\mcb^\prime}+m_{\mcb})^2}|F_2(0)|^2 \quad \text{with}\quad |\vec{q}| = \frac{(m_{\mcb^\prime}^2-m_{\mcb}^2)}{2 m_{\mcb^\prime}}.
		 \label{eq:decayrate}
\end{equation}
Since the relation in \Cref{eq:decayrate} is defined in continuum formalism, it is evaluated by using the experimental masses of $\Xi_c$ and $\Xi_c^\prime$. $F_2(0)$ is determined by solving the Sachs form factors $G_E(Q^2)$ and $G_M(Q^2)$ simultaneously for all lattice data. At zero momentum transfer, $G_E(0) = F_1(0)$ so $F_1(0)$ must have a very small value, if not zero, for $\mcb \ne \mcb^\prime$. Since $\Xi_c \gamma \rightarrow \Xi_c^\prime$ cannot occur through electric transition, this implies $G_M(0) \simeq F_2(0)$. Consequently, it is found that
\begin{align}
	\begin{split}
	F_2(0)=2.036(280)\quad \text{for} &\quad \Xi_c^+ \gamma \rightarrow \Xi_c^{\prime +}, \\
	F_2(0)=0.039(46)\quad \text{for} &\quad \Xi_c^0 \gamma \rightarrow \Xi_c^{\prime 0}.
\end{split}
\end{align}
Using the formula in \Cref{eq:decayrate}, the decay widths of the charged and neutral $\Xi_c$ baryons are estimated as
\begin{equation}
	\Gamma_{\Xi_c^{\prime +}}=5.468(1.500)~\text{keV}, \quad \Gamma_{\Xi_c^{\prime 0}}=0.002(4)~\text{keV}.
\end{equation} 
The decay width of $\Xi_c^+$ can be translated into a lifetime using $\tau = 1/\Gamma$,
\begin{equation}
	\tau_{\Xi_c^{\prime +}} = 1.148(322)\times10^{-19}~\text{s}.
\end{equation}

Both neutral and charged transitions of $\Xi_c \gamma \rightarrow \Xi_c^\prime $ have been studied using QCD sum rules~\cite{Aliev:2016xvq}, heavy baryon chiral perturbation theory~\cite{Cheng:1992xi,Banuls:1999br,Jiang:2015xqa,Wang:2018cre}, quark models~\cite{Ivanov:1999bk,Dey:1994qi,Wang:2017kfr}, bag model~\cite{Bernotas:2013eia}, chiral model~\cite{Kawakami:2019hpp} and pion mean-field approach~\cite{Yang:2019tst}. For the charged transition, lattice results for the transition form factor and decay width are in agreement with those from QCD sum rules~\cite{Aliev:2016xvq} and heavy baryon chiral perturbation theory~\cite{Wang:2018cre}, while other methods predict higher values. Curiously, the pion mean-field approach result~\cite{Yang:2019tst} has the opposite sign of the lattice value. In the case of neutral transition, their predictions are small but finite, while the lattice prediction is consistent with zero.

\subsubsection{Transitions for \texorpdfstring{$\ocs, \; \xccs, \; and \; \occs$}{Omegac, Xicc and Omegacc}}
The $\omc^0 (css)$ has the quantum numbers $J^P=\frac{1}{2}^+$ and is the heaviest known singly charmed baryon that decays weakly. Within the multiplet structure of flavor SU(4), $\omc$ belongs to a sextet of flavor symmetric states, which sits on the second layer of the flavor mixed-symmetric 20-plet. Belle Collaboration has made a rigorous experimental study of $\omc$ using the decay $\omc^0\rightarrow \Omega^-\pi^+$~\cite{Solovieva:2008fw}. The excited $\Omega_c^{\ast 0}(css)$ baryon was first observed by BABAR Collaboration in the radiative decay channel $\ogo$~\cite{Aubert:2006je}. Belle Collaboration has confirmed their observation by reconstructing $\ocs$ in the same radiative decay mode~\cite{Solovieva:2008fw}. They measured the relative mass difference with respect to the ground state $m_{\ocs} - m_{\omc} = 70.7 \pm 0.9^{+0.1}_{-0.9} \; {\rm MeV}$ in very good agreement with the BABAR observation. The quantum numbers have not been measured but the natural assignment is that it completes the ground state $J^P=\frac{3}{2}^+$ sextet, which sits on the second layer of the flavor-symmetric 20-plet of SU(4). Lattice QCD calculations of the mass of $\ocs$ support this assignment. The mass difference with respect to the ground state is too small for any strong decay to occur, therefore the radiative channel $\ogo$ is the dominant decay mode. 

Observation of the doubly-charmed baryons, on the other hand, have been challenging for experiments. First observation of the doubly charmed baryon was reported by SELEX collaboration in 2002 \cite{Mattson:2002vu}. Mass of the $\Xi_{cc}^{+}$ (ccd) baryon was reported as $3519 \pm 1$ MeV/c$^2$. However, none of the following experiments could confirm this result~\cite{Ratti:2003ez, Aubert:2006qw, Chistov:2006zj, Aaij:2013voa}, until the LHCb Collaboration discovered the isospin partner of $\Xi_{cc}^{+}$, namely $\Xi_{cc}^{++}$~\cite{Aaij:2017ueg}, containing two $c$ quarks and one $u$ quark. Mass of $\Xi_{cc}^{++}$ reported by LHCb is $3621.40 \pm 0.72 \pm 0.27 \pm 0.14$ MeV/c$^2$, approximately $100$ MeV larger than the SELEX finding and in agreement with lattice QCD predictions. No experimental evidence for $\Xi_{cc}^\ast$, $\Omega_{cc}$ or $\Omega_{cc}^\ast$ baryons exist, however it is reasonable to assume that their radiative decay modes would dominate due to the tightening phase space expected from the heavy quark spin symmetry.  

Radiative decay modes of $\ocs$, $\xccs$, and $\occs$ are of interest since those decays are heavy sector analogs of the $N \gamma \rightarrow \Delta$ transition. Experimentally, pure single spin-flip $M1$ transition has been found to dominate the $N \gamma \rightarrow \Delta$. Of special interest is the small but non-vanishing values of $E2$ and $C2$ moments, implying that the shapes of $N$ and $\Delta$ deviate from spherical symmetry~\cite{Buchmann:2001gj}. Quark models predict a nonzero value for $E2$ and $C2$~\cite{Isgur:1981yz}, which has also been confirmed experimentally~\cite{Mertz:1999hp, Joo:2001tw}. The experimental results for $\oog$, on the other hand, are not yet precise enough to allow a determination of the transition strengths. Experimental evidence for $\xcctoxccs$ and $\occtooccs$ is nonexistent. Contrasting the results of these transitions to that of light sector would help us to understand the heavy-quark dynamics better. It would also be highly relevant for experimental facilities such as LHCb, PANDA, Belle II and BESIII in their searches for further states.

The $\Omega_c \gamma \rightarrow \Omega_c^\ast$ transition has been first studied in lattice QCD in Ref.~\citen{Bahtiyar2015281} and later updated in Ref.~\citen{PhysRevD.98.114505}. A lattice QCD study of the doubly charmed $\xcctoxccs$ and $\occtooccs$ transitions are presented in Ref.~\citen{PhysRevD.98.114505}. Gauge ensembles (\ensl) and the lattice setups for both works are same as the $\Xi_c \gamma \rightarrow \Xi_c^\prime$ work except for two notable differences. First, the form factors are obtained on only one $Q^2$ value corresponding to the lowest-allowed lattice three momentum ${\bf q}^2 = 0.183 \; {\rm GeV}^2$. The $Q^2 = 0$ form factors are estimated via the scaling method (\Cref{eq:scaling}) discussed in \Cref{sec:diagres}. Secondly, a notable improvement concerning these results is the use of a relativistic heavy quark action for the charm quark~\cite{PhysRevD.98.114505}, following the prescription discussed in \Cref{sec:fa}, which suppresses the discretization errors. 

\begin{figure}[htb]
	\centerline{\hspace{6mm}\includegraphics[width=1.085\textwidth]{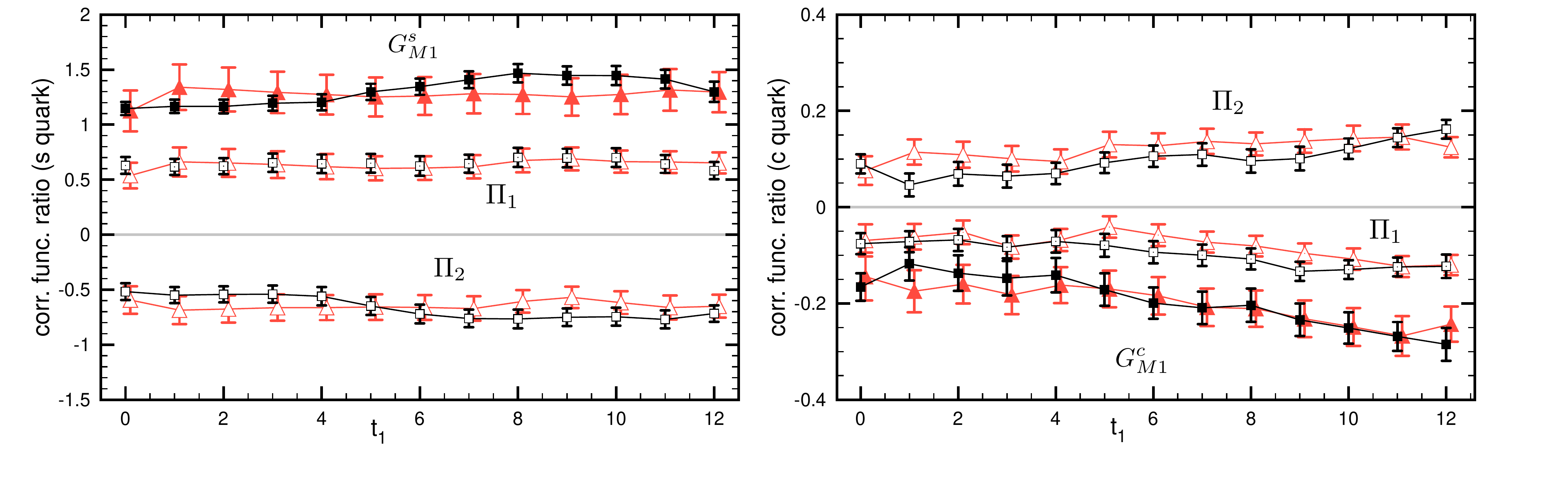}}
	\centerline{\includegraphics[width=.5\textwidth]{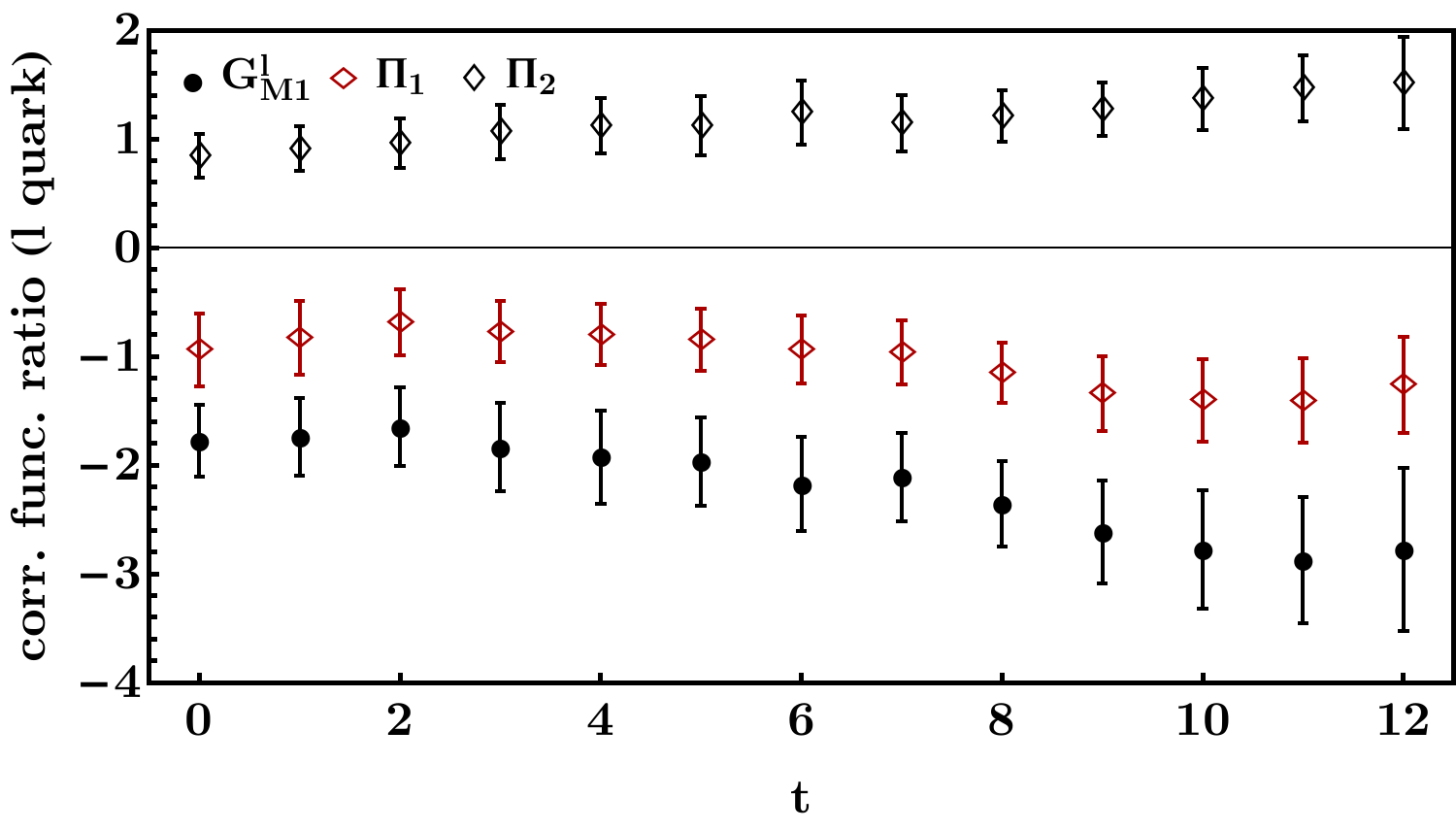} \includegraphics[width=.5\textwidth]{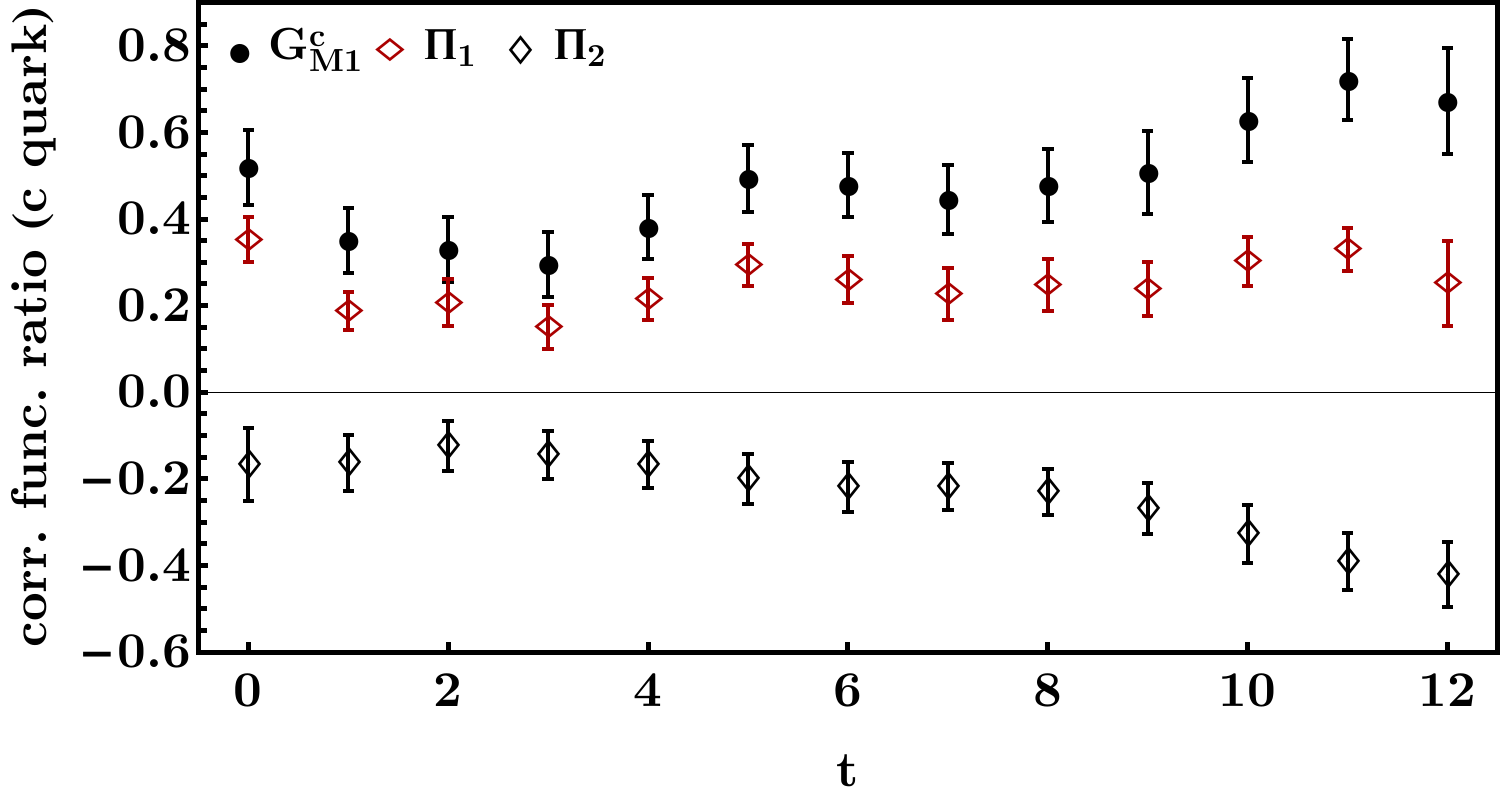}}
	\centerline{\includegraphics[width=.5\textwidth]{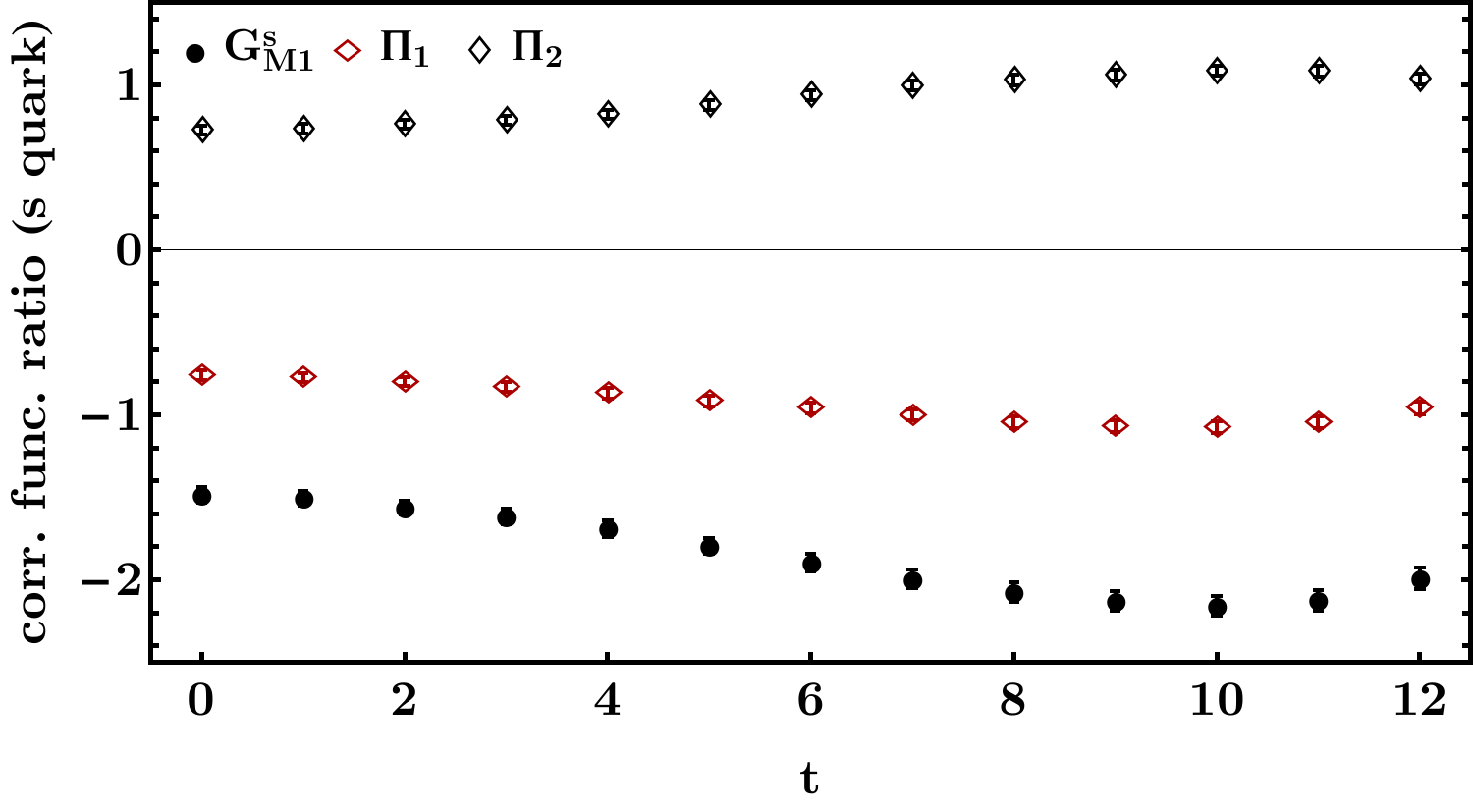} \includegraphics[width=.5\textwidth]{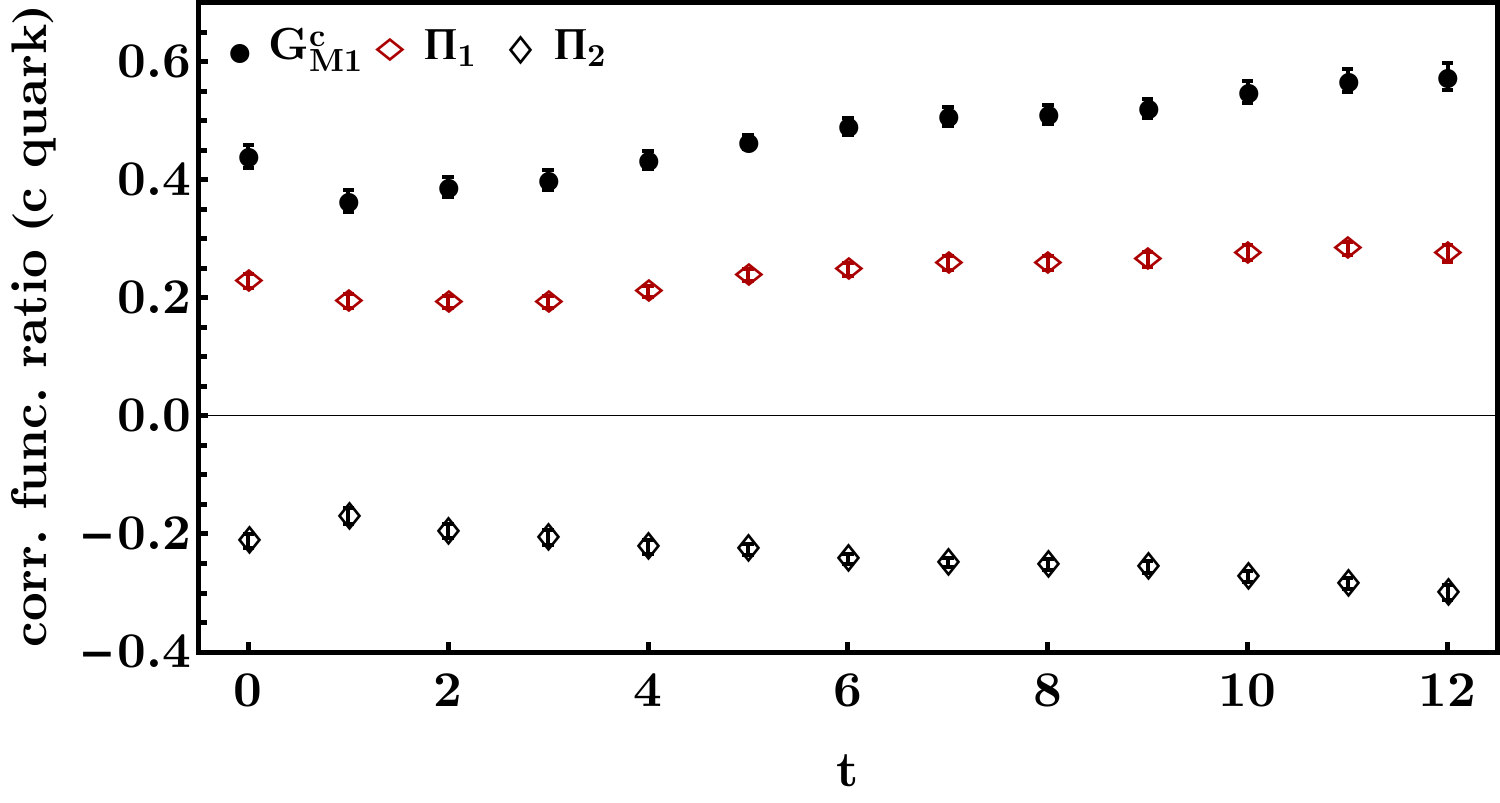}}
	\caption{The correlation function ratios $\Pi_1$ and $\Pi_2$ in \Cref{eq:pi1pi2} as functions of the current insertion time ($t_1$) for light and charm quark sectors. $G_{M1}^{u/d,s,c}$ obtained using \Cref{eq:mffavg} is shown as well. Plots in the first row are for the $\Omega_c \gamma \rightarrow \Omega_c^\ast$ transition and the squares (triangles) denote the kinematical case when $\ocs$ ($\omc$) at rest. $\xcctoxccs$ and $\occtooccs$ transitions are shown in the middle and last rows, respectively, where the signals are obtained in the $\mcbp$ rest frame. Figures taken from Refs.~\citen{Bahtiyar2015281} and \citen{PhysRevD.98.114505}. \label{fig:radoc_ff}}
\end{figure}

Signals for the strange and charm quark contributions to the lattice ratios $\Pi_1$ and $\Pi_2$ are shown in \Cref{fig:radoc_ff}. 
%
\begin{table}[ht]
	\tbl{The results for $G_{M1}$ and $G_{E2}$ form factors at the lowest allowed four-momentum transfer and at zero momentum transfer in the rest frame of $\ocs$. The quark sector contributions to each form factor, $\ell \equiv u/d$ for $\Xi_{cc}$ and $\ell \equiv s$ for $\Omega_{c}$ or $\Omega_{cc}$, are given separately. Note that the statistical uncertainty is large in $G_{E2}$ and results are consistent with zero. We have normalized these results to unit valence quark contribution as opposed to those reported in Refs.~\citen{Bahtiyar2015281} and \citen{PhysRevD.98.114505}, for a cleaner comparison. }
	{
	\begin{tabular}{@{}cccccccc@{}} \toprule
	       & $Q^2$[GeV$^2$]   & $G^{\ell}_{M1}(Q^2)$ & $G^c_{M1}(Q^2)$ & $G_{M1}(Q^2)$ & $G^{\ell}_{E2}(Q^2)$ & $G^c_{E2}(Q^2)$ & $G_{E2}(Q^2)$  \\
	\colrule
	$\octoocs$     & 0.180       & \pneg0.728(51) & -0.209(30)      & -0.625(43)\pz     & -0.098(6)\pz\pz & \pneg0.010(23)   & \pneg0.059(43)\pz \\
	               & 0\pzp\pz\pz & \pneg0.874(61) & -0.215(31)      & -0.725(50)\pz     & -0.117(67)\pz   & \pneg0.010(24)   & \pneg0.071(52)\pz \\ 
	\\	
	$\occtooccs$   & 0.181       & -1.252(27)	  & \pneg0.269(12)  & \pneg0.775(24)\pz	& -0.034(30)\pz   & \pneg0.001(7)\pz & \pneg0.013(14)\pz \\
	 			   & 0\pzp\pz\pz & -1.504(32)  	  & \pneg0.286(12)  & \pneg0.882(27)\pz & -0.040(36)\pz   & \pneg0.002(7)\pz & \pneg0.015(16)\pz \\
	\\
	$\xcctoxccsP$  & 0.180       & -1.398(50) 	  & \pneg0.252(54)	& \pneg0.774(94)\pz	& \pneg0.069(301) & -0.003(36)	     & -0.026(108) 	     \\
	 			   & 0\pzp\pz\pz & -1.763(64)	  & \pneg0.264(56)	& \pneg0.906(103)	& \pneg0.087(380) & -0.003(38)	     & -0.033(133) 	     \\
	\\
	$\xcctoxccsPP$ & 0.180       & -1.398(50) 	  & \pneg0.252(54)	&  -0.552(113)	    & \pneg0.069(301) & -0.003(36)	     & \pneg0.043(210) 	 \\
	 			   & 0\pzp\pz\pz & -1.763(64)	  & \pneg0.264(56)	&  -0.772(127)	    & \pneg0.087(380) & -0.003(38)	     & \pneg0.054(269) 	 \\ 
    \botrule
	\end{tabular} \label{tab:radoc_mom}
	}
\end{table}
%
The $M1$ and $E2$ form factors are then extracted by combining the ratios using \Cref{eq:mffavg,eq:qffavg}. The $\Pi_1$ and $\Pi_2$ have opposite signs and they add constructively when they are subtracted for $M1$. However they are of similar magnitudes, which result in a vanishing value for $G_{E2}$ when they are added. The numerical results for the $G_{M1}$ and $G_{E2}$ form factors at the lowest allowed four momentum transfer and at zero momentum transfer are collected in \Cref{tab:radoc_mom}. The quark sector contributions to each form factor are given separately. The form factors can be inferred from individual quark contributions by adding them weighted with their electric charges and the number of valence quarks.  

\begin{figure}[htb]
	\centerline{\includegraphics[width=.7\textwidth]{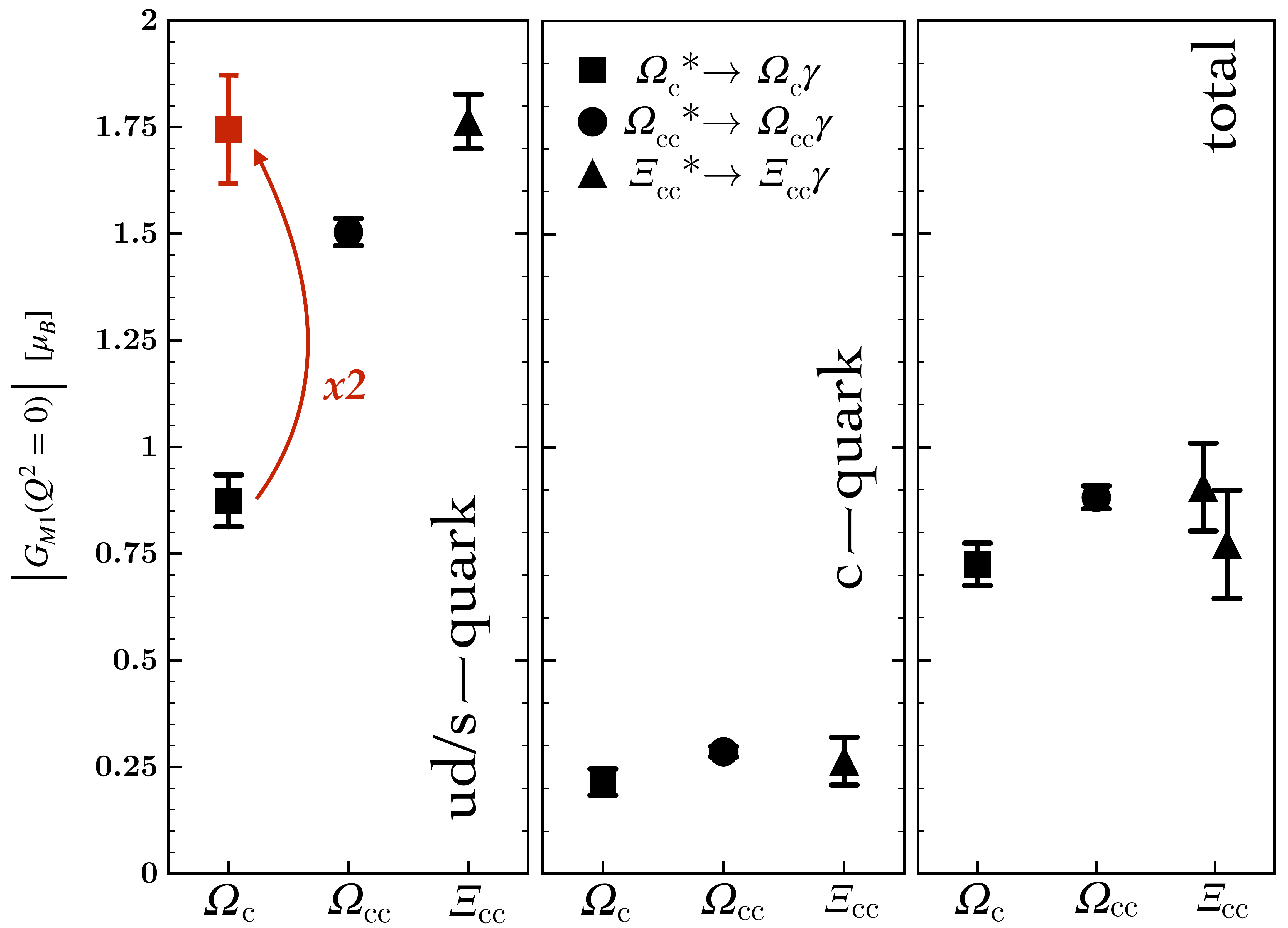}}
	\caption{Comparison of the quark sector contributions to the magnetic form factors of $\oog$, $\occtooccs$, and $\xcctoxccs$ transitions. Values are taken from \Cref{tab:radoc_mom}. Absolute values are plotted for a better comparison. $G_{M1}$ are given in units of natural magnetons $\mu_\mathcal{B}$ \label{fig:oc_occ_xcc_m1_comp}}
\end{figure}

Similar to what has been observed in the case of elastic form factors (\Cref{sec:diagres}), $M1$ form factors are dominantly determined by the contribution of the light ($u/d$ or $s$) quark sector, which is approximately one order of magnitude larger than that of the $c$-quark. This pattern is consistent with hyperon transition form factors~\cite{Leinweber:1992pv}: The heavier quark contribution is systematically smaller than that of the light quarks. Charm quark contributions are similar in magnitude in all transitions but the sign of the singly ($[c]$) and doubly represented ($[cc]$) sectors are opposite. Although the individual light quark contributions seem to differ for singly and doubly represented cases, \emph{e.g.} the $s$-quark contributions in $\occtooccs$ and $\oog$, they are similar to each other when one multiplies the contribution by a factor of two to account for the number of valence quarks in $\oog$. This is illustrated in \Cref{fig:oc_occ_xcc_m1_comp}. One again sees that the relative signs in singly and doubly represented sectors are opposite. This may be interpreted as the parallel (anti-parallel) alignment of the spin of the doubly (singly) represented sector with the total spin of the baryon.   

From a quark-model point of view, the coupling of the photon to the light quarks prevails in the heavy-quark limit and the heavy quark acts as a spectator. In this limit, the transition proceeds dominantly through the spin flip of the light degrees of freedom and only $M1$ transition is allowed. Only finite mass effects of the heavy quark may lead to a nonzero value of $E2$ form factor. $M1$ results show that the two quark sectors contribute with opposite signs and yield a value with a statistical error of approximately $5\%$ when combined. 

In contrast, the extracted values of $G_{E2}$ at finite and zero momentum transfer are small and consistent with zero within their error bars. The quadrupole transition moments arise from the tensor-induced D-state admixtures of the single-quark wavefunctions~\cite{Isgur:1981yz} and the two-quark exchange currents~\cite{Buchmann:1991cy, Buchmann:1996bd}. For the former, the spins of the quarks remain the same but an $S$-state quark is changed into a $D$-state. The latter can be interpreted as the spin flip of a diquark inside the baryon. Given the dependence of the tensor force on the inverse quark mass, one would expect to obtain a smaller $G_{E2}$ value for heavy baryons as compared to that in the light-baryon sector, in consistency with the quoted values. The smallness of the $E2$ form factor can also be understood as a chiral suppression. The $E2$ amplitude is dominated by pion loops and the leading contribution comes from chiral logs which can be computed in heavy-baryon chiral perturbation theory~\cite{Butler:1993ht, Savage:1994wa}.

The Sachs form factors studied above can be related to phenomenological observables such as the helicity amplitudes and the decay width. The relation between the Sachs form factors and the standard definitions of electromagnetic transition amplitudes $f_{M1}$ and $f_{E2}$ in the rest frame of $\mcbp$ are given by~\cite{Nozawa:1989pu, Sato:2000jf}
\begin{align}
		f_{M1}(q^2) &= \frac{\sqrt{4\pi\alpha}}{2m} \left( \frac{|{\bm{q}}|m_\ast}{m} \right)^{1/2} \frac{G_{M1}(q^2)}{[1-q^2/(m+m_\ast)^2]^{1/2}}, \\
		f_{E2}(q^2) &= \frac{\sqrt{4\pi\alpha}}{2m} \left( \frac{|{\bm{q}}|m_\ast}{m} \right)^{1/2} \frac{G_{E2}(q^2)}{[1-q^2/(m+m_\ast)^2]^{1/2}},
\end{align}
where $\alpha=1/137$ is the fine structure constant. The helicity amplitudes $A_{1/2}$ and $A_{3/2}$ can be deduced from the transition amplitudes as,
\begin{align}
	\label{eq:hamp12}
	A_{1/2}(q^2) &= -1/2[f_{M1}(q^2)+3f_{E2}(q^2)], \\
	\label{eq:hamp32}
	A_{3/2}(q^2) &= -\sqrt{3}/2[f_{M1}(q^2)-f_{E2}(q^2)].
\end{align}
Then the decay width is given by~\cite{PDG:2020}
\begin{equation}
	\Gamma = \frac{m_\ast m}{8\pi} \left( 1-\frac{m^2}{m_\ast^2} \right)^2 \left\{|A_{1/2}(0)|^2 + |A_{3/2}(0)|^2 \right\},
\end{equation} 
where ${\bf q} = (m_\ast^2-m^2)/2m_\ast$ at $q^2=0$. Alternatively the decay width can also be obtained from the Sachs form factors,
\begin{equation}\label{eq:dwG}
	\Gamma = \frac{\alpha}{16} \frac{(m_\ast^2-m^2)^3}{m^2 m_\ast^3} \left\{ \left| G_{M1}(0) \right|^2 + 3 \left| G_{E2}(0) \right|^2 \right\}.
\end{equation}
Since the above formulas are continuum relations, experimental values of the baryon masses are used in calculating the helicity amplitudes and decay widths. The numerical results for the helicity amplitudes in the rest frame of $\mcbp$ at finite and zero momentum transfer and the decay widths are reported in \Cref{tab:radoc_hamp}. Since mass splittings between these baryons kinematically forbid an on-shell strong decay channel, the total decay rates are almost entirely determined in terms of the electromagnetic mode.
%
\begin{table}[ht]
	\tbl{Results for the helicity amplitudes, decay widths and lifetimes. Zero-momentum values are obtained using the simple scaling assumption given in \Cref{eq:scaling}. }
	{
	\begin{tabular}{@{}cccccccc@{}} \toprule
	& $Q^2$ & $f_{M1}$ & $f_{E2}$ & $A_{1/2}$ & $A_{3/2}$ & $\Gamma$ & $\tau$ \\   
	& [GeV$^2$]	& \scriptsize{$10^{-2}$[GeV$^{-1/2}$]} & \scriptsize{$10^{-2}$[GeV$^{-1/2}$]} & \scriptsize{$10^{-2}$[GeV$^{-1/2}$]} & \scriptsize{$10^{-2}$[GeV$^{-1/2}$]} & \scriptsize{[keV]}& \scriptsize{[$10^{-18}$ s]} \\
	\colrule
	$\octoocs$     & 0.180 	     & -0.951(66)\pz     & -0.090(65)\pz     & \pneg0.341(99)\pz & \pneg0.901(85)\pz & ---        & ---             \\
	               & 0\pzp\pz\pz & -1.104(76)\pz     & \pneg0.109(79)\pz & \pneg0.389(119)   & \pneg1.050(101)   & 0.096(14)\pz  & 6.889(997)  \\
	\\
	$\occtooccs$   & 0.181       & \pneg0.812(26)\pz & \pneg0.013(15)\pz & -0.429(13)\pz	 & -0.690(22)\pz     & --- 		  & --- 			\\
	 			   & 0\pzp\pz\pz & \pneg0.924(28)\pz & \pneg0.016(17)\pz & -0.489(14)\pz     & -0.785(25)\pz     & 0.0565(4)\pz  & 11.66(3.83)\pz 	\\
	\\
	$\xcctoxccsP$  & 0.180       & \pneg0.838(101)   & -0.027(118) 	     & -0.419(51)\pz  	 & -0.726(88)\pz 	 & --- 		  & --- 			\\
	 			   & 0\pzp\pz\pz & \pneg0.982(111)   & -0.034(145)  	 & -0.491(56)\pz  	 & -0.850(96)\pz  	 & 0.0648(38) & 10.28(3.30)\pz 	\\
	\\
	$\xcctoxccsPP$ & 0.180       & -0.597(123)	     & \pneg0.048(229) 	 & \pneg0.298(61)\pz & \pneg0.517(106)   & --- 		  & --- 		    \\
				   & 0\pzp\pz\pz & -0.835(137) 	     & \pneg0.061(293) 	 & \pneg0.417(69)\pz & \pneg0.723(119)   & 0.0518(56) & 12.70(2.04)\pz 	\\ 
	\botrule
	\end{tabular} \label{tab:radoc_hamp}
	}
\end{table}
%

In comparison to $N \gamma \to\Delta$ transition~\cite{PDG:2020}, the helicity amplitudes are suppressed by roughly two orders of magnitude due to the diminishing contribution of the heavy quark, the overall reduction in the transition form factors and the larger baryon masses. Considering that the form factors are directly related to the transition matrix elements and thus to the interesting internal dynamics, it is desirable to compare the form factors as well. One can derive the dominant $M1$ form factor of the $N \gamma \to\Delta$ transition by inserting the PDG quoted $A_{1/2}$ and $A_{3/2}$ helicity amplitudes into \Cref{eq:hamp12,eq:hamp32} and following the calculation steps backwards. This calculation returns $G^{M1}_{N\gamma\to\Delta}(0) = 3.063^{+0.102}_{-0.096}$, which is approximately four times greater than the $M1$ form factors of the $\ocs$, $\occs$ and $\xccs$ transitions. Assuming the $u$- and $d$-quark have the same contribution within the $\Delta^+$ baryon, individual quark contributions (without electric charge and quark number factors) can be deduced as $G^{M1,u}_{N\gamma\to\Delta}(0) = G^{M1,d}_{N\gamma\to\Delta}(0) = G^{M1}_{N\gamma\to\Delta}(0)$. In contrast to the charm quark contributions, this reveals a suppression of around one order of magnitude in $G_{M1}^{c}(0)$. Decay widths are smaller by almost four orders of magnitude, three orders of which are directly related to the similar decrease in the kinematical factor of \Cref{eq:dwG}.

The electromagnetic transitions of singly charmed baryons have also been studied within bag model~\cite{Simonis:2018rld}, chiral perturbation theory~\cite{Cheng:1992xi,Banuls:1999br,Jiang:2015xqa,Wang:2018cre}, quark models~\cite{Dey:1994qi,Ivanov:1996fj,Ivanov:1999bk,Wang:2017hej,Wang:2017kfr,Gandhi:2018lez}, pion mean-field approach~\cite{Yang:2019tst,Kim:2021xpp}, and QCD sum rules~\cite{Zhu:1998ih,Aliev:2014bma,Aliev:2016xvq}. It has been found that the singly charmed baryon electromagnetic decays are suppressed, in qualitative agreement with the lattice results. However there is up to two orders of magnitude discrepancy between the lattice and non-lattice predictions of the $\ocs$ decay width. Curiously, an earlier pion mean-field result~\cite{Yang:2019tst} has the opposite sign of the lattice value but the situation is improved in a recent work~\cite{Kim:2021xpp}.

For the doubly charmed case, the transitions have been studied within the chiral perturbation theories~\cite{Meng:2017dni,Li:2017pxa,Li:2017cfz,Liu:2018euh,Meng:2018zbl}, bag model~\cite{Hackman:1977am,Bernotas:2013eia}, quark models~\cite{SilvestreBrac:1996bg,Lichtenberg:1976fi,JuliaDiaz:2004vh,Faessler:2006ft,Branz:2010pq,Oh:1991ws} and QCD sum rules~\cite{Sharma:2010vv,Aliev:2021hqq}. Similar to the singly charmed ones, decays of doubly charmed baryons are found to be suppressed. Bag model predictions~\cite{Hackman:1977am,Bernotas:2013eia} for decay widths are one order of magnitude larger than the lattice results. Quark model predictions are even larger by two orders of magnitude~\cite{Xiao:2017udy,Lu:2017meb,Branz:2010pq} similar to those of the chiral perturbation theory~\cite{Li:2017pxa} and QCD sum rules~\cite{Cui:2017udv}.   

The reason of this discrepancy between the lattice and non-lattice methods for singly and doubly charmed transitions both is studied in Ref.~\citen{PhysRevD.98.114505} further and traced to $M1$ form factors, i.e. the magnetic moments. It is found that the $G_{M1}$ values of the non-lattice methods are close to or larger than the $N\gamma\to\Delta$ value, which is highly unlikely since the lattice results clearly indicate that the heavy-quark contribution to $M1$ transition is heavily suppressed and the light quark contribution is not enhanced enough to compensate the change. $E2$ transitions, on the other hand, almost vanish so that they do not play a significant role. Additionally, the decay widths are highly sensitive to the mass splittings where a change of $\sim50 \, {\rm MeV}$ leads to an order of magnitude difference.~\cite{PhysRevD.98.114505,Aliev:2021hqq}.

It is plausible that there may be uncontrolled systematic errors affecting the lattice results presented in Ref.~\citen{PhysRevD.98.114505} and quoted here. However most of the systematic effects are already controlled: there are no errors due to chiral extrapolations since the ensembles are near the physical-quark point; any discretization error arising from the charm-quark action is suppressed and controlled since a relativistic heavy quark action is employed; excited states contamination is accounted for by employing multi-exponential fits in the analysis; and finally the finite size effects on these configurations are expected to be less than $1\%$ for charmed and strange quark observables~\cite{Can:2015exa}. Systematics that might arise from continuum extrapolation remains unchecked, however, an order of magnitude effect would be surprising. Therefore the discrepancies between lattice and non-lattice results are an open issue that needs to be understood better from both sides.

\section{Summary and Conclusions} \label{sec:sum}

	Electromagnetic form factors are one of the elements that play an important role in describing the internal dynamics of hadrons. They reveal valuable information about the size and the shape of the hadrons. Determining such form factors is an important step in our understanding of the hadron properties in terms of quark-gluon degrees of freedom. There have been enormous efforts to determine the electromagnetic form factors of light hadrons. The theoretical challenge is to understand these quantities from QCD. One intriguing question is how the structure of the hadrons gets modified in the heavy quark region, like in the case of charmed baryons. While there exist experimental results for the light baryons revealing their spectrum and electromagnetic properties, the situation for the charmed baryons is not as extensive for the time being. Future experimental efforts at facilities like \emph{e.g.} J-PARC, SuperKEKB, BES-III \emph{etc.}, are expected to provide a wealth of information, which calls for a better understanding of the heavy-sector dynamics from theoretical grounds. Combined with the available information on the light sector, insights on the heavy-flavored baryons would reveal the quark-gluon dynamics better.

	The electromagnetic form factors of light baryons and mesons have been (and still are) extensively studied in the framework of lattice QCD. Main challenges for the lattice QCD form factor calculations have been the pseudoscalar/vector-meson states and the nucleon while the octet and decuplet baryon structure had less attention. Lattice literature on charmed baryon electromagnetic form factors have been non-existent. Along this perspective, members of the TRJQCD Collaboration have built upon their initial experience on open-charmed meson sector to extend their calculations to charmed baryons. 

	This review has covered two kinds of electromagnetic form factors, namely the elastic and transition form factors, of baryons that contain at least one charm quark. After reviewing the lattice techniques to study the non-perturbative aspects of QCD, we have presented the formalism for accessing the form factors. All of the works collected here have utilized gauge ensembles generated by the PACS-CS Collaboration incorporating the dynamical effects of $u/d$ and $s$ quarks with varying light-quark masses down to almost their physical values. The gauge configurations are of $(32a)^3 \times 64 a$ in size with a fine enough lattice spacing of $a=0.0907(13)$ fm capable of resolving the inner structure of charmed baryons and corresponding to a spacious enough volume of approximately $(2.9 \, {\rm fm})^3 \times 5.9 \, {\rm fm}$ to accommodate them while minimizing the finite-size effects to a negligible extent except for the lightest-pion-mass ensemble (\ensl) which has $m_\pi L < 4$. Strange and charm quark masses are tuned to their respective physical values, albeit not precisely and effects of any mistuning has been studied. The Clover action is employed for all the valance quarks. For the charm quark the Fermilab approach has been followed to determine the improvement coefficients. A relativistic heavy-quark action is used in studying the radiative transitions of doubly charmed baryons to further suppress the discretization errors on the charm quark.     

	The elastic form factors are related to the static properties, such as electric charge radii, magnetization densities and the magnetic and higher order moments of the baryons, each providing an aspect of the internal structure. One important conclusion, which confirms the earlier non-lattice studies, is that the charmed baryons are compact--the magnitude of their observables is decreased--compared to their light-sector counterparts. This is unambiguously reflected in the individual quark contributions to the electromagnetic observables, where the charm quark contributions are systematically smaller than the light quark contributions. Comparing different spin and flavor compositions of the charmed baryons one sees that the electric properties are mildly sensitive to the flavor of the quark and the spin composition of the baryon. This can be inferred from the similarity of the charge radii of $\Xi_{cc}^{++} \, (\frac{1}{2}^+)$, $\Omega_{cc}^{+} \, (\frac{1}{2}^+)$ and ${\Omega_{cc}^{\ast}}^+ \, (\frac{3}{2}^+)$ baryons. Magnetization density follows a similar trend as well but its spread is larger compared to the electrical distribution. Furthermore there is evidence from their electric-quadrupole moments that the charge distribution of charmed \spinth baryons are not spherical but distorted. 

	A phenomenologically more interesting quantity is the magnetic moment. Once again, due to the charm quark the magnetic moments of charmed baryons are decreased compared to light baryons. Individual quark sector analyses reveal that the doubly represented quarks have a significant effect on the magnetic moment of the baryon. Interestingly, even though the charm quark contribution is small it cannot be neglected when there are two charm valence quarks. Alignment of the quark spins with respect to each other can be deduced from the sign of their magnetic moments and the results indicate that the quark sectors are anti-aligned in \spinoh but aligned in \spinth baryons. Additionally, the doubly represented quark sector in a \spinoh baryon is aligned with the total spin of the baryon. Contribution of the charm quark is systematically smaller than that of light quark, but its magnitude changes in different spin configurations where it is enhanced in \spinth baryons. 

	In general, results are found to be consonant with the qualitative expectations of quark model and heavy-quark symmetry, although there are apparent quantitative differences. When a quantitative comparison is made, non-lattice estimations of magnetic moments overestimate the lattice results. There are indications from covariant baryon chiral perturbation theory calculations~\cite{Blin:2018pmj,Liu:2018euh,Shi:2021kmm}, which incorporate the reviewed lattice results, that the quark model predictions fail to describe the pion-mass dependence of the charmed baryon magnetic moments accurately, which further stress the discrepancy between the lattice and quark model predictions.   

	The radiative decays of charmed baryons are particularly important because this decay mode is the dominant transition between the low-lying states due to the reduced phase space that forbids a strong decay. The $\Xi_c^\prime \rightarrow \Xi_c \gamma$ and $\Omega_c^\ast \rightarrow \Omega_c \gamma$ decays are both observed in experiments but their widths are not measured. No experimental evidence for the radiative decays of doubly charmed baryons, \emph{e.g.} $\xcctoxccs$ and $\occtooccs$, exist however on the grounds of heavy-quark symmetry one would expect the electromagnetic transition to be the dominant channel. One then has a unique \spinoh $\rightarrow$ \spinoh transition in the $\Xi_c^\prime$ case and \spinoh $\rightarrow$ \spinth transitions as heavy sector counterparts of the $N \gamma \rightarrow \Delta$ channel. The lattice results provide predictions for the transition magnetic moments, transition and helicity amplitudes and consequentially the decay widths. 

	For the $\Xi_c^\prime$ case one sees a realization of the U-spin symmetry which forbids the neutral $\Xi_c^0 \gamma \rightarrow \Xi_c^{\prime 0}$ transition and the lattice prediction of the $\Xi_c^{\prime 0}$ decay width is consistent with zero. Similar to the elastic case, light quarks dominate the dynamics where the charm quark acts as a spectator with a negligible contribution. There is again a tension between the lattice and non-lattice estimations for $\Xi_c^{\prime 0}$ and $\Xi_c^{\prime +}$ both. 

	The charmed \spinth $\rightarrow$ \spinoh transitions, \emph{e.g.} $\Omega_c \gamma \rightarrow \Omega_c^\ast$, $\xcctoxccs$ and $\occtooccs$,  are especially informative as they can be compared to the $N \gamma \rightarrow \Delta$ transition. As one expects, the light quarks dominate the dynamics and the transitions proceed through the magnetic interaction---the spin flip of the light degrees of freedom. Any tensor-force induced interaction is negligibly small. In comparison to $N \gamma \to\Delta$, the helicity amplitudes are suppressed by roughly two orders of magnitude due to the diminishing contribution of the heavy quark, the overall reduction in the transition form factors and the larger baryon masses. 
	
	Insightes gained from non-lattice calculations works well in this case also where there is a qualitative agreement between the lattice and non-lattice estimations of the amplitudes and decay widths. However the lattice results are one or two orders of magnitude smaller than the values estimated by non-lattice methods. The reason of this discrepancy lies in the transition magnetic moments. The magnetic moments used by the non-lattice methods are close to or larger than the $N\gamma\to\Delta$ transition moment, which is highly unlikely since the lattice results clearly indicate that the heavy-quark contribution to the $M1$ transition is heavily suppressed and the light quark contribution is not enhanced enough to compensate the change.

	In conclusion, the works reviewed here only scratch the surface and already reveal some interesting phenomena for the charmed sector seen through an electromagnetic perspective. It is crucial that the ongoing and anticipated experimental programs are complemented and driven with rigorous, first-principles calculations to advance our understanding of the dynamics of the heavy quarks and strong interactions. Discrepancies between the lattice and non-lattice calculations need to be understood better to have a solid insight into the dynamics of the heavy sector. Eventually, reliably determined charmed baryon observables would be input to further works that investigate the nature of the exotic states, some of which already incorporate the available data~\cite{Liu:2019zvb,Pan:2020xek,Lyu:2021qsh}.

\section*{Acknowledgments}
I thank my colleagues M. Oka, G. Erkol, T. T. Takahashi, and H. Bahtiyar from the TRJQCD Collaboration for their endless support and the stimulating environment that they have provided throughout the years. This work would not have been possible without their invaluable contributions. I would like to additionally thank E. Hiyama for her hospitality during my time at RIKEN where parts of this research have been conducted. I appreciate the support from the Special Postdoctoral Researcher (SPDR) program of RIKEN during my time there. I also acknowledge the support by the Australian Research Council Grant DP190100297.   

\appendix

\section{Baryon masses}

	We collect the baryon masses used in each work for the interested reader. Details of the extractions can be found in the respective references.
	\begin{table}[ht]
		\tbl{Baryon masses as obtained on lattice simulations in units of GeV.}
		{
		\begin{tabular}{@{}cccccccc@{}} \toprule
			Ref. & \texttt{ID} & $\sgc$ & $\omc$ & $\xcc$ & $\occ$	\\
			\colrule
			[\citen{Can:2013tna}] & \ensh & 2.841(18) & 2.959(24) & 3.810(12) & 3.861(17) \\
								& \enh	& 2.753(19) & 2.834(19) & 3.740(13) & 3.806(12) \\
								& \enm 	& 2.647(19) & 2.815(26) & 3.708(16) & 3.788(16)	\\
								& \enl 	& 2.584(28) & 2.781(26) & 3.689(18) & 3.781(28) \\
			\\
			\colrule
			Ref. & \texttt{ID} & $\om$ & $\ocs$ & $\occs$ & $\occc$\\
			\colrule
			[\citen{Can:2015exa}] & \ensl & 1.790(17) & 2.837(18) & 3.819(10) & 4.769(6) \\
			\\
			\colrule
			Ref. & \texttt{ID} & $\Xi_c$ & $\Xi_c^\prime$ \\
			\colrule
			[\citen{Bahtiyar:2016dom}] & \ensl & 2.519(15) & 2.646(17) \\
			\\
			\colrule
			Ref. & \texttt{ID} & $\omc$ & $\ocs$ & $\xcc$ & $\xccs$ & $\occ$ & $\occs$ \\
			$[$\citen{PhysRevD.98.114505}$]$ & \ensl & 2.707(11) & 2.798(24) & 3.626(30) & 3.693(48) & 3.719(10) & 3.788(11) \\
			\botrule
		\end{tabular} \label{tab:masses_t2}
		}
	\end{table}

\end{document}